%% file: sph_final.tex
\documentclass{aa}

\usepackage{graphicx}
\usepackage{amssymb}
\usepackage{natbib}
\usepackage{amsmath}

\usepackage{bm}
\usepackage{epstopdf}
\usepackage{pgfplots}
\usepackage{subcaption}

\def\d{{\rm d}}
\let\vec\boldsymbol

\def\Qd{{Q_{\rm D}^\star}}

\long\def\cb#1\ce{}
\def\todo#1{{\bf TODO: #1}}

\def\planss{\ref@jnl{Planet.~Space~Sci.}}   

\long\def\combeg#1\comend{}

\begin{document}


\title{Impacts into rotating targets: angular momentum draining and efficient formation of synthetic families}
    \titlerunning{Impacts into rotating targets}

\author{P.~\v Seve\v cek$^{\rm a,\star}$
    \and
    M.~Bro\v z$^{\rm a}$ \and
    M.~Jutzi$^{\rm b}$ }

\institute{$^{\rm a}$ Institute of Astronomy, Charles University, Prague, V Hole\v sovi\v ck\'ach 2, 18000 Prague 8, Czech Republic \\
        \email{sevecek@sirrah.troja.mff.cuni.cz}
    \and
    $^{\rm b}$ Physics Institute, University of Bern, NCCR PlanetS, Sidlerstrasse 5, 3012 Bern, Switzerland}


    \abstract{%
    About 10\% of the observed asteroids have rotational periods lower than $P=3\,\rm h$
    and they seem to be relatively close to the spin barrier.
    Yet, the rotation has often been neglected in simulations of asteroid collisions. 
    To determine the effect of rotation, we perform a large number of SPH/N-body 
    impact simulations
    with rotating targets. We developed a new unified SPH/N-body code with self-gravity, 
    suitable for simulations of both fragmentation phase and gravitational reaccumulation. 
    The code has been verified against previous ones
    \citep{Benz_Asphaug_1994}, but we also tested new features, e.g.~rotational stability, 
    tensile stability, etc. Using the new code, we ran simulations with 
    $D_{\rm pb}=10\,\rm km$ and $100\,\rm km$ monolithic targets and compared  
    synthetic asteroid families created by these impacts with families
    corresponding to non-rotating targets. The rotation affects mostly cratering events 
    at oblique impact angles.
    The total mass ejected by these collision can be up to five  
    times larger for rotating targets.
    We further compute the transfer of the angular momentum and determine conditions under 
    which impacts accelerate or decelerate the target.
    While individual cratering collisions can cause both acceleration and deceleration,
    the deceleration prevails on average, collisions thus cause a systematic 
    spin-down of asteroid population.
}


    \maketitle





\renewcommand{\thefootnote}{\fnsymbol{footnote}}

\section{Asteroid collisions in the Main Belt}\label{sec:introduction}

The Main Belt of asteroids is a collisional system. 
The breakups of asteroids have been recorded in the form of asteroid families
\citep{Nesvorny_etal_2015, Vinogradova_2019}, we can also see impact features, such as craters or boulders, 
on surfaces of asteroids. These features can be observed directly during a satellite 
flyby or even with ground-based instruments \citep{Vernazza_2018,Fetick_2019}.


Physical processes during asteroid collisions are rather complicated for 
purely analytical estimates to yield precise quantitative results;
it is necessary to model a propagation of shock wave in the target, 
crack growth and consequent fragmentation, gravitational reaccumulation 
of ejected fragments, etc.
Fragmentation of targets due to hyper-velocity impacts has been 
studied using laboratory experiments \citep{Nakamura_Fujiwara_1991, Morris_2017, Wickham_Eade_2018}.
While the experiments can produce valuable constraints, the results cannot be 
directly compared with the breakups of asteroids, as the sizes of targets and 
kinetic energies of the impact would have to be extrapolated over several orders 
of magnitude. Experiments also do not take into account a gravitational reaccumulation.
The collisions of asteroids are therefore commonly studied using numerical methods;
the experiments then provide the calibration for the respective numerical codes.

Common methods for studying the collisions can be divided into particle-based 
\citep[for example the N-body code \texttt{pkdgrav}, see][]{Richardson_2000},
and shock-physics ones, such as mesh-based methods \citep[used by code \texttt{iSALE}, see][]{Suetsugu_2018} 
or the smoothed particle hydrodynamics
\citep{Jutzi_2015}, used in this work. This is a Lagrangian, grid-less method, which makes it suitable for impact simulations,
as the computational domain is not a priori known.
Especially in a hit-and-run impact, fragments of the projectile can travel to 
considerable distances from the target.
In SPH, the distant fragments do not require any special handling (they only might 
affect the performance of the code). 
SPH is also versatile, allowing to relatively easily implement new physics.
The model of fragmentation can be straightforwardly incorporated into SPH,
but it would be a difficult task for grid-based methods.

An outcome of a collision depends on a number of parameters, namely 
the diameter~$D_{\rm pb}$ of the target, 
the diameter~$d_{\rm imp}$ of the projectile,
the specific impact energy~$Q$, 
the impact angle~$\phi_{\rm imp}$, 
but also the rotational periods of the colliding 
bodies, their shapes, material properties, etc.
For completeness, we should also include parameters introduced by the 
numerical scheme, such as the spatial resolution, the time step, the artificial viscosity~$\Pi_{\rm AV}$, etc.
The extent of this parameter space prohibits a thorough analysis of every collision
as a function of all the mentioned parameters,
we thus have to restrict ourselves to a particular set of simulations, varying some parameters
and keeping the others constant.

Asteroidal targets have been considered non-rotating 
in most previous studies of asteroid families 
\citep{Durda_2007, Benavidez_2012, Sevecek_2017, Benavidez_2018, Jutzi_2019}. 
\cite{Jutzi_2013} considered a rotating Vesta, 
 rotating bodies have also been studied by \cite{Jutzi_Benz_2017}.
\cite{Cuk_2012} and \cite{Canup_2008} take the rotation into account for 
simulations of the Moon-forming impact, \cite{Canup_2005} for the 
impact event forming Pluto and Charon. 
\cite{Ballouz_2015} used an N-body code to simulate collisions of rotating
rubble-pile asteroids, \cite{Takeda_2007, Takeda_2009} studied 
the angular momentum transfer for both stationary and rotating rubble-piles.
In this work, we study a formation of asteroid families from monolithic targets,
extending the parameters 
of the simulation by the initial rotational period~$P_{\rm pb}$ of the target,
including cases close to the spin barrier.

\cb
While there are some indications that suggest the initial rotation of the target 
does not affect the outcome of the collision greatly and can be neglected without introducing 
a significant uncertainty, this assumption has not been tested to date. 
Studying the effect of the rotation of the target is the main goal of this work.

To estimate the importance of the initial rotation of the target, 
we can compute the ratio of the angular frequency $\omega_{\rm pb}$ of the target and 
angular velocity $\omega_{\rm imp}$ of the impactor with respect to the target:
\begin{equation}
    \frac{\omega_{\rm imp}}{\omega_{\rm pb}} \sim \frac{v_{\rm imp} P_{\rm pb} \sin \phi_{\rm imp}}{ D_{\rm pb}}\,,
\end{equation}
where $v_{\rm imp}$ is the impact speed, $P_{\rm pb}$ is the rotational period of the target, 
$\phi_{\rm imp}$ is the impact angle and $D_{\rm pb}$ is the diameter of the target.
This corresponds to the ratio of the timescale of rotation and the timescale of impact;
if $\omega_{\rm imp} \simeq \omega_{\rm pb}$, the target is rotated considerably during the impact
and cannot be considered stationary,
on the other hand if $\omega_{\rm imp} \gg \omega_{\rm pb}$, the target rotates 
only negligibly.
For a typical collision in the Main-belt with $v_{\rm imp} = 5\,\rm km/s$, 
$\phi_{\rm imp} = 45^\circ$, $P_{\rm pb} = 6\,\rm h$, 
$D_{\rm pb} = 10\,\rm km$, we get $\omega_{\rm imp}/\omega_{\rm pb} \sim 10^3$, 
suggesting that the target can be considered stationary during the fragmentation
for most considered collisional events.


Another quantity characterizing the impact is the ratio of the angular momentum $L_{\rm imp}$ 
transferred from the impactor into the target (or into the largest remnant, to be exact) 
and the initial angular momentum $L_{\rm pb}$ of the target:
\begin{equation}
    \frac{L_{\rm imp}}{L_{\rm pb}} \sim \gamma \frac{d_{\rm imp}^3 v_{\rm imp} P_{\rm pb}\sin \phi_{\rm imp}}{ D_{\rm pb}^4 }\,,
\end{equation}
where $\gamma$ is the non-dimensional effectivity of the momentum transfer. The coefficient
is not a constant, but rather a function of impact parameters $\gamma = \gamma(v_{\rm imp}, \phi_{\rm imp}, ...)$;
we determine the functional dependence of $\gamma$ later from the results of the simulation.

\todo{estimate the value}

From these ballpark estimates, it would seem that the initial rotation of the target is 
irrelevant the majority of realized collisions, but this conclusion might be a bit 
premature.
If the target rotates close to the critical frequency:
\begin{equation}
    \omega_{\rm crit} = \sqrt{\frac{4}{3}\pi G \rho}\,,
\end{equation}
even a low-energy impact cause ejection of fragments
that would otherwise be reaccumulated if the target was stationary.
We can therefore intuitively predict that the rotation of the target
will affect the outcome of the collision significantly if the 
angular frequency of the target is close to $\omega_{\rm crit}$.
On the other hand, the rotation will be less importance or even negligible 
for asteroids with frequency $\omega \ll \omega_{\rm crit}$.
Testing this prediction is the primary goal of this work.
\ce

The paper is organized as follows.
In Section~2, we describe our numerical model. 
Section~3 analyzes
differences between synthetic families created from parent bodies 
with various rotational periods.
Section~4 is focused on largest remnants, specifically 
on their accelerations or decelerations and the angular momentum transfer.
Finally, we summarize our results in Section~5. 

\section{Numerical model}
We developed a new SPH/N-body code. 
The code is publicly available, see Appendix~\ref{sec:code}.
In this section, we do not attempt to present a thorough review of the SPH method
\citep[as e.g.][]{Cossins_2010}, 
but instead we summarize the exact equations used in the code, 
emphasizing the modifications introduced in order to properly deal with rotating bodies.

\subsection{Set of equations}
\label{sec:equations}

The set of hydrodynamical equations is solved with a smoothed particle hydrodynamics \citep{Monaghan_1985}.
The continuum is discretized into particles co-moving with the continuum, with the density
profile of the particles given by a kernel $W$, which is a cubic spline in our case: 
\def\arraystretch{1.4}
\begin{equation}
    W(r, h) = \frac{\sigma}{h^3} \left\{\begin{array}{ll}
            \frac{1}{4}(2-q)^3-(1-q)^3\,, & 0\leq q < 1\,, \\
            \frac{1}{4}(2-q)^3\,, & 1\leq q < 2 \,, \\
            0  & q \geq 2 \,,
        \end{array}
            \right.
\end{equation}
where $q \equiv r/h$.

Below, we denote indices of particles with Latin subscripts (usually $i$, $j$, ...),
the components of vector and tensor quantities with Greek superscripts (usually $\alpha$, $\beta$, ...). We also use Einstein notation to sum over components (but not for particles, of course).

The equation of motion for $i$-th particle reads:
\begin{multline}
    \frac{\d v_i^\alpha}{\d t} = \sum_j m_j 
    \left( \frac{\sigma^{\alpha \beta}_i}{\rho_i^2} + \frac{\sigma^{\alpha \beta}_j}{\rho_j^2} + \Pi_{ij} \delta^{\alpha\beta} + \zeta^{\alpha\beta}_{ij} f^n\right) 
    \frac{\partial W_{ij}}{\partial x^\beta}  + \\ +\nabla \Phi
    -  \left[\vec\omega \times (\vec \omega \times \vec r_i)\right]^\alpha - \left[2\vec \omega \times \vec v_i\right]
    ^\alpha    \,,
    \label{eq:motion}
\end{multline}
where $\sigma^{\alpha\beta} = -P\delta^{\alpha\beta} + S^{\alpha\beta}$ is the total stress tensor,
$\Pi$ is the artificial viscosity \citep{Monaghan_Gingold_1983}, 
$\zeta^{\alpha\beta}f^n$ is the artificial stress \citep{Monaghan_2000},
$\Phi$ is the gravitational potential (including both external fields and self-gravity of particles),
and $\vec\omega$ is the angular velocity of the reference frame. 

The respective terms in the Eq.~(\ref{eq:motion}) correspond to the stress divergence, 
gravitational acceleration, centrifugal force and Coriolis force.
Inertial forces are only applied if the simulation is carried out in a non-inertial 
reference frame, corotating with the target asteroid (see Appendix~\ref{sec:inertial}).

We use the standard artificial viscosity with linear and quadratic
velocity divergence terms and coefficients $\alpha_{\rm AV}$ and $\beta_{\rm AV}$,
respectively. This term is essential for a proper shock propagation
and thus is always enabled in our simulations.
Optionally, it is possible to enable the Balsara switch \citep{Balsara_1995},
which reduces the artificial viscosity in shear motions
in order to reduce an unphysical angular momentum transfer.
Additionally, the code includes the artificial stress term $\zeta^{\alpha\beta}f^n$, 
which reduces the tensile instability, i.e.~an unphysical clustering of particles
due to negative pressure. We tested the effects of this term using the ``colliding
rubber cylinders'' test \citep[c.f.][]{Schafer_2016}.

We use a different discretization of the equation than in \cite{Sevecek_2017}, as we 
found the above equation to be more robust, avoiding a high-frequency oscillation 
in the pressure field. This is a recurring problem in high-velocity impact simulations
and while it can be suppressed by a larger kernel support, additional modifications of the 
method have been suggested to address the issue, 
for example the $\delta$-SPH modification \citep{Marrone_2011}.

The density is evolved using the continuity equation:
\begin{equation}
    \frac{\d \rho_i}{\d t} = \sum_j m_j \frac{ \partial v_i^\alpha}{\partial x^\alpha} \,. 
    \label{eq:continuity}
\end{equation}
We solve the evolution equation instead of direct summation of particle masses to avoid 
the artificial low-density layer at the surface of the asteroid \citep{Reinhardt_2017}.
The velocity gradient at the right-hand side of Eq.~(\ref{eq:continuity}) is computed as:
\begin{equation}
    \rho_i \frac{ \partial v_i^\alpha}{\partial x^\beta} = 
    \sum_j m_j (v_j^\alpha -  v_i^\alpha) \, C_i^{\beta\gamma} \frac{\partial W_{ij}}{\partial x^\gamma} \,,
\end{equation}
where the correction tensor $C^{\alpha\beta}$ is defined as \citep{Schafer_2016}:
\begin{equation}
    C_i^{\alpha\beta} \equiv \left[ \sum_j \frac{m_j}{\rho_j} (r_j^\alpha - r_i^\alpha) \frac{\partial W_{ij}}{\partial x^\beta} \right]^{-1}\,.
    \label{eq:correction_tensor}
\end{equation}
In case the bracketed matrix is not invertible, we use the Moore-Penrose pseudo-inverse
instead. The correction tensor is further set to 1 for fully damaged material.

The correction tensor has been introduced to tackle the linear inconsistency of 
the standard SPH formulation. It is a fundamental term in the velocity gradient 
that allows for a stable bulk rotation of the simulated body and significantly improves 
the conservation of the total angular momentum.

We evolve the internal energy $u$ using the energy equation:
\begin{multline}    
    {\d u_i\over\d t} = -{\sigma^{\alpha\beta}\over\rho_i}
     \frac{ \partial v_i^\alpha}{\partial x^\beta} 
     +
    \sum_j \frac{1}{2}m_j\Pi_{ij}(v^\alpha_i - v_j^\alpha) \frac{\partial W_{ij}}{\partial x^\alpha} 
    +\\\sum_j \frac{1}{2}m_j\zeta^{\alpha\beta}_{ij}f^n (v^\beta_i - v_j^\beta) \frac{\partial W_{ij}}{\partial x^\alpha}
    \,.
    \label{eq:energy}
\end{multline}
In this equation the velocity gradient is also corrected by the tensor $C^{\alpha\beta}$.
While this is required for a consistent handling of rotation, the inequality of kernel gradients used in
the energy equation~(\ref{eq:energy}) and in the equation of motion~(\ref{eq:motion}) implies 
the total energy is generally \emph{not} conserved in the simulations.
This is usually not an issue, as the total energy does not increase by more than 5\%.

However, in some cases (weak cratering impacts or exceedingly long integration time),
the energy growth can be prohibitive.
For such cases, the code also offers an alternative way to evolve the internal energy,
using a compatibly-differenced scheme \citep{Owen_2014}. Instead of computing the energy derivative,
the energy change is computed directly from particle 
pair-wise accelerations~$a_{ij}^\alpha$
and half-step velocities $w_i^\alpha = v_i^\alpha + \frac{1}{2} a_i^\alpha \Delta t$,
using the equation:
\begin{equation}
    \Delta u_i = \sum_j f_{ij} (w_j^\alpha - w_j^\beta) \, a_{ij}^\alpha \Delta t\,,
\end{equation}
where the energy partitioning factors $f_{ij}$ can be chosen arbitrarily,
provided they fulfil constraint $f_{ij} + f_{ji} = 1$.
With this form of SPH,
the total energy can be conserved to machine precision, at a cost of performance overhead.
However, this does not solve the inconsistency mentioned above.

The listed set of equations is supplemented by the Tillotson equation of state \citep{Tillotson_1962}.
To close the set of equation, we have to specify the constitutive equation.
We use the Hooke's law, evolving the deviatoric stress tensor in time using:
\begin{equation}
    {\d S^{\alpha\beta}_i\over\d t} = 2\mu\left(\frac{ \partial v_i^\alpha}{\partial x^\beta} - {{1\over3}}\delta^{\alpha\beta} \frac{\partial v_i^\gamma}{\partial x^\gamma} \right) \,,
\end{equation}
where $\mu$ denotes the shear modulus. To account for plasticity of the material, 
we further use the von Mises criterion, which reduces the deviatoric stress tensor by the factor:
\begin{equation}
    f = \operatorname{min}\left[ \frac{Y_0^2(1-u/u_{\rm melt})^2}{\frac{3}{2} S^{\alpha\beta} S^{\alpha\beta}}, 1 \right]\,,
\end{equation}
where $Y_0$ is the yield stress and $u_{\rm melt}$ is the specific melting energy. 
While more complex, pressure-dependent rheology models exist \citep{Jutzi_2015}, 
von Mises rheology is reasonable for monolithic asteroids 
and still consistent with laboratory experiments \citep{Remington_2018}.
The effects of friction have been studied by 
\cite{Jutzi_2015} or \cite{Kurosawa_Genda_2018};
 and we do not discuss such effect in this work.


Additionally, we integrate the fragmentation model to model an activation
of flaws and a propagation of fractures in the material. Following \cite{Benz_Asphaug_1994},
we define a scalar quantity damage $D$, modifying the total stress tensor as
\begin{equation}
    \sigma_{ D}^{\alpha\beta} = - \left(1-DH(-P)\right)P\delta^{\alpha\beta} + (1- D)S^{\alpha\beta}\,,
\end{equation}
where $H(x)$ is the Heaviside step function. A fully damaged material ($D = 1$)
has no shear nor tensile strength, it only resists to compressions.

Smoothing lengths of particles are evolved to balance the changes of particle concentration.
We thus derive the equation directly from the continuity equation:
\begin{equation}
    \frac{\d h_i}{\d t} = \frac{h_i}{3} \frac{\partial v_i^\alpha}{\partial x^\alpha}\,.
\end{equation}
Since it is also an evolution equation, we need to specify the initial conditions for smoothing lengths: 
\begin{equation}
    h =\eta \left( \frac{V}{N} \right)^{\frac{1}{3}}\,,
\end{equation}
where $V$ is the volume of the body, $N$ is the number of particles in the body 
and $\eta$ is a free non-dimensional parameter, which we set to $\eta = 1.3$,
corresponding to an average number of neighbouring particles $N_{\rm neigh} \simeq 65$.

\subsection{Gravity}
\label{sec:gravity}

Beside hydrodynamics, the code also computes accelerations of SPH particles
due to self-gravity. To compute it efficiently (albeit approximately),
we employ the Barnes-Hut algorithm
\citep{Barnes_Hut_1986}. Instead of computing pair-wise interactions of particles,
we first cluster the particles hierarchically and evaluate gravitational moments 
of particles in each node of the constructed tree. The accelerations are then computed by 
a tree-walk; if the evaluated node is distant enough, the acceleration can 
be approximated by evaluating the multipole moments up to the octupole order,
otherwise we descend into child nodes. 
The precision of the method is controlled by an opening radius~$r_{\rm open}$.
For an extensive description of the method, see \cite{Stadel_2001}.

As our SPH particles are spherically symmetric, they can be replaced by point masses,
provided they do not intersect each other (the corresponding kernel $W_{ij}$ is zero). 
However, it is absolutely necessary to account for softening of the gravitational potential for 
neighbouring particles. We follow \cite{Cossins_2010} by introducing a gravitational softening kernel 
$\phi$ (associated with the SPH smoothing kernel $W$) using the equation:
\begin{equation} 
    \frac{\partial\phi}{\partial r} = \frac{4\pi}{r}\int\limits_0^r r'^2 W(r')\,\d r' + \frac{h}{r^2}\,.
\end{equation}
The gravitational kernel $\phi$ corresponding to our M4 spline kernel $W$ is then:

\def\arraystretch{1.4}
\begin{equation}
    \phi(r, h) = \left\{\begin{array}{ll}
    \frac{2}{3} q^2 - \frac{3}{10} q^4 + \frac{1}{10} q^5 - \frac{7}{5} \,,& 0 \leq q < 1 \,,\\
    \frac{4}{3} q^2  - q^3  + \frac{3}{10} q^4  -\frac{1}{10} q^5 
        \\ \hskip50pt -\frac{8}{5} + \frac{1}{15q} \,,&  1 \leq q < 2 \,,\\
        -\frac{1}{q} \,, &  q \geq 2\,,
    \end{array}
    \right.
    \label{eq:softening_kernel}
\end{equation}
where $q \equiv r/h$. This kernel does not have a compact support, though.


\subsection{Temporal discretization}
Using derivatives computed at each time step, the equations are integrated 
using explicit timestepping. The scheme used in this work is the predictor-corrector,
however other schemes are implemented in the code, such as the leapfrog, 4th order Runge-Kutta
or Bulirsch-Stoer.

The value of the time step is determined by the CFL criterion:
\begin{equation}
    \Delta t \leq C_{\rm CFL} \min_i \frac{h_i}{c_{\rm s}}\,,
    \label{eq:CFL}
\end{equation}
where $h_i$ is the smoothing length of $i$-th particle, $c_{\rm s}$ the local
 sound speed and $C_{\rm CFL}$ is the Courant number.
In our simulations, we usually use $C_{\rm CFL}=0.25$, as higher values can lead to 
instabilities in some cases. Moreover, we restrict the time step by 
the value-to-derivative ratio of all time-dependent quantities in the simulation
to control the discretization errors.
The upper limit of the time step is therefore:
\begin{equation}
    \Delta t \leq C_{\rm d} \frac{| x |+s_{x}}{|\dot x| }\,,
    \label{eq:derivative_criterion}
\end{equation}
where $s_x$ is a parameter with the same dimensions as $x$, assigned to each quantity
in order to avoid zero time step if the quantity $x$ is zero. 
Constant $C_{\rm d}$ is 0.2 for all quantities.


\subsection{Equilibrium initial conditions}
Setting up the initial conditions for the impact simulation is not a trivial task.
It is necessary to assign a particular value of the density $\rho$, internal energy $u$
and deviatoric stress tensor $S$ to each particle, so that the configuration
is stable when the impact simulation starts and the particles do not oscillate.

This problem is not restricted to simulations with rotating targets.
A proper handling of initial conditions is essential in simulations of 
the Moon formation, collisions of planetary embryos, etc. 
If neglected, the initial gravitational compression would introduce 
macroscopic radial oscillations in the target. 

For small and stationary asteroids 
with $D\simeq 10\,\rm km$, the self-gravity is much less important, 
if fact it is often completely neglected during 
the fragmentation phase \citep[as in][]{Sevecek_2017}. These asteroids
are assumed undifferentiated, hence it is reasonable to set up  
homogeneous bulk density $\rho_0 = \rm const$.
For these stationary targets, the stress tensor of 
particles is zero in equilibrium.

For rotating targets, however, such initial conditions are unstable due to 
the emerged centrifugal force (in the corotating frame). 
To prevent any unphysical fractures in the target,
the configuration of particles has to be set up carefully,
especially for asteroids rotating close to the critical spin rate.
For this reason, we run a stabilization phase before the actual 
impact simulation, with an artificial damping of particle velocities:
\begin{equation}
    \vec v_{\rm damp} = \frac{\vec v - \vec \omega \times \vec r}{ 1 + \delta \Delta t } + 
    \vec \omega\times \vec r\,,
\end{equation}
where $\vec v$ is the undamped velocity, $\vec\omega$ the angular frequency of the target,
$\vec r$ the position vector of the particle, $\delta$ an arbitrary damping coefficient
(gradually being decreased during the stabilization phase) and $\Delta t$ the actual time step.
In this equation, we need to subtract and re-add the bulk rotation velocity, otherwise the damping would cause 
the target to slow down.
We also do not integrate the fragmentation model during this phase, as the oscillations 
of the particles might damage the target prematurely.

While more general approaches for setting up the initial conditions exist 
\citep{Reinhardt_2017}, the presented method is simple, robust and sufficient 
for our purposes. A disadvantage of our method is a significant computational 
overhead, as for some simulations the time needed to converge into a stable solution is 
comparable to the duration of the actual impact simulation. However, here we 
perform many simulations with fixed target diameter $D_{\rm pb}$ and period $P$, 
so we have to precompute the initial conditions only once and then use the cached particle
configuration for other runs.

\subsection{Reaccumulation phase}

As our numerical model contains both the hydrodynamics and the gravitational interactions,
it could be used for the \emph{entire} simulation --- from the stabilization, pre-impact flight,
fragmentation phase  
until the gravitational reaccumulation of all fragments.
However, the time step is often severely limited by the CFL criterion
(Eq. \ref{eq:CFL}).

We can increase the time step by several orders of magnitude and hence speed up
the simulation considerably by changing the numerical scheme from SPH to N-body
integration. This is a common approach in studies of asteroid 
families \citep{Durda_2007, Michel_2015, Sevecek_2017, Jutzi_2019};
once the target is fully fractured and the fragments start 
to recede, we convert all SPH particles into hard spheres
and replace the complexity of hydrodynamic equations with a simple collision detection. 
This allows us to overcome the time step limitation.


We can further optimize the simulation by merging the collided particles into larger spheres.
By doing so, we lose information about the shape of the fragments;
to preserve the shapes, it is necessary to either form rigid aggregates of particles 
instead of mergers \citep{Michel_Richardson_2013} or simulate the entire 
reaccumulation using the SPH \citep[as in][]{Sugiura_2018}. Here, we are mainly interested
in distribution of sizes and rotational periods, merging the particles is thus a viable option.

Merging does not only affect shape, but also dynamics of fragments.
As it modifies the moment of inertia, the merger has generally different rotational period 
than a real non-spherical fragment would have. Merging also removes higher gravitational moments,
thus altering motion of near fragments. This is a slight limitation of the presented model.

Hard spheres are created directly from SPH particles. Their mass is unchanged, 
the radius of the formed spheres is computed as:
\begin{equation}
    r_i = \sqrt[3]{\frac{3m_i}{4\pi\rho_i}} \,,
    \label{eq:handoff}
\end{equation}
so that the volume of spheres is equal to the volume of SPH particles.
As the total volume is conserved, created spheres inevitably overlap;
Appendix~\ref{sec:overlaps} describes how the code handles such overlaps.

When two spheres collide, they are merged only if their relative speed
is lower than the mutual escape speed:
\begin{equation}
    v < v_{\rm esc} \equiv \sqrt{\frac{2 G (m_i + m_j)}{r_i + r_j}}
    \label{eq:esc_velocity}
\end{equation}
and if the rotational angular frequency of the merger does not exceed 
the critical frequency:
\begin{equation}
    \omega < \omega_{\rm crit}  \equiv \sqrt{\frac{G  (m_i + m_j)}{r_{\rm merger}^3}}\,.
    \label{eq:critical_freq}
\end{equation}
This way we prevent a formation of unphysically fast rotators.
If any of these conditions is not fulfilled, particles undergo 
an inelastic bounce. The damping of velocities is determined by 
the normal $\eta_{\rm n}$ and tangential $\eta_{\rm t}$ coefficient of restitution, 
which we set to $0.5$ and $1$, respectively.

When merging the particles, we determine the mass, radius, velocity and angular frequency 
of the merger, so that the total mass, volume, momentum and angular momentum is conserved.
As the tangential components of velocities are not damped by a bounce, 
merging is the only way to spin up fragments in our simulations.


\section{Synthetic families created from rotating targets}
\label{sec:sfd}

\def\snapshot#1{%
     \includegraphics[width=0.1111\textwidth]{\detokenize{#1}}%
}
\def\xlabel#1{%
    \hfil\hbox to 20pt{#1\,min}\hfil%
}%

\begin{figure*}
    \hbox to \textwidth{%
   \snapshot{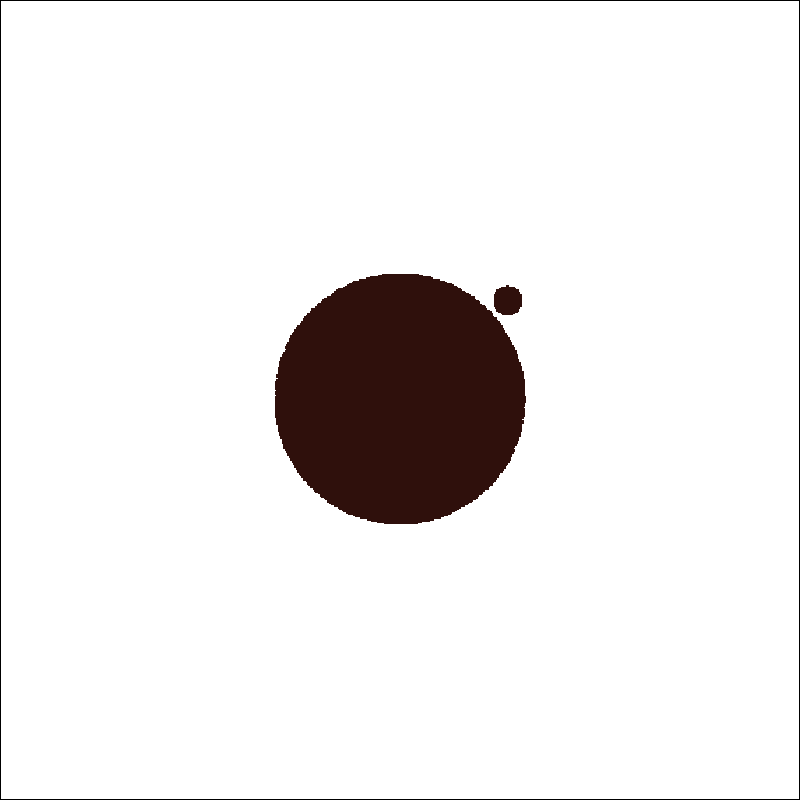}%
   \snapshot{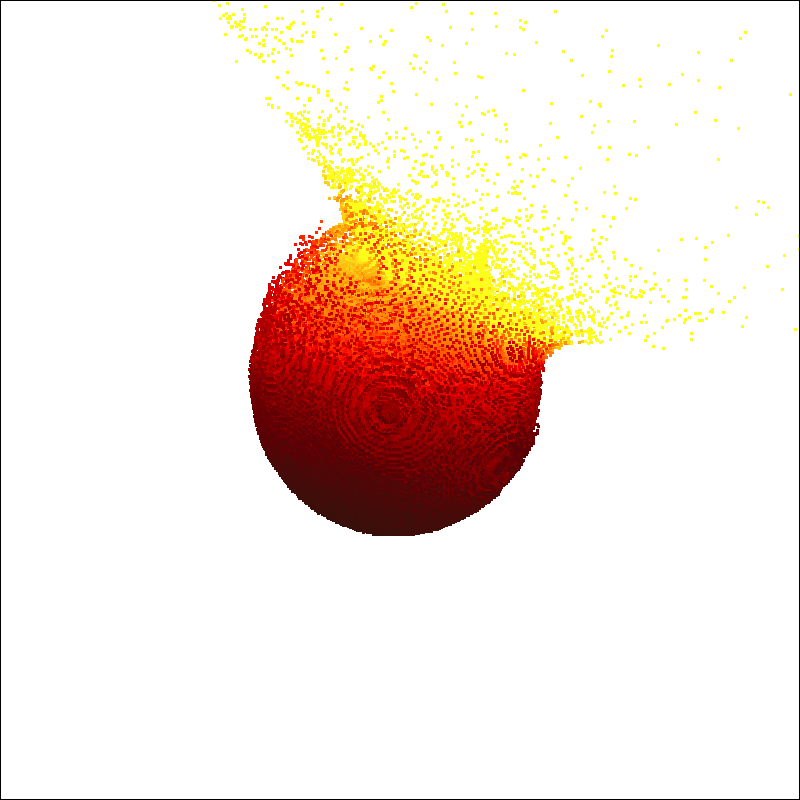}%
   \snapshot{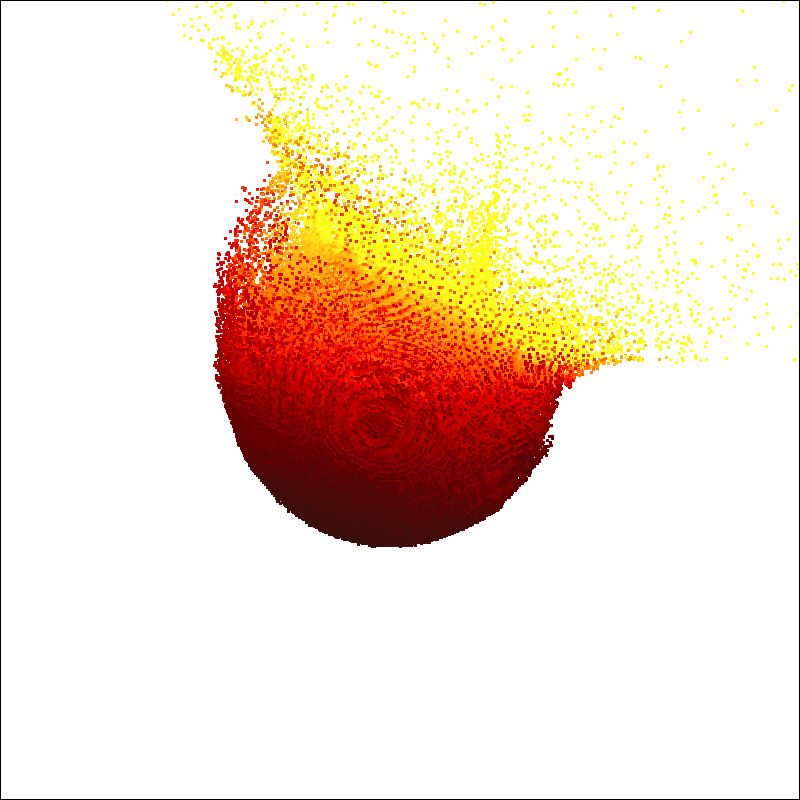}%
   \snapshot{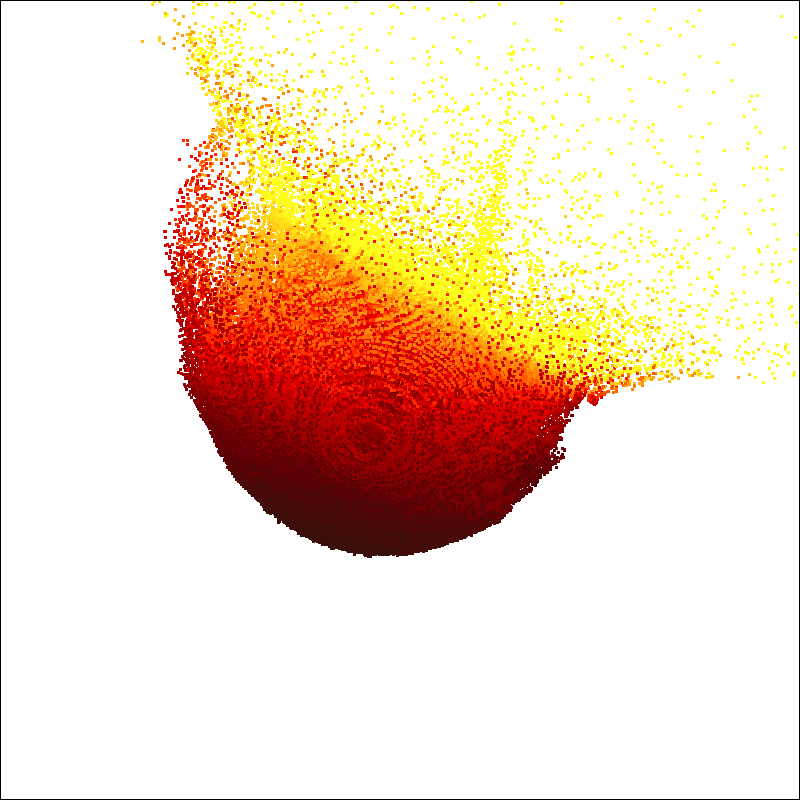}%
   \snapshot{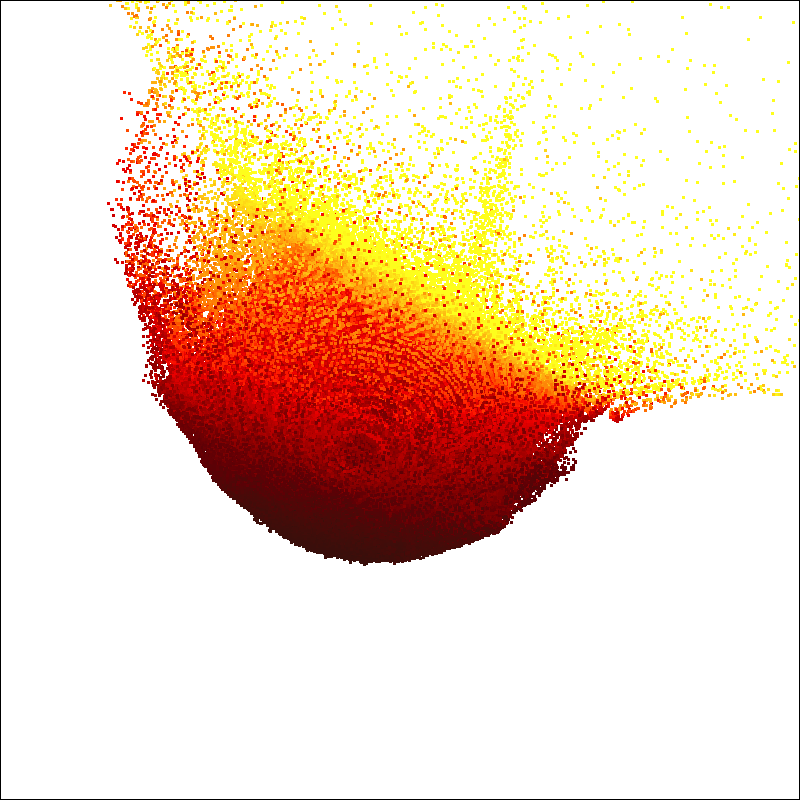}%
   \snapshot{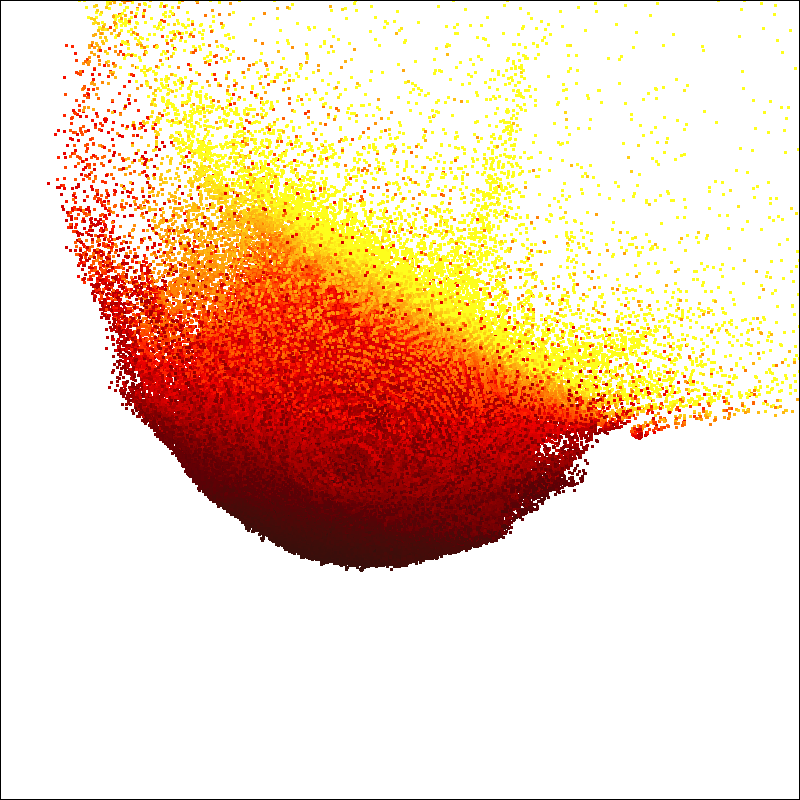}%
   \snapshot{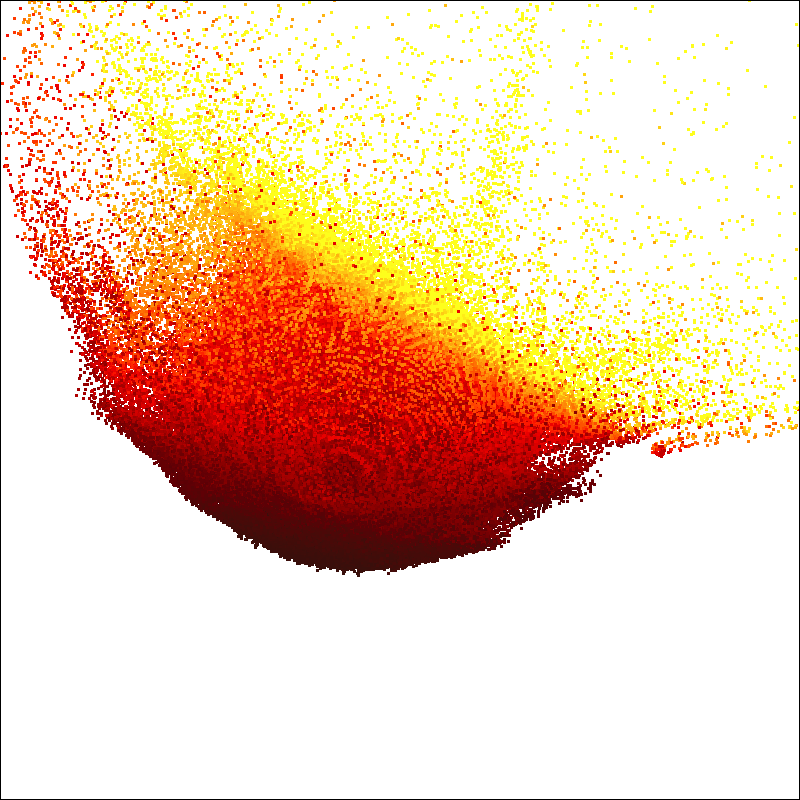}%
   \snapshot{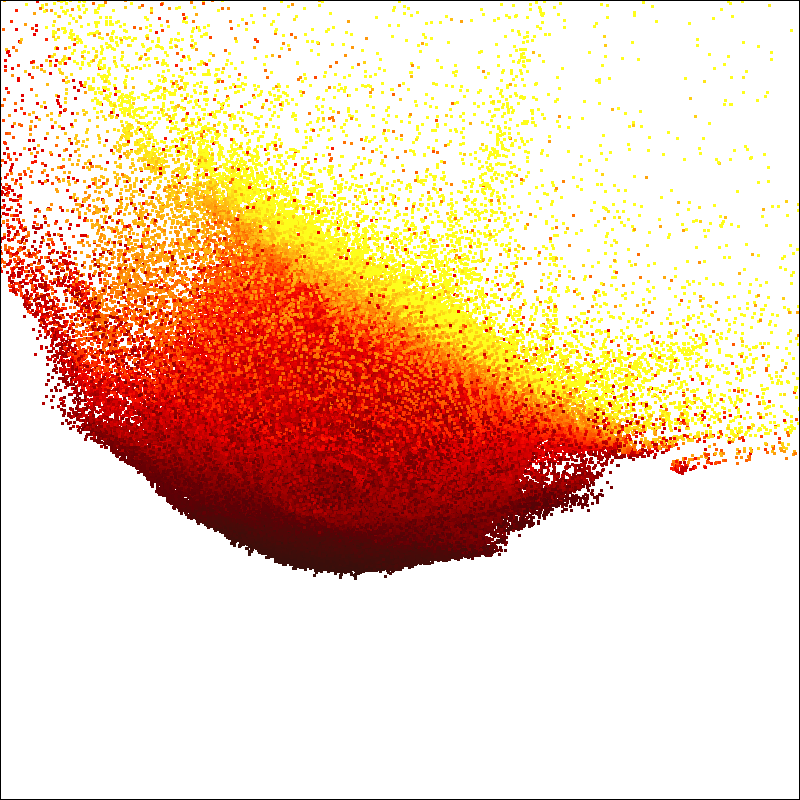}%
   \snapshot{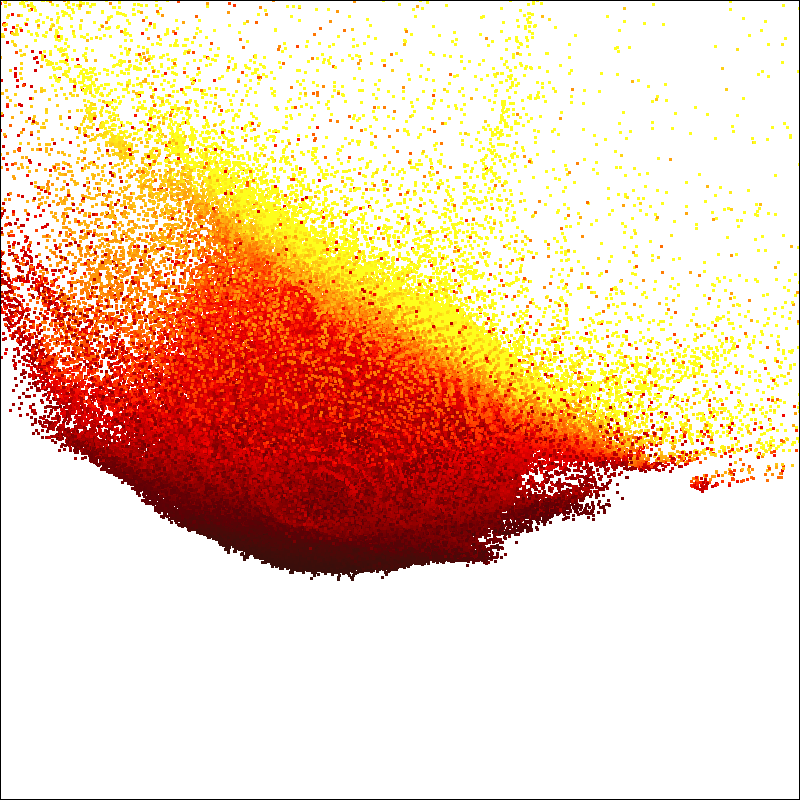}%
    }
    \lineskip=0pt%
    \hbox to \textwidth{%
   \snapshot{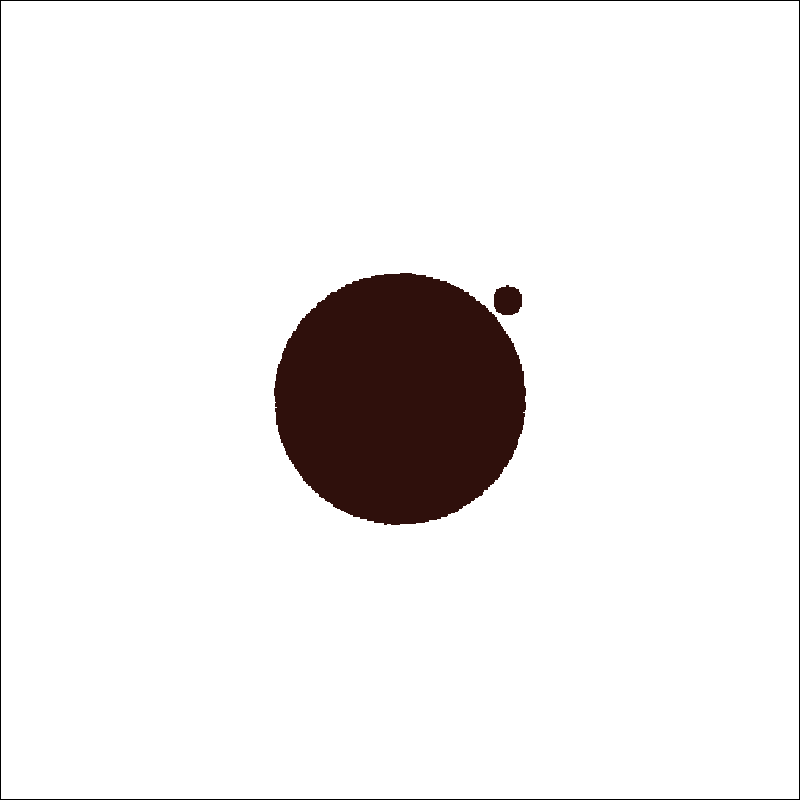}%
   \snapshot{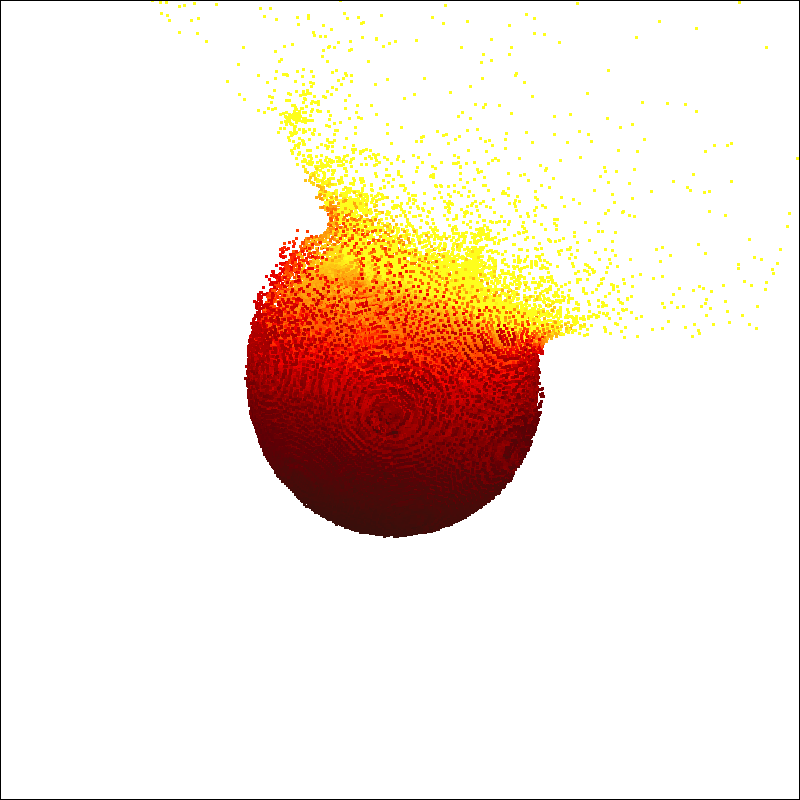}%
   \snapshot{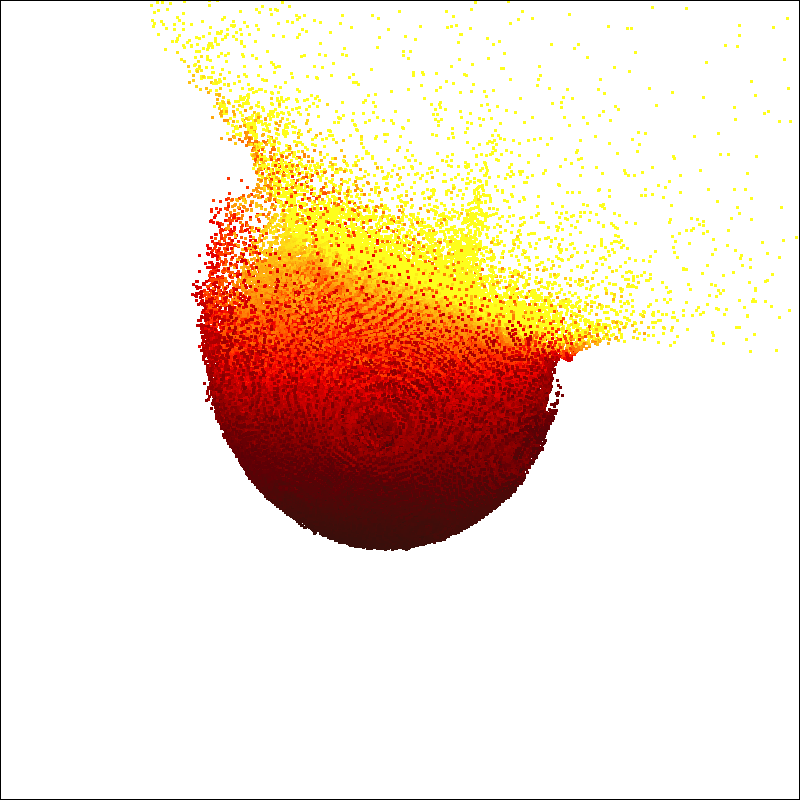}%
   \snapshot{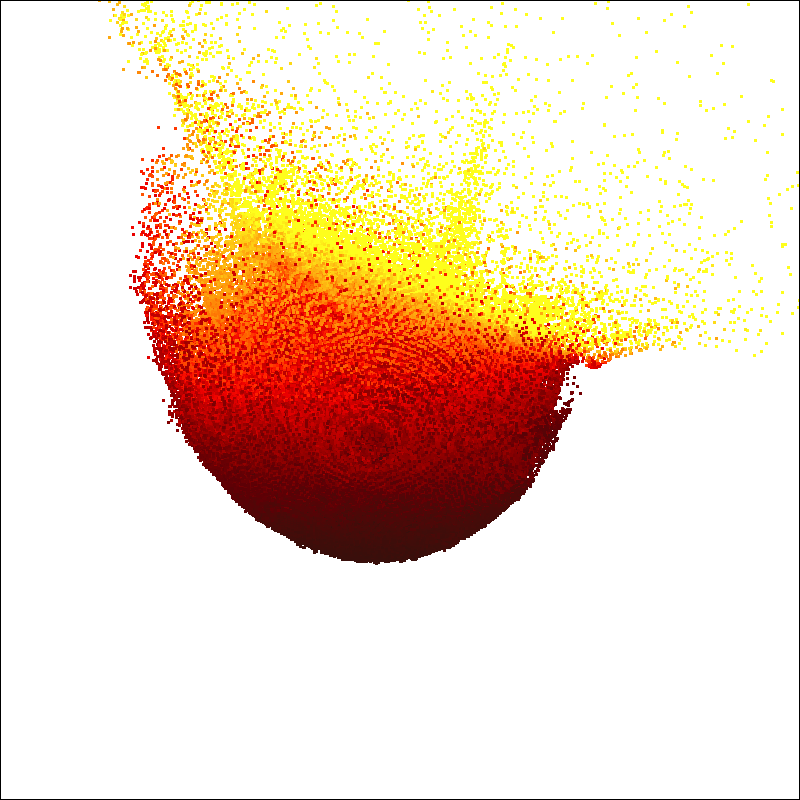}%
   \snapshot{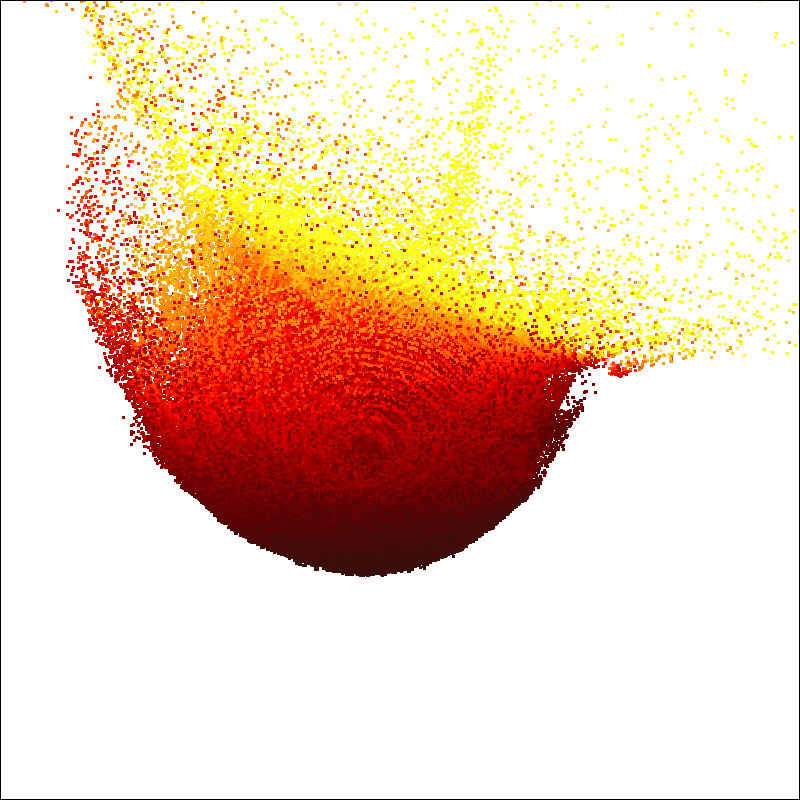}%
   \snapshot{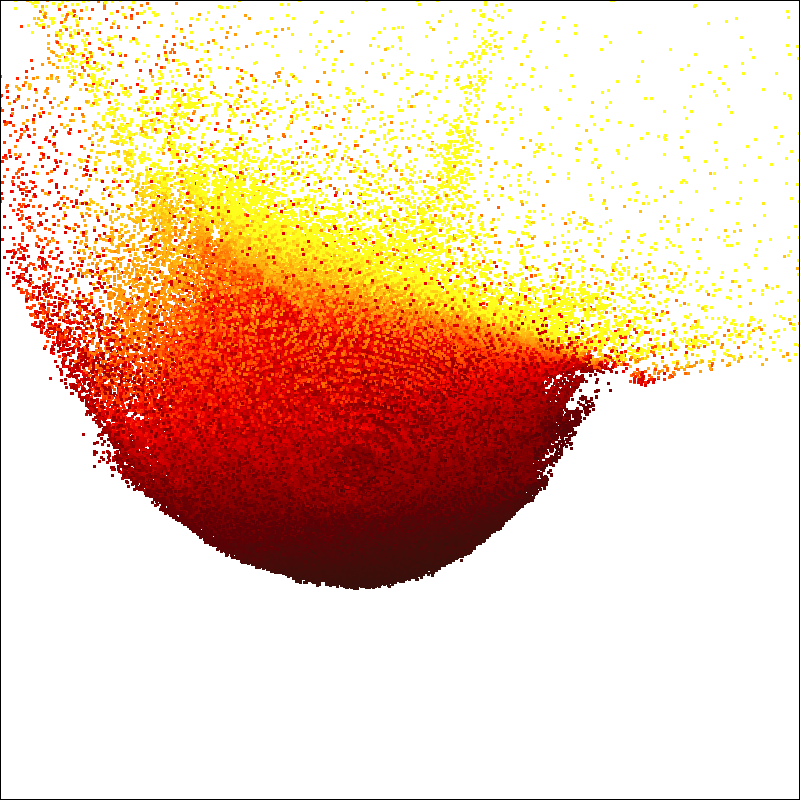}%
   \snapshot{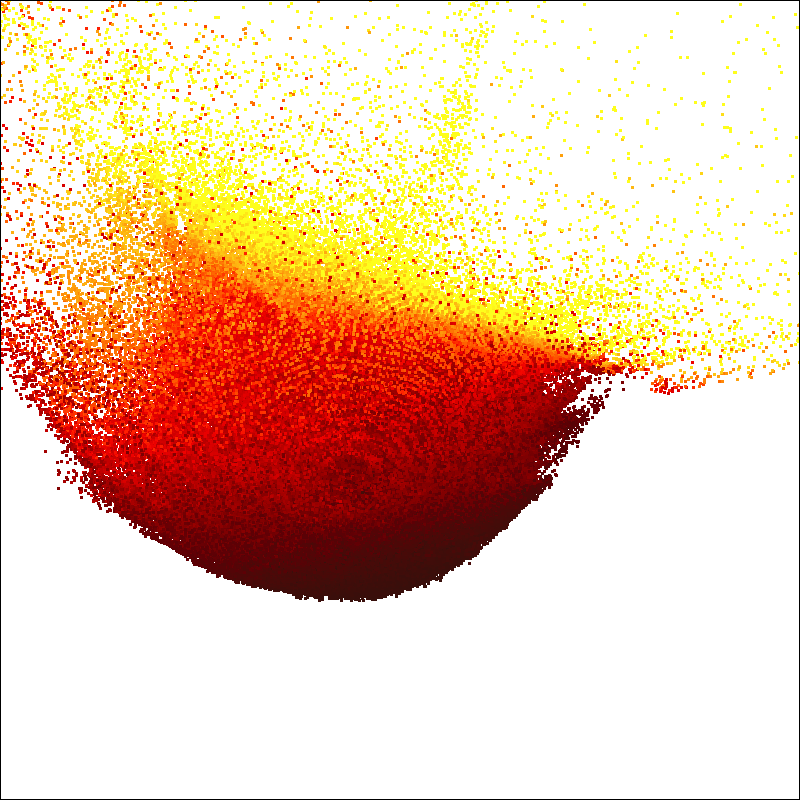}%
   \snapshot{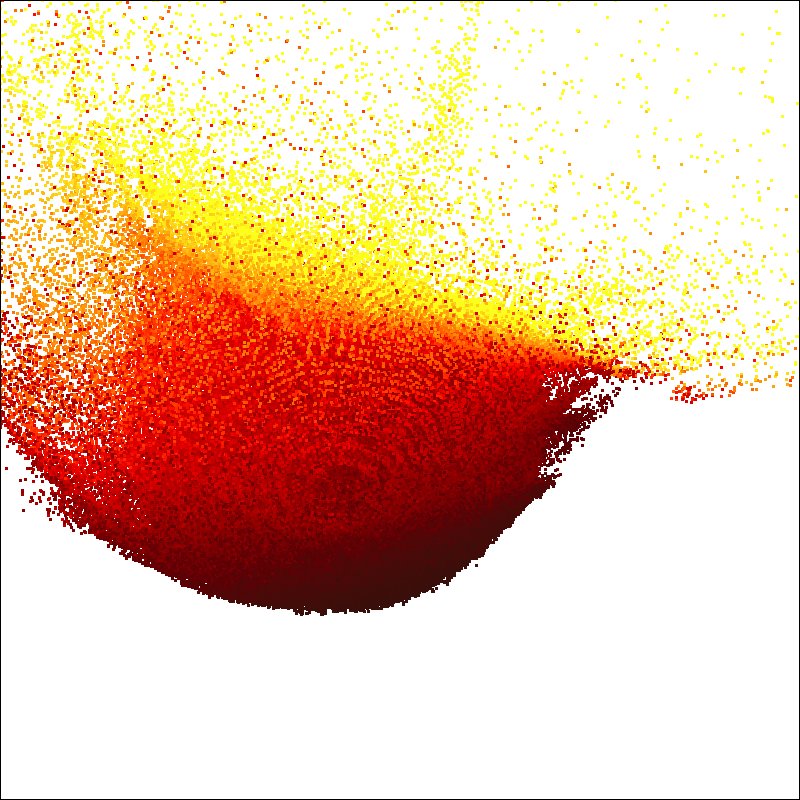}%
   \snapshot{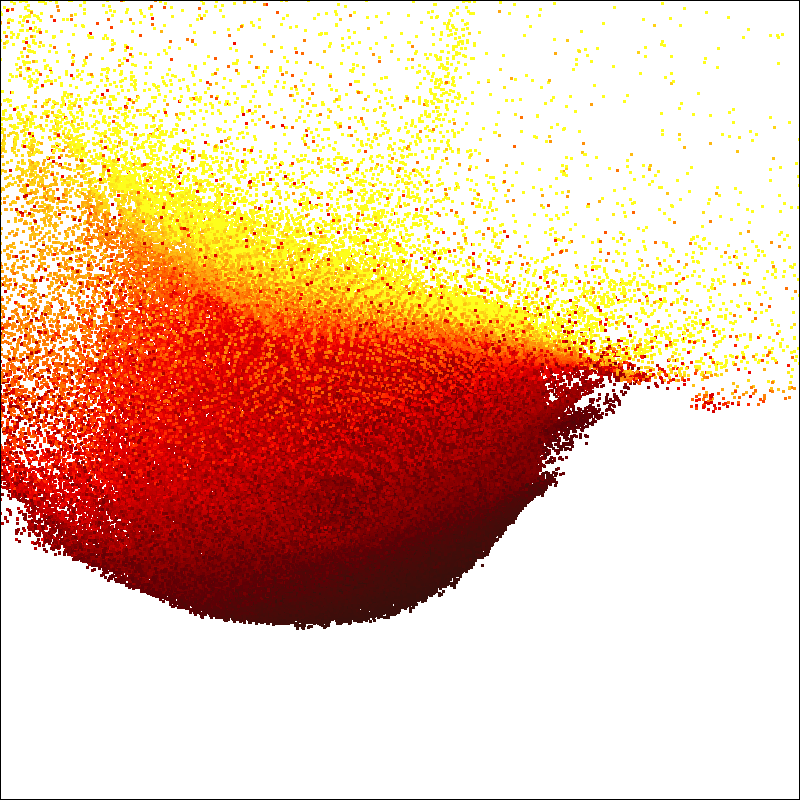}%
    }
    \hbox to \textwidth{%
   \snapshot{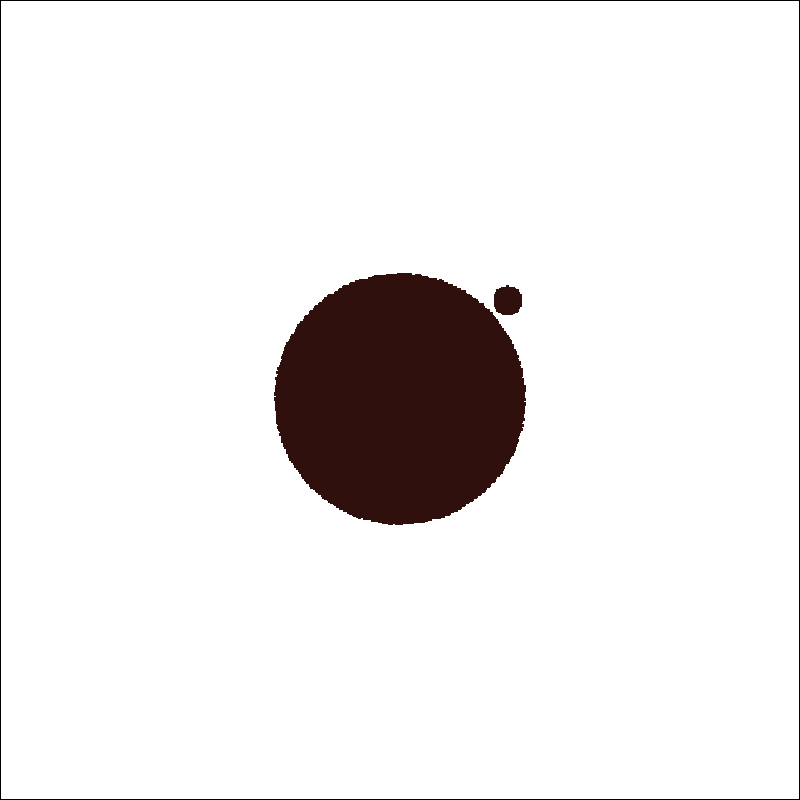}%
   \snapshot{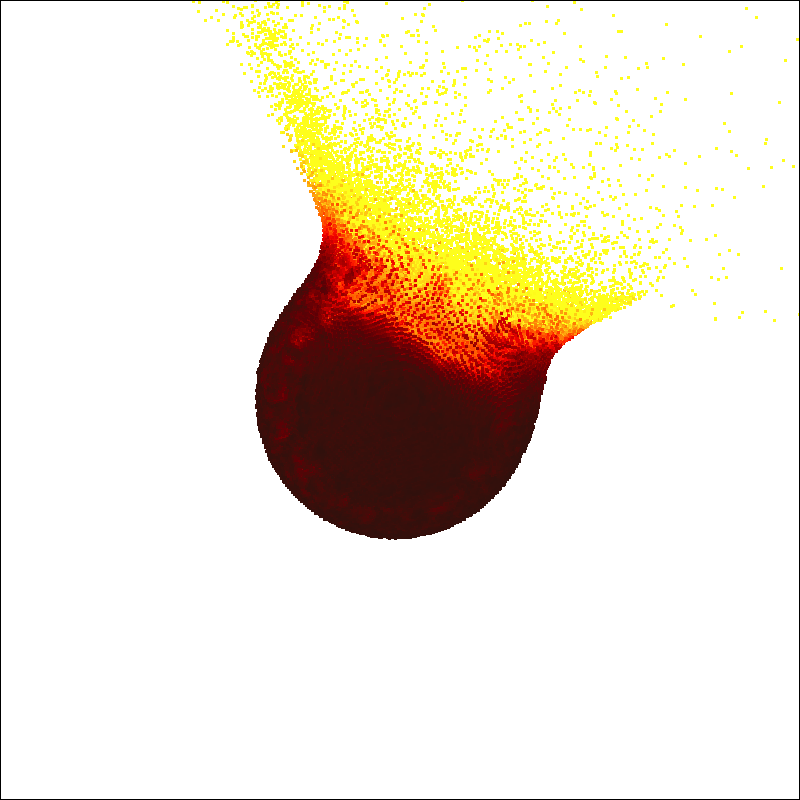}%
   \snapshot{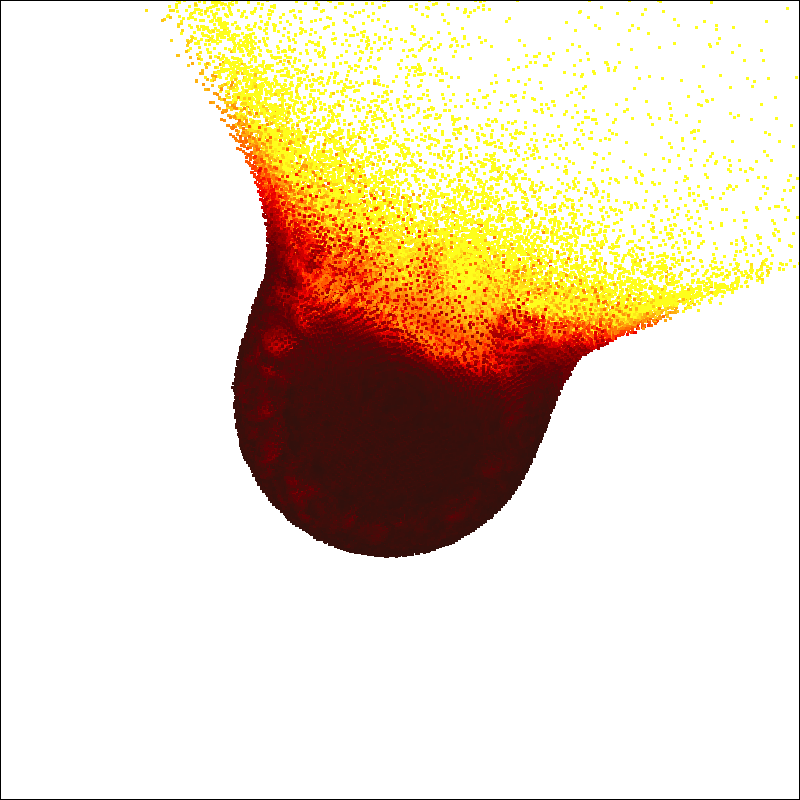}%
   \snapshot{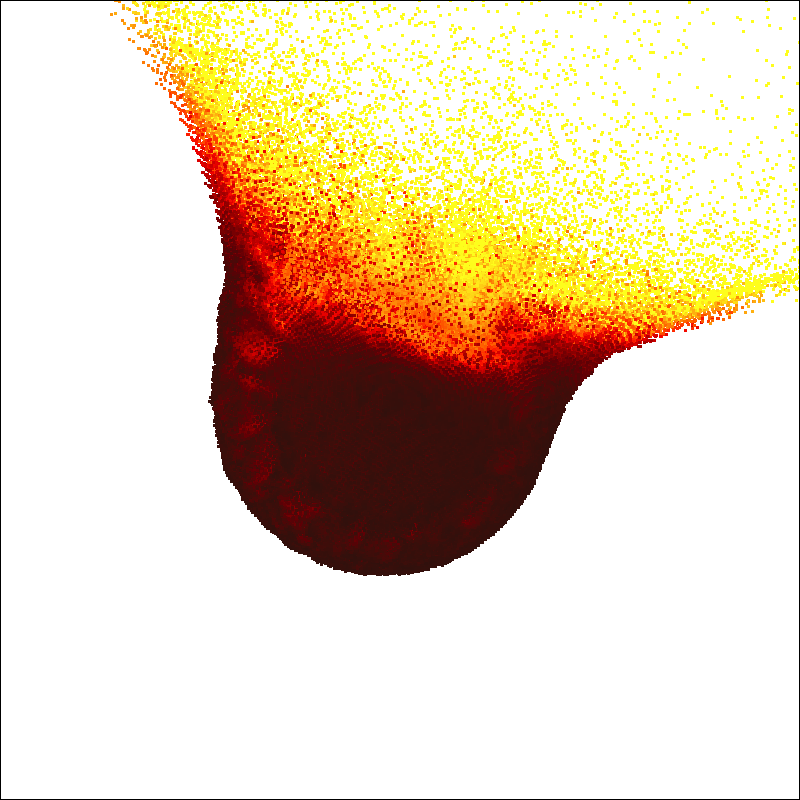}%
   \snapshot{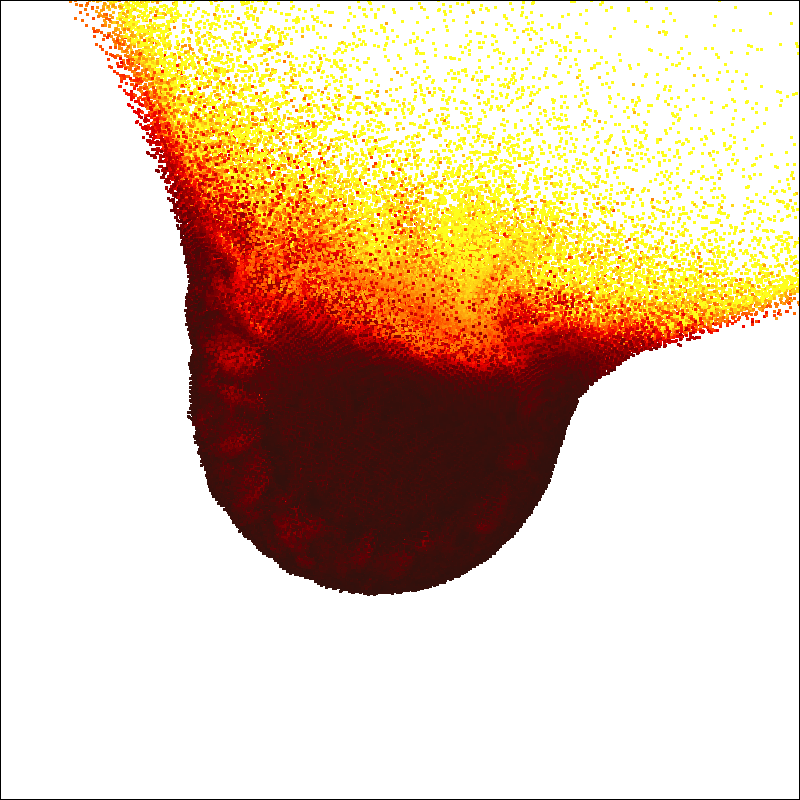}%
   \snapshot{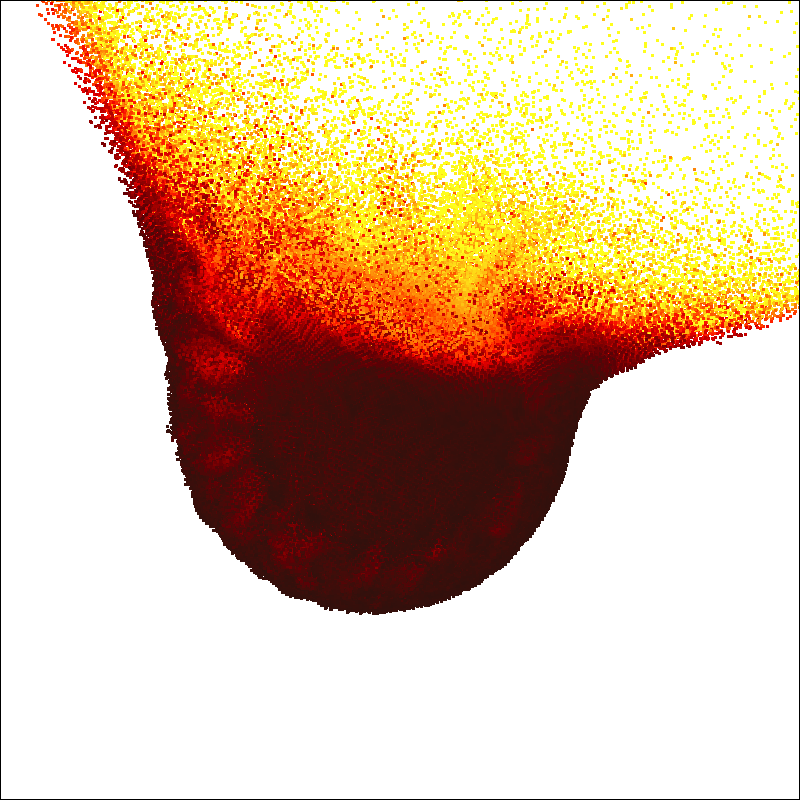}%
   \snapshot{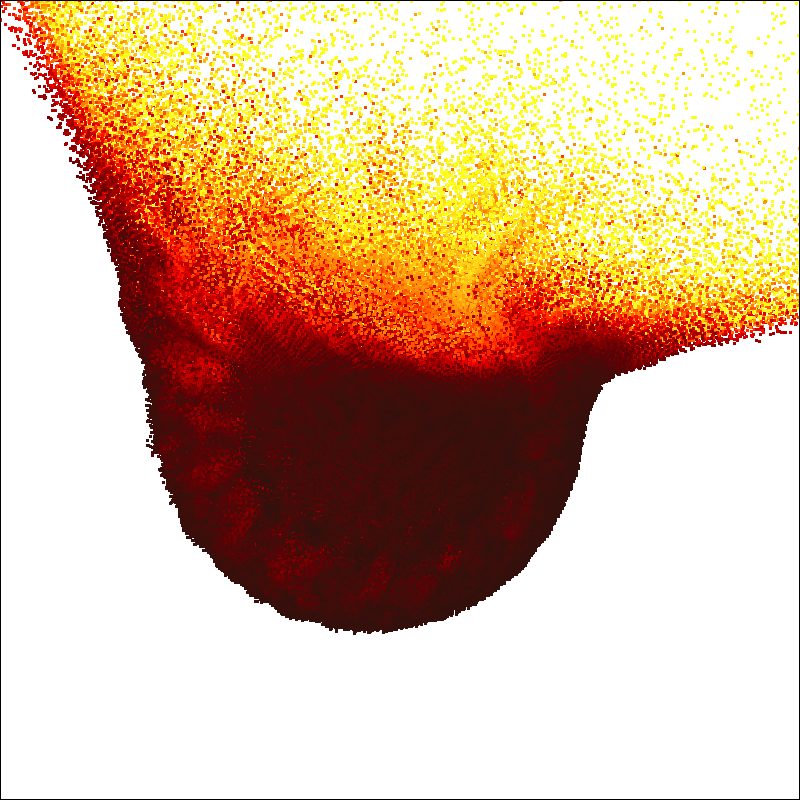}%
   \snapshot{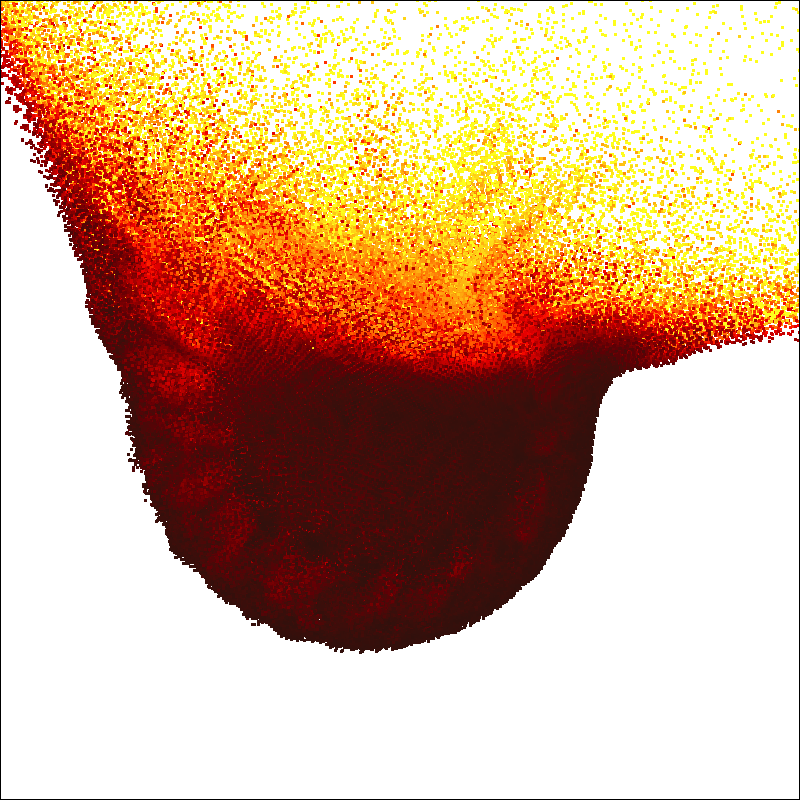}%
   \snapshot{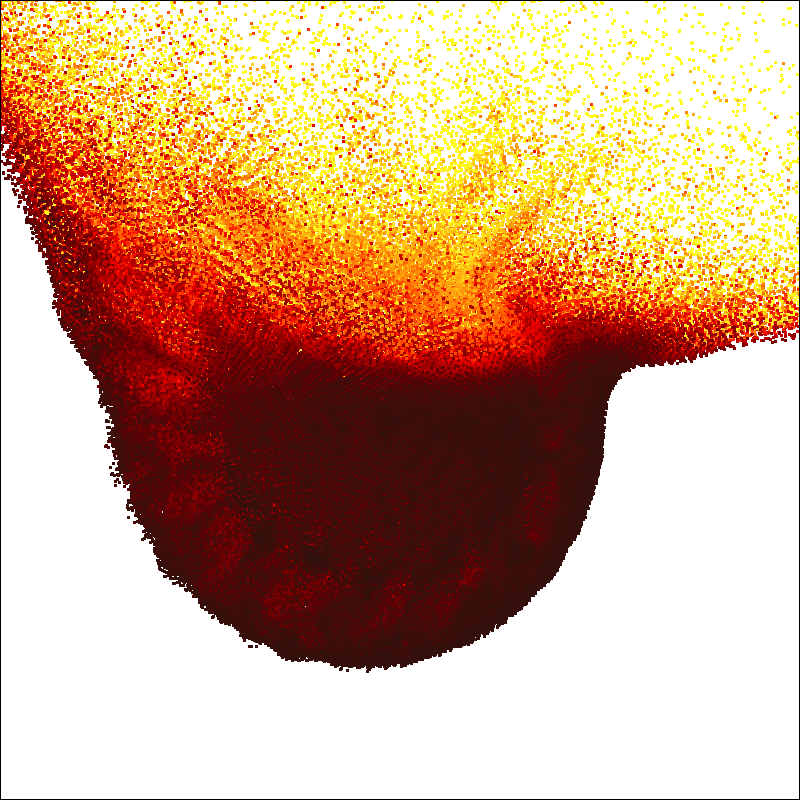}%
    }
    \hbox to \textwidth{\footnotesize
    \xlabel{0}\xlabel{3}\xlabel{6}\xlabel{9}\xlabel{12}\xlabel{15}\xlabel{18}\xlabel{21}\xlabel{24}%
    }
    \caption{Impacts into $D_{\rm pb}=100\,\rm km$ targets 
        without rotation (first row), rotational period $P_{\rm pb}=3\,\rm h$ (second row) and 
    $P_{\rm pb}=2\,\rm h$ (third row). 
    The period of two hours is approximately the critical period of the target.
    The color brightness corresponds to specific internal energy 
    of the particles.}
    \label{fig:timelapse}
\end{figure*}

To better understand how the rotation influences impact events, 
we decided to compute a matrix of simulations for various impact parameters.
We ran simulations for two different target sizes, $D_{\rm pb}=10\,\rm km$ and 
$D_{\rm pb}=100\,\rm km$, in order to ascertain the scaling of the rotational effect.
We tested head-on impacts, having the impact angle $\phi_{\rm imp}=15^\circ$, 
intermediate cases with $\phi_{\rm imp}=45^\circ$ and oblique impacts with $\phi_{\rm imp}=75^\circ$.
We have to further distinguish \emph{prograde} events, 
i.e.~impacts where the orbital velocity has 
the same direction as the impact velocity, and \emph{retrograde} events, 
where the orbital velocity has the opposite direction.
In the following, the prograde impacts have positive values of impact angles,
while the retrograde impacts have negative.

In all of our simulations, we set the impact velocity to $v_{\rm imp} = 5\,\rm km/s$,
which is close to the mean velocity for Main-belt collisions \citep{Dahlgren_1998}.
The simulation matrix covers both the cratering and the catastrophic events;
we ran simulations with relative impact energies $Q/\Qd = 0.1, 0.3, 1 \mbox{ and } 3$,
where the critical energy $\Qd$ is given by the scaling law of \cite{Benz_Asphaug_1999}.
As~$\Qd$ is defined as the specific impact energy (relative to mass of the target) required  
to eject 50\% of the target's mass as fragments, it necessarily depends on 
the rotational period of the target. However, we consider $\Qd$ to be independent 
of rotation and use the same value for all performed simulations, 
as it provides a convenient dimensionless measure of the impact energy.


We assume that both the target and the impactor are monolithic bodies, 
the material parameters are summarized in Table~\ref{tab:constants}.
The spatial resolution of the target was approximately $N=500{,}000$ SPH particles, 
the number of projectile particles was selected to match the particle density of the target. Three simulations with different periods $P_{\rm pb}$ 
are shown in Fig.~\ref{fig:timelapse}.


\begin{table}[t]
    \centering
    \renewcommand{\arraystretch}{1.1}
\begin{tabular}{ll}
    \multicolumn{2}{c}{Material parameters}\\
    \hline
    density at zero pressure & $\rho = 2700\,\rm kg/m^3$ \\
    bulk modulus & $A = 2.67\times 10^{10} \,\rm Pa$ \\
    non-linear Tillotson term & $B =  2.67 \times 10^{10}\,\rm Pa$ \\
    sublimation energy & $u_0 = 4.87 \times 10^8\,\rm J/kg$ \\ 
    energy of incipient vaporization & $u_{\rm iv} = 4.72\times 10^6\,\rm  J/kg$ \\
    energy of complete vaporization  & $u_{\rm cv} = 1.82 \times 10^7\,\rm J/kg$ \\
    shear modulus & $\mu = 2.27\times 10^{10}\,\rm Pa$ \\
    von Mises elasticity limit & $Y_0=  3.50\times 10^{9} \,\rm Pa$\\
    melting energy & $u_{\rm melt} = 3.4 \times 10^6 \,\rm J/kg$\\
    Weibull coefficient & $k = 4.00\times 10^{29}$ \\
    Weibull exponent & $m = 9$ \\
\end{tabular}
    \caption{Material parameters used in simulations.}
    \label{tab:constants}
\end{table}


\subsection{Coordinate system}
\label{sec:coords}
Due to the rotation, the impact geometry is more complex compared to the stationary case,
where it was determined by a single parameter --- the impact angle $\phi_{\rm imp}$ between
the normal at the impact point and the velocity vector of the impactor.
To describe the impact into a rotating target, we first define a coordinate system of the simulations.
We place the target at origin with zero velocity, 
the impactor has velocity $[-v_{\rm imp}; 0; 0]$ and its position in $x$-$y$ plane is given 
by $\phi_{\rm imp}$; specifically:
$$\vec r_0 = \left[ \begin{array}{l}
    x_0 + 0.5 (D_{\rm pb} + d_{\rm imp})\cos \phi_{\rm imp}, \\
    0.5 (D_{\rm pb} + d_{\rm imp})\sin \phi_{\rm imp}, \\
    0
\end{array}
    \right]\,,$$
where $x_0$ is the distance of the impactor from the impact point. 
These initial conditions have a mirror symmetry in~$z$.

The rotation vector $\vec\omega_{\rm pb}$ of the target adds additional three free parameters into 
the simulation setup. Generally, the vector does not have to be aligned with any coordinate axis.
We reduce the number of free parameters and thus simplify the analysis 
by aligning the vector with $z$-axis,
meaning we only consider impacts in the equatorial plane of the target.
We expect these impacts will be affected by the rotation the most,
as the centrifugal force is the largest on the equator.
Furthermore, the angular momentum of the target is aligned with 
the angular momentum of the impactor, we can thus expect the largest 
changes of the angular momentum. These expectations have been confirmed
for rubble-pile bodies by N-body simulations of \cite{Takeda_2009}.

\subsection{Size-frequency distributions for $D_{\rm pb}=10\,\rm km$ targets}
\label{sec:sfd_10km}

\begin{figure*}[t]
    \centering
    \includegraphics[width=\textwidth]{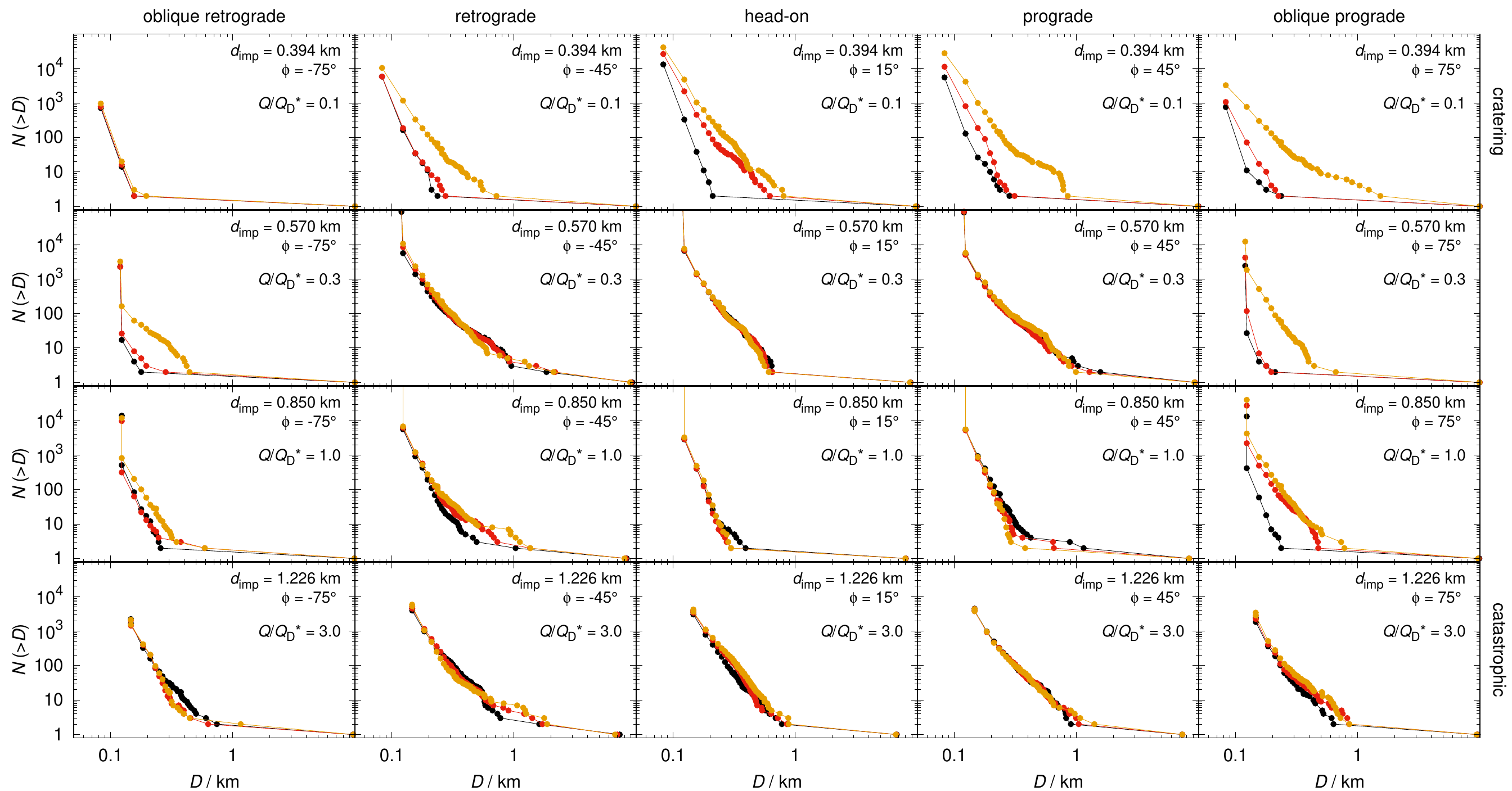}
    \caption{Cumulative size-frequency distributions $N(>D)$ of the synthetic families
    for $D_{\rm pb}=10\,\rm km$ targets.
    Stationary targets are plotted in black, red and yellow plots correspond to targets 
    with rotational period $P=3\,\rm h$ and $P=2\,\rm h$, respectively. 
    Columns 3 to 5 of the plot show prograde impacts (positive impact angles), columns 1 and 2
    are retrograde impacts (negative impact angles).
    }

    \label{fig:sfd_10km}
\end{figure*}

%

First set of simulations was carried out with the target size $D_{\rm pb}=10\,\rm km$.
The diameters of impactors were $d_{\rm imp}= 394, 570, 850 \mbox{ and } 1226\,\rm m$,
respectively. 
We ran a number of simulations for different $P_{\rm pb}$, $d_{\rm imp}$ and $\phi_{\rm imp}$
and compared the size-frequency distribution of a family created by an impact
into a rotating target with corresponding impact into a stationary target.
The resulting distributions are plotted in Fig.~\ref{fig:sfd_10km}.

At first glance, the differences between the targets rotating with period the $P_{\rm pb}=3\,\rm h$
and the non-rotating targets are relatively small. The slope of the SFD is almost unchanged
in most simulations, it is only shifted as more mass is ejected from the rotating target.
In several simulations, e.g.~for $\phi_{\rm imp}=-45^\circ$ and $d_{\rm imp}=0.85\,\rm km$,
we can see a larger number of fragments in the middle part of the SFD
for rotating targets; fragments that would reaccumulate to the largest remnant 
in the stationary case now escape due to the extra speed from rotation and 
contribute to the family.

Much larger differences in SFDs can be seen for the target with period $P_{\rm pb}=2\,\rm h$,
which is rather expected;
for $\rho_0=2700\,\rm kg/m^3$ the critical period is $P_{\rm crit}\simeq 2.009 \,\rm h$,
so a $P_{\rm pb}=2\,\rm h$ target actually rotates very \emph{slightly} supercritically,
although is is held stable by the material strength.
The difference is the most prominent for oblique $\phi_{\rm imp}=\pm 75^\circ$ impacts.
In several cases, the rotation seems to make the SFD less steep,
although this might be partially attributed to a numerical artifact, 
as the synthetic families of
non-rotating targets are already very close to the resolution limit.

On the other hand, energetic impacts produce practically the 
same SFDs regardless of $P_{\rm pb}$.
In this regime, angular momentum of projectiles is larger than the rotational
angular momentum of the target; for $Q/\Qd = 3$ and $P_{\rm pb}=2\,\rm h$, 
the angular momentum of a projectile is larger 5x. 
Ejection velocities are also considerably larger than 
orbital velocities, hence it is not surprising that the rotation does not make 
a substantial difference.

It is not probable that these differences come from different fragmentation
pattern, as targets are fully damaged by the impact.
In our model, such damaged material is strengthless and it essentially behaves like a fluid. 
Since there is no internal friction or a mechanism to regain the material strength,
this model is insufficient to determine shapes of the fragments; however, here we 
are only interested in size distributions and using a simplified model is therefore 
justified.

\subsection{Size-frequency distributions for $D_{\rm pb}=100\,\rm km$ targets}
\label{sec:sfd_100km}

\begin{figure*}
    \centering
    \includegraphics[width=\textwidth]{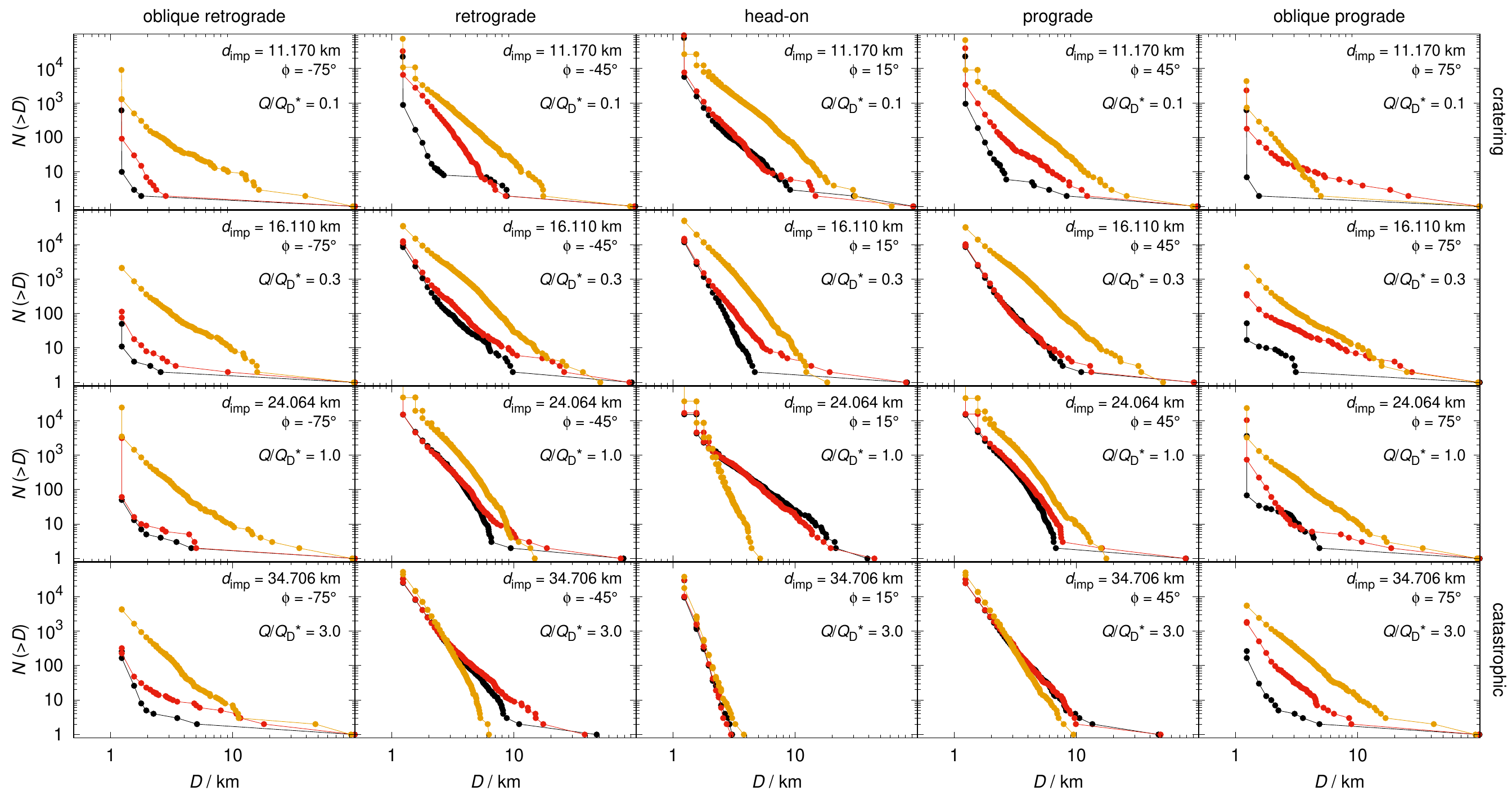}
    \caption{Cumulative size-frequency distribution for $D_{\rm pb}=100\,\rm km$ targets.
            The notation is the same as in Fig.~\ref{fig:sfd_10km}.
   }
    \label{fig:sfd_100km}
\end{figure*}

It is \emph{a priori} not clear how rotation affects targets of different sizes.
To preliminary estimate the importance of initial rotation, 
we compute the ratio of the angular frequency $\omega_{\rm pb}$ of the target and 
$\omega_{\rm imp}$ of the impactor with respect to the target:
\begin{equation}
    \frac{\omega_{\rm pb}}{\omega_{\rm imp}} \sim \frac{ D_{\rm pb}}{v_{\rm imp} P_{\rm pb} \sin \phi_{\rm imp}}\,.
\end{equation}
Because the ratio scales linearly with the target size $D_{\rm pb}$, we expect 
that the rotation will play a bigger role for impacts into larger targets;
however, this back-of-the-envelope computation is by no means a definite proof 
and it needs to be tested.

To this point, we ran a set of simulations with target size $D_{\rm pb}=100\,\rm km$.
The set is analogous to the one in Sec.~\ref{sec:sfd_10km} --- we use 
the same impact angles and rotational periods, the impactor diameters 
were $d_{\rm imp}=11.170, 16.110, 24.064 \mbox{ and } 34.706 \,\rm km$ in order 
to obtain the required relative energies $Q/\Qd$. The size-frequency distributions
of the synthetic families are plotted in Fig.~\ref{fig:sfd_100km}.

As expected, the differences between rotating and non-rotating targets 
are indeed substantially larger than for $D_{\rm pb}=10\,\rm km$.
The rotation can completely change the impact regime from cratering to catastrophic;
see e.g.~the impact with $\phi_{\rm imp}=15^\circ$ and $d_{\rm imp}=16.110\,\rm km$,
where a cratering gradually changes to a catastrophic disruption as we decrease $P_{\rm pb}$.
Focussing on $P_{\rm pb}=3\,\rm h$ targets, 
they produce very shallow SFD in case of oblique prograde impacts.
For $\phi_{\rm imp}=\pm 45^\circ$ cratering impacts, 
we see numerous intermediate-sized fragments (somewhat separated in the SFD) 
if the target is non-rotating, but the SFD becomes continuous when rotation is introduced.

The effects are even stronger for critically rotating bodies with $P_{\rm pb}=2\,\rm h$, of course.
Generally, SFDs of formally cratering events are more similar to catastrophic ones.
It also seems that oblique retrograde craterings produce more fragments than prograde ones.
For the impact $\phi_{\rm imp}=15^\circ$ and $d_{\rm imp} = 24.064\,\rm km$, the 
SFD is well below the non-rotating case and most of the mass is contained in 
smallest fragments.


Although large ($D \gg 10\,\rm km$) asteroids typically rotate much slower 
than smaller bodies, there are few that rotate close to the critical 
spin rate for elongated bodies, such as (216)\,Kleopatra \citep{Hirabayashi_2014}. 
Rotation in collisional simulations of such bodies should therefore not be neglected.

\subsection{Total ejected mass}
\label{sec:mass}

\begin{figure*}
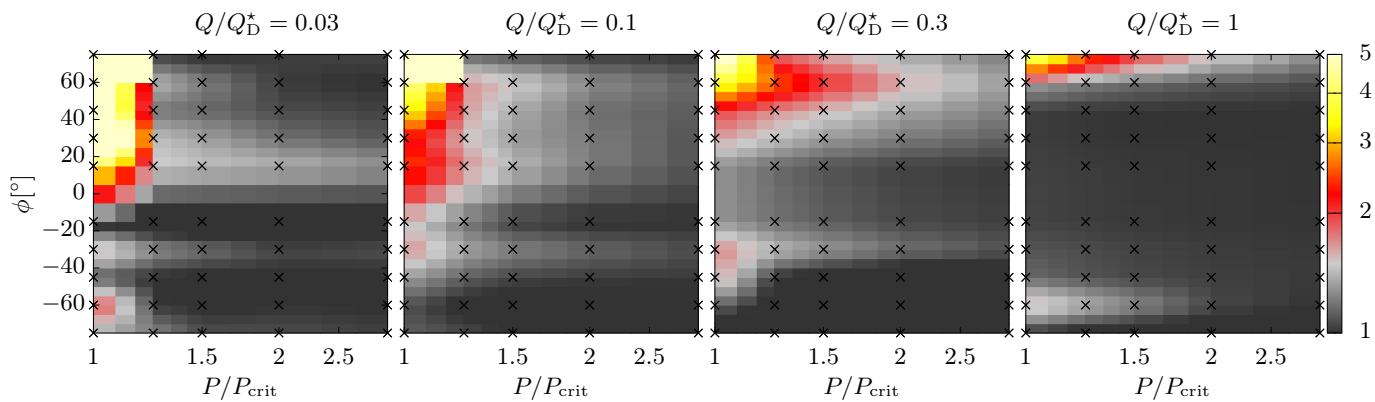

    \footnotesize
   \input mass.tex

    \caption{The total mass of fragments $M_{\rm ej}(P)/M_{\rm ej}(\infty)$ ejected by collisions, 
    normalized by the mass of fragments from a corresponding collision into a stationary target.
    Four figures correspond to different relative impact energies,
    from cratering (left) to mid-energy (right) events.
    The ejected mass is plotted as a function of the impact angle $\phi_{\rm imp}$ and 
    initial period $P_{\rm pb}$ of the target.
    }

    \label{fig:mass}

\end{figure*}

While Figs.~\ref{fig:sfd_10km} and \ref{fig:sfd_100km} 
clearly show
the differences between the SFDs, it is quite difficult to read the total mass ejected 
from the target during the impact. Even when the SFDs of a rotating and a stationary
target seem to differ only negligibly, the total integrated mass of fragments may 
be significantly different. 

To show the effect of the initial rotation on the ejected mass clearly, 
we performed over 400 simulations with the target size $D_{\rm pb} = 10\,\rm km$ 
and various impact angles $\phi_{\rm imp}$, projectile diameters $d_{\rm imp}$ and 
initial periods $P_{\rm pb}$ of the target.
These simulations have a lower spatial resolution compared to the simulations 
of family formation in previous sections, as here we need not to resolve individual
fragments in detail. The target is resolved by approximately $N=100000$ particles.

We ran simulations for $\phi_{\rm imp}$ ranging from $15^\circ$
to $75^\circ$ (both prograde and retrograde).
To capture the dependence on $P_{\rm pb}$, we selected nine different values 
from $P_{\rm pb} = P_{\rm crit}$ to $50 P_{\rm crit}$.
The impact energies of the simulations were $Q/\Qd=0.03,0.1,\,0.3 \mbox{ and } 1$, 
meaning the simulations range from cratering events to mid-energy events.

Our goal is to compute the total mass of the fragments as a function of 
the impact angle $\phi_{\rm imp}$,
the initial rotational period $P_{\rm pb}$ and 
the diameter $d_{\rm imp}$.
We are actually not interested in the absolute value of the ejected mass, but rather in the 
ejected mass \emph{relative} to the mass that would be ejected if the targets were stationary.
Therefore, we compute the ratio:
\begin{equation}
    \mu_{\rm ej} = \frac{M_{\rm ej}(\phi_{\rm imp}, d_{\rm imp}, P_{\rm pb})}{ M_{\rm ej}(\phi_{\rm imp}, d_{\rm imp}, \infty)}
\end{equation}
and plot the result in Fig.~\ref{fig:mass}.
Values $\mu_{\rm ej}<1$ would mean that the impact into the rotating target ejected fewer fragments,
compared to the stationary target; no such result was found in the performed simulations. 

Generally, the rotation amplifies the ejection by several tens of percent.
However, the increase is significantly higher if the following conditions are satisfied:
\begin{itemize}
    \item Target rotates near the critical period. The effect of rotation
        decreases rapidly with increasing period of the target, as expected.
    \item The impact results in a cratering rather than a catastrophic event.
        While high-energy impacts eject more fragments in an absolute measure, the initial rotation
        does not affect the value notably in this regime.
    \item The impact is oblique and has a prograde direction. Head-on impacts and 
        the impacts in retrograde directions are not affected by the rotation
        to the same degree.
\end{itemize}

In extreme cases, the rotation can amplify the ejected mass by a factor of 5.
On the other hand, the ejection ratio $\mu_{\rm ej}$ does not exceed 1.6 for rotational periods 
$P_{\rm pb} > 2P_{\rm crit}$ in any of the performed simulations.

Although it is a different rheology, rubble-pile bodies
also exhibit a minor effect of rotation (on $Q^\star_{\rm D}$
as well as $\mu_{\rm ej}$) in this range of~$P_{\rm pb}$
\citep[][see Fig.~2 therein]{Takeda_2009}. However,
their strength is an order of magnitude lower than for
monoliths of \cite{Benz_Asphaug_1999}, so the comparison
is not straightforward.

\combeg

\subsection{Angular distribution of fragment velocities}

Using the same set of simulations as in the previous section,
we analyzed the velocity field of the fragments.
The angular histograms of velocity directions are plotted 
in Fig.~\ref{fig:angular}.
Analogously to the previous section, we compare a simulation
with non-rotating target and simulations with rotational period $P=2 \mbox{ and } 3\,\rm h$.
The histogram is constructed from velocity directions projected into $x$-$y$ plane, i.e.
from values $\theta = \operatorname{atan2}(v_y, v_x)$, with bin size of~$5^\circ$.

The differences between the corresponding simulations are again more prominent for oblique impacts.
Impacts into rotating targets produce more isotropic velocity field, 
although this can be a consequence of generally bigger ejection of mass;
as \cite{Sevecek_2017} showed, more energetic impacts produce more isotropic velocity field.
Furthermore, the rotation causes a shift of the velocity field in the direction of the rotation,
especially for oblique impacts (with $\phi_{\rm imp} =\pm 75^\circ$).

\comend






\section{Embedding and draining the angular momentum}

\begin{figure*}
    \includegraphics[width=0.195\textwidth]{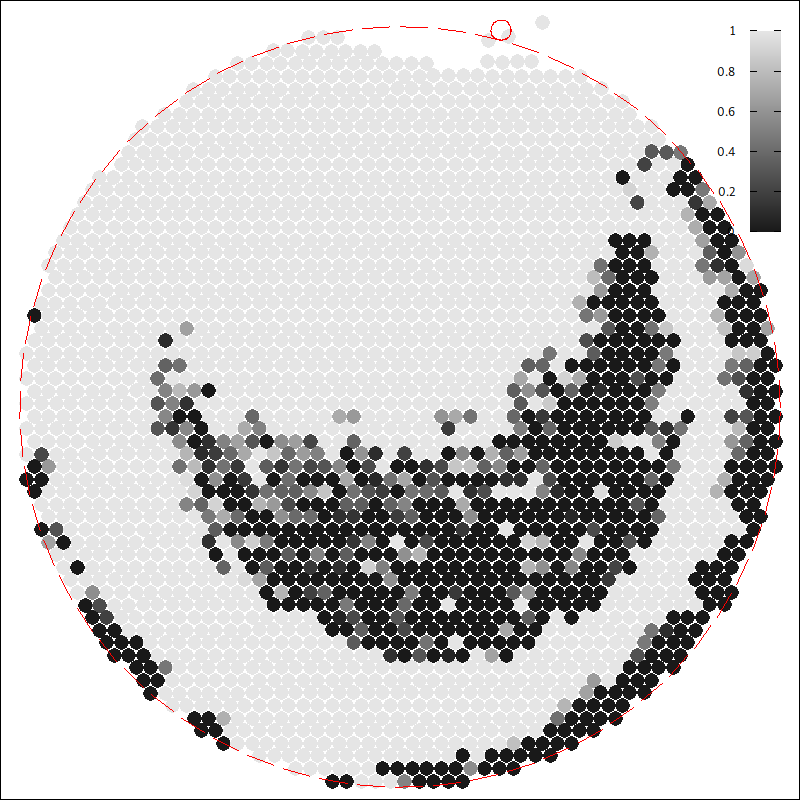}
    \includegraphics[width=0.195\textwidth]{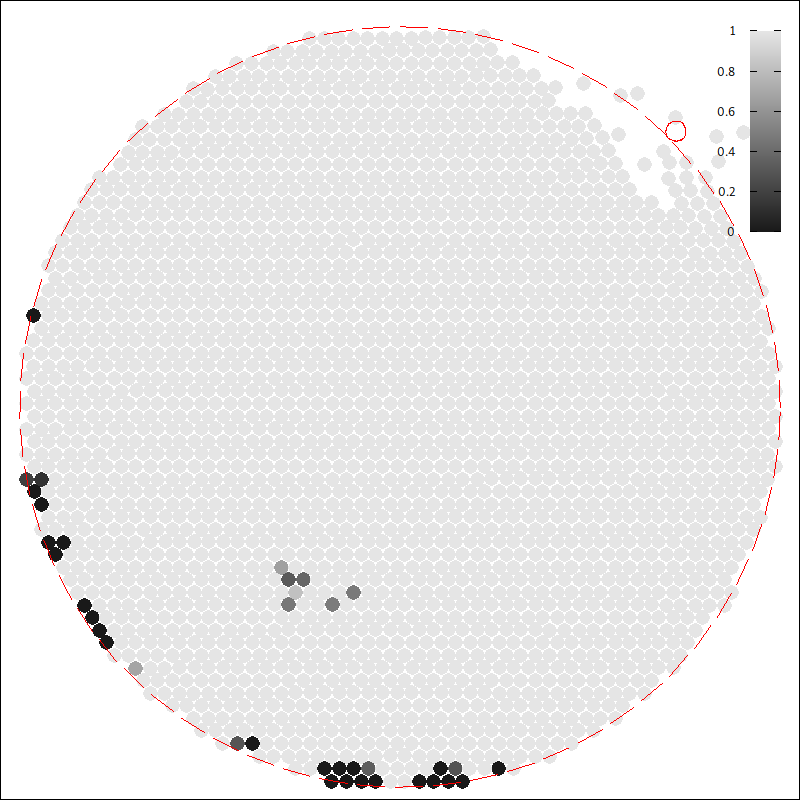}
    \includegraphics[width=0.195\textwidth]{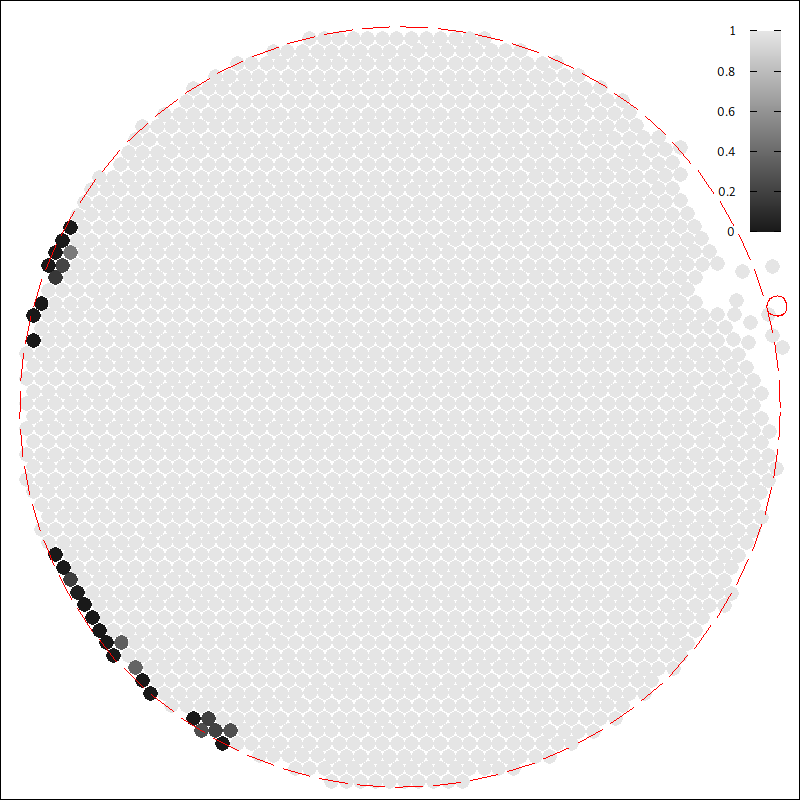}
    \includegraphics[width=0.195\textwidth]{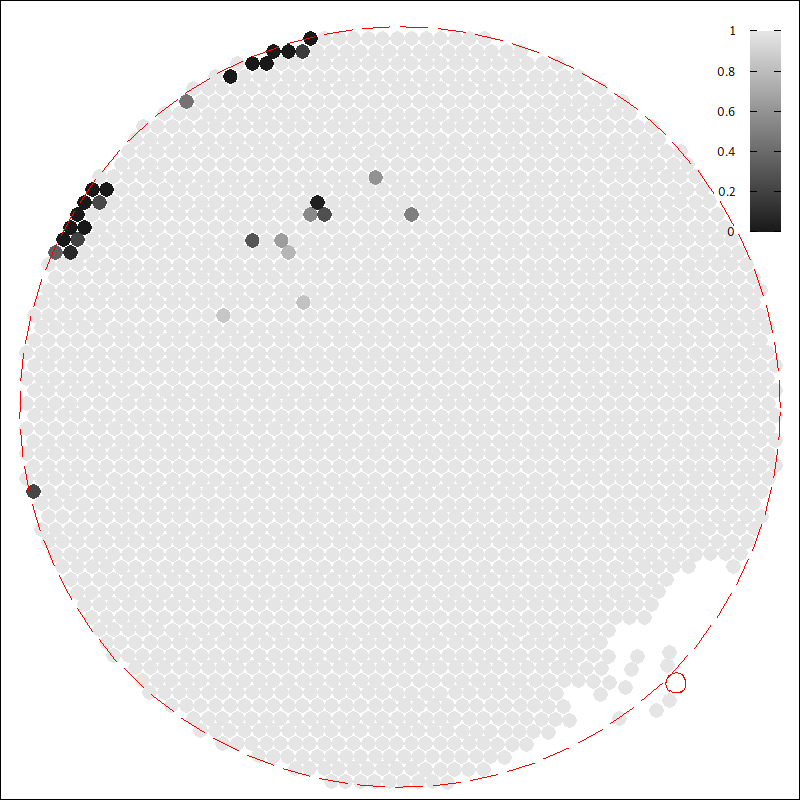}
    \includegraphics[width=0.195\textwidth]{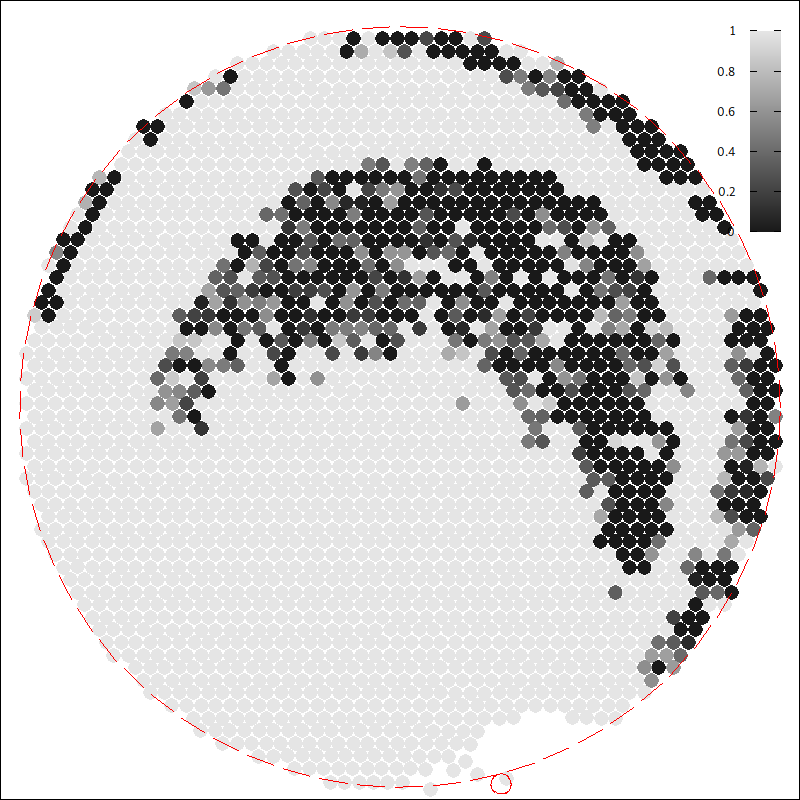}
    \caption{Damage of the target at time $t=10\,\rm s$ after the impact
    for various impact angles; 
        from left to right, $\phi_{\rm imp} = 75^\circ, 45^\circ, 15^\circ,
    -45^\circ, -75^\circ$. Simulations were carried out with the impactor of 
    size $d_{\rm imp} = 314\,\rm m$ and speed $v_{\rm imp}=5\,\rm km/s$, 
    target was not rotating. Red outline shows the original position of the 
    target and the impactor.
    There is an undamaged cavity only for oblique impacts, otherwise 
    the target is fully damaged by the impact.}
    \label{fig:damage}
\end{figure*}
\begin{figure*}
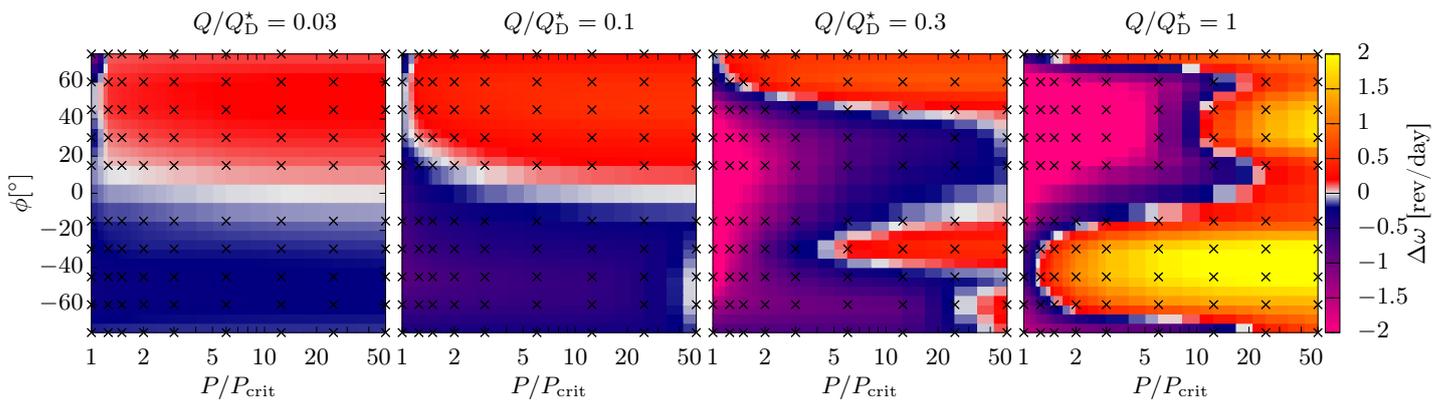

    \footnotesize
    \input period.tex
    \vskip-10pt

    \caption{Change of the spin rate $\Delta \omega$ of the target 
    (or the largest remnant in case $Q/\Qd \simeq 1$) as a function of the initial period $P_{\rm pb}$ 
    of the target (here in units of the critical period $P_{\rm crit}$) and the impact angle $\phi_{\rm imp}$.
    Four images correspond to different sizes of the impactor, the energy of the impact 
    is from left to right $Q/\Qd = 0.03, 0.1, 0.3 \hbox{ and } 1$. 
    The impact velocity was $v_{\rm imp} = 5\,\rm km/s$ in all cases.} 
    \label{fig:period}
\end{figure*}
\begin{figure*}
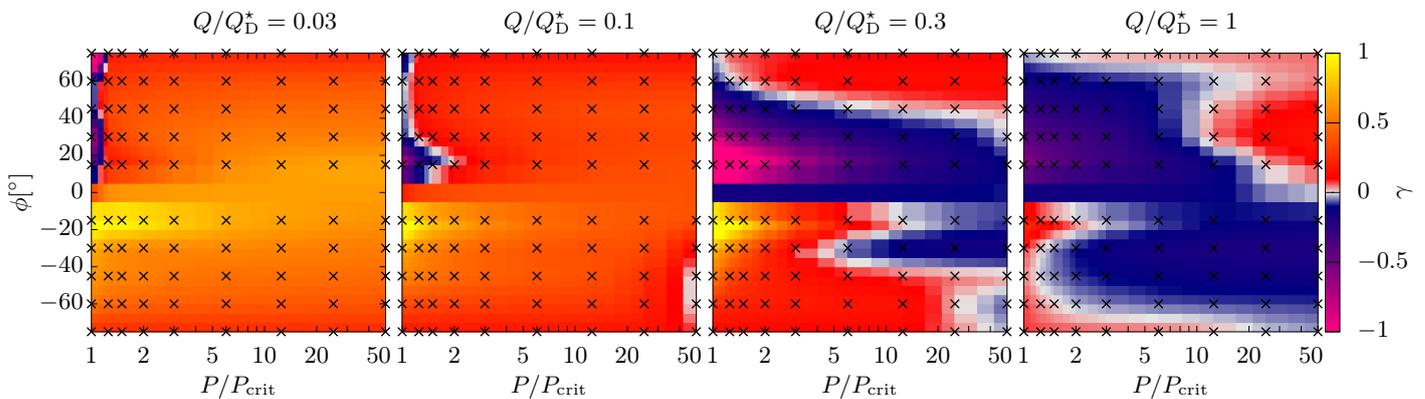

    \footnotesize
    \input gamma.tex
    \vskip-10pt

    \caption{Dimensionless effectivity $\gamma$ of the angular momentum transfer.
     Other quantities are the same as in Fig.~\ref{fig:period}.
    }
    \label{fig:gamma}
\end{figure*}

Impact into a rotating target can cause either an acceleration or a deceleration 
of the target's rotation. This can be immediately seen from two limit cases: 
a stationary target is always spun by the impact, on the other hand 
a target rotating at the breakup limit cannot be accelerated any further 
and the collision thus always causes a deceleration.

It has been proposed that rotating asteroids are decelerated over time 
by numerous subsequent cratering collisions, as a fraction of the angular
momentum is carried away by fragments.
Coined the angular momentum drain \citep{Dobrovolskis_1984},
this effect could explain the excess of slow rotators in the Main Belt.
In this section, we examine whether this effect emerges in our simulations and 
we determine the functional dependence of the deceleration on the impact parameters.


We analyze the angular momentum transfer as a function of impact parameters,
using the set of simulations described in Sec.~\ref{sec:mass}.
The impacts range from cratering ($Q/\Qd \sim 0.03$) to mid-energy ($Q/\Qd \sim 1$) events.
For $Q/\Qd \sim 1$, the whole target asteroid is disintegrated by the collision
and fragments with mass of about $0.5M_{\rm pb}$ are reaccumulated later, forming the largest remnant. 
This can no longer be viewed as a cratering event that merely modifies the rotational state of 
the target, nevertheless we can still formally compute the relation between 
the period of the target and the largest remnant.

In majority of performed simulations, the target is completely damaged by the impact.
Only for the weakest oblique impacts with $Q/\Qd = 0.03$, 
there remained an undamaged cavity, as shown in Fig.~\ref{fig:damage},
otherwise all particles of the target have damage $D=1$ after the fragmentation phase.

The change of spin rate $\Delta \omega$ of the target
is plotted in Fig.~\ref{fig:period}. 
We plot the change of frequency rather than period,
as the period is formally infinite for a non-rotating body;
change of period is thus not a meaningful quantity.

For cratering events,
the prograde events (denoted with positive $\phi_{\rm imp}$) 
mostly accelerate the target, while retrograde
events cause deceleration. The two exceptions from this rule are:
\begin{enumerate}
    \item a prograde impact into a critically rotating body, in which case it cannot be accelerated any more and some deceleration is expected,
    \item a retrograde impact to an almost stationary body.
\end{enumerate}
Impacts with higher energies show a different pattern. It seems that the two regions 
described above expand.
Prograde impacts into fast rotators actually decelerate the target, 
while retrograde impacts start to accelerate it.  
For the most energetic event we studied, $Q/\Qd \simeq 1$, it seems that the two regions completely swapped 
--- most prograde impacts cause a deceleration while retrograde ones an acceleration.

The despinning (or angular momentum draining) on rubble-piles
in catastrophic disruptions was confirmed by \cite{Takeda_2009}. 
In our simulations, the pattern is more complex, likely
    because the parameter space ($Q/Q^\star_{\rm D}$, $P_{\rm pb}$)
        is significantly more extended; also rubble-piles cannot initially
            rotate critically.


\subsection{Effectivity of the angular momentum transfer}
\label{sec:gamma}

Let us define the effectivity $\gamma$ of the angular momentum transfer as:
\begin{equation}
    \gamma \equiv \frac{L_{\rm lr} - L_{\rm pb}}{L_{\rm imp}}\,,
\end{equation}
where $L_{\rm pb}$ is the rotational angular momentum of the target before the impact,
$L_{\rm pb}$ is the rotational angular momentum of the largest remnant 
and $L_{\rm imp}$ is the angular momentum of the impactor with respect to the target.
As these values are scalars, we assign a negative sign to the value $L_{\rm imp}$
for retrograde impacts.

We emphasise that the effectivity $\gamma$ is not necessarily in the unit interval $\langle 0; 1\rangle$.
Specifically, it may be significantly larger than 1 for head-on impacts, as the delivered angular momentum
is very low; in fact, $L_{\rm imp}$ approaches zero for $\phi_{\rm imp} = 0$. 
The effectivity can also be a negative number for impacts to critically rotating targets,
as in these cases, the target cannot accelerate over the breakup limit, so a zero or 
even negative values of $\gamma$ are expected.

The effectivity $\gamma$ as a function of the initial period $P_{\rm pb}$, the impact angle $\phi_{\rm imp}$
and the impactor diameter $d_{\rm imp}$ is plotted in Fig.~\ref{fig:gamma}.
We can see that cratering impacts have generally higher effectivity than high-energy
impacts. This result might have been expected, as the cratering impacts eject less mass and 
thus transfer less angular momentum to fragments, compared to the catastrophic impacts.
A less expected outcome is the negative effectivity for the high-energy impacts.
We predicted the negative values only for prograde impacts into critically rotating targets,
but for $d_{\rm imp} = 850\,\rm m$ the effectivity is negative for the majority of performed simulations.

Finally, the highest effectivity is achieved for retrograde impacts into 
critically rotating targets. However, this results is a bit fake,
because in this regime the target is always decelerated, $\gamma$ can therefore exceed 1 
as the angular momentum lost in the collision is higher than $L_{\rm imp}$ (in absolute value).
Impacts into slower rotators have values of $\gamma$ around 0.5.

\subsection{Angle-averaged ejection and momentum transfer}
\begin{figure*}
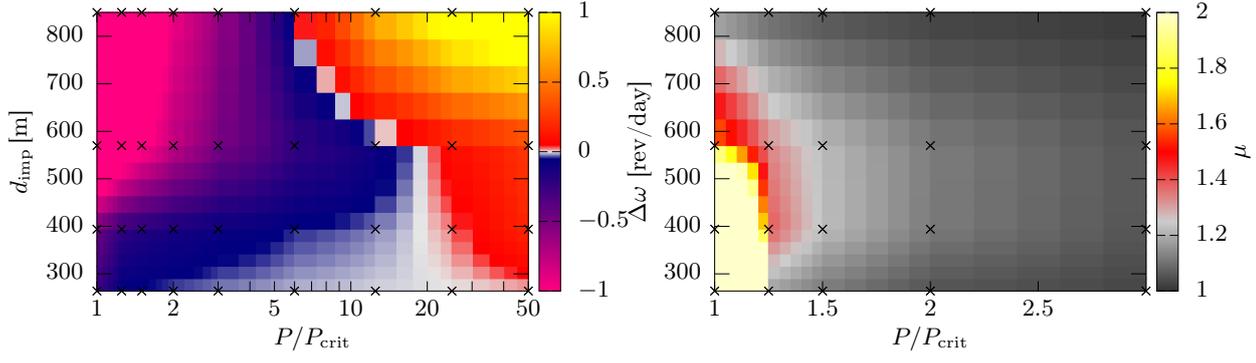

    \footnotesize
    \input integral_period.tex
    \input integral_mass.tex

    \caption{Quantities averaged over impact angles using Eq.~(\ref{eq:domega_avg})
    as a function of period $P_{\rm pb}$ of the target and the impactor diameter $d_{\rm imp}$.
    The left figure shows the change of spin rate $\Delta \omega$, the right
    one the total mass of fragments, normalized 
    by the mass of fragments from corresponding collision into a stationary target.
    }
    \label{fig:integral}
\end{figure*}

To express an overall effect of collisions on a rotational state of a target,
it is useful to consider a large number of collisions at random impact angles
and compute the average change of spin rate:
\begin{equation}
    \label{eq:domega_avg}
    \overline {\Delta \omega} \equiv \int\limits_0^{\pi/2} \Delta\omega \sin 2\phi \, \d \phi 
    \approx \frac{\sum_i \Delta\omega_i \sin 2\phi_i}{\sum_i\sin 2\phi_i} \,,
\end{equation}
where $\sin 2\phi$ is the probability of the impact with the impact angle $\phi$.
The averaged change $\overline\Delta \omega$ is plotted in Fig.~\ref{fig:integral},
together with a plot of the relative mass ejection $\overline\mu_{\rm ej}$, 
defined in Sec.~\ref{sec:mass}.

The figures show that targets are on average decelerated 
for both low-energy and mid-energy impacts.
A target is only accelerated if it was originally a very slow rotator, since it cannot 
be decelerated any more. The transition between these two regimes (plotted in white) 
is where the target does not change its angular frequency upon the impact. 
It seems to depend on the energy of the impact; for $Q/\Qd \sim 0.1$, 
the transition occurs at around $P \sim 20P_{\rm crit}$, while for $Q/\Qd \sim 1$, 
it is shifted to around $P\sim 6P_{\rm crit}$.

\combeg
\subsection{Dependency of the scaling law on rotation}
Generalize $\Qd$ , $\Qd = \Qd(D_{\rm pb}, P_{\rm pb})$.

Impact energy so that $M_{\rm lr} = 0.5 \,M_{\rm pb}$.
For fixed $\phi_{\rm imp}$ and $v_{\rm imp}$, we run a simulation
with several $d_{\rm imp}$, so that few of them are $Q > \Qd$ and few
$Q < \Qd$. We then interpolate to $Q = \Qd$ using quadratic fit. 

We follow \cite{Benz_Asphaug_1999}.

\begin{figure}
    \footnotesize
    \input scaling.tex
    \caption{Mass $M_{\rm lr}$ of the largest remnant as a function of 
    impact energy $Q$ for various rotational periods $P_{\rm pb}$ of the target.
    Impact angle $\phi_{\rm imp}=45^\circ$, impact speed $v_{\rm imp}=5\,\rm km/s$.
    As above, large difference for $Q\ll \Qd$ and $P\simeq P_{\rm crit}$,
    however for $Q \simeq \Qd$ and $P > 2.5\,\rm h$, barely changes,
    rotation causes difference about 5\%.
    Scaling law mostly unaffected by rotation.
   }

\end{figure}
\comend




\section{Conclusions and future work}

In this paper, we showed that a fast initial rotation of targets may 
significantly affect resulting synthetic families.
The effect is more prominent for larger target bodies and for 
oblique impact angles.
Generally, more fragments is ejected 
from prograde ($\phi_{\rm imp}>0$) compared to the retrograde ($\phi_{\rm
imp}<0$) targets.

In extreme cases, the mass ejection can be amplified by a factor 5.
Neglecting the rotation would therefore introduce a considerable bias.
Other parameters of the simulation do introduce similar (or sometimes larger) uncertainties,
for example the Weibull parameters of the fragmentation model \citep{Sevecek_2017},
the rheological model of the target, etc.
As shown in \cite{Jutzi_Benz_2017}, the initial shape can also have a significant
effect.

Throughout this paper, we assumed that both the targets and impactors 
are monolithic bodies. It is \emph{a priori} not clear whether 
the rotation would be more important for rubble-pile bodies (with macro-porosity),
or when a rheological model with crushing (micro-porosity) is used in the simulations
\citep{Jutzi_2019}. It should also be explored how initial shapes relate 
to spin rates of fragments.
We postpone such studies to future works.

In the future, we also plan to determine the scaling law as a function of 
both $D_{\rm pb}$ and $P_{\rm pb}$.
It is clear that the critical energy $\Qd$
is a steep function of $P_{\rm pb}$ close to the critical spin rate.
Finding a functional dependence $\Qd=\Qd(P_{\rm pb})$ might be a valuable result 
for studies of Main Belt evolution, as it could be used to 
construct a combined model that includes both collisions and rotations.

\section*{Acknowledgements}
\label{sec:acknowledgements}
    The work of P.Š.~and M.B.~has been supported by 
     the Czech Science Foundation (GACR 18-04514J)
    and Charles University (GAUK 1584517).
    M.J.~acknowledges support from the Swiss National
    Centre of Competence in Research PlanetS.

\appendix

\section{Handling particle overlaps}
\label{sec:overlaps}
Since we treat particles as solid spheres during the reaccumulation phase,
particle overlaps are unavoidable and need to be handled by our N-body
integrator. There are two main reasons why particle overlaps occur.

First, the spheres overlap initially after the hand-off (Eq.~\ref{eq:handoff}). 
In SPH, the particles naturally overlap as they describe a continuum rather than point masses. 
After converting them to solid spheres, particles belonging to the same body will necessarily overlap,
unless their radius is decreased significantly.

    Second, overlaps occur when particles are being merged.
     When two spherical particles collide, they merge into a larger
        particle with volume equal to the sum of volumes of the colliders. This merging is an 
        atomic operation, particles are converted into the merger in an instant rather than 
        over several timesteps, so any other particles located close to the colliders
        potentially overlap the particle merger.
   
Our code allows for several options to resolve overlaps.
One straightforward solution is to always merge the overlapping particles.
While this is a simple and robust solution, it can potentially create unphysical,
supercritically rotating bodies.
Alternatively, we can repel the overlapping particles, so that they are 
in contact rather than overlap. However, this causes an ``inflation'' 
of the largest remnant after the hand-off. Even worse, the angular momentum is 
no longer conserved.

Another option is to abandon the 1-1 conversion of spheres
and instead construct a new set of spheres inside the alpha-shape 
of the largest remnant \citep{Ballouz_2018}.
Such approach allows to place spheres onto a regular grid 
and thus avoid overlaps by construction.
However, it is more suitable when collided particles form rigid aggregates
instead of mergers.
As spheres never fill the entire volume (filling factor of hexagonal
close packing is about 0.74) and the merging conserves volume,
fragments would shrink considerably.

We decided to merge particles only if the spin rate of the would-be merger is 
lower than the critical spin rate, otherwise we allow particles to pass 
through each other. Of course, such handling is only applied 
to resolve overlaps, particles that collide are always treated as solid.

\vfill
\section{Comparison of inertial and co-rotating reference frames}
\label{sec:inertial}

\begin{figure*}
    \includegraphics[width=0.246\textwidth]{\detokenize{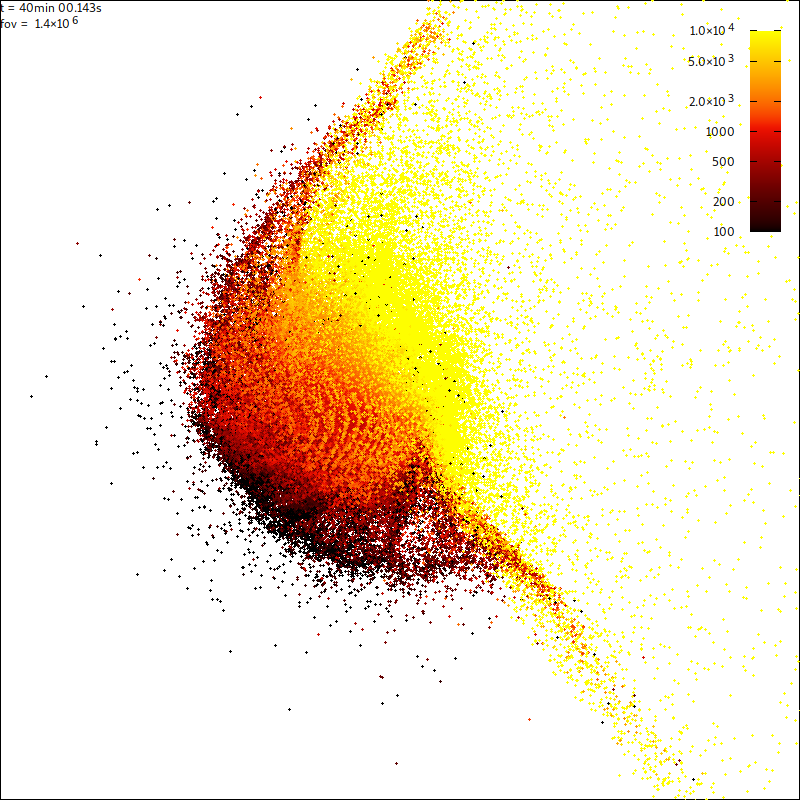}}
    \includegraphics[width=0.246\textwidth]{\detokenize{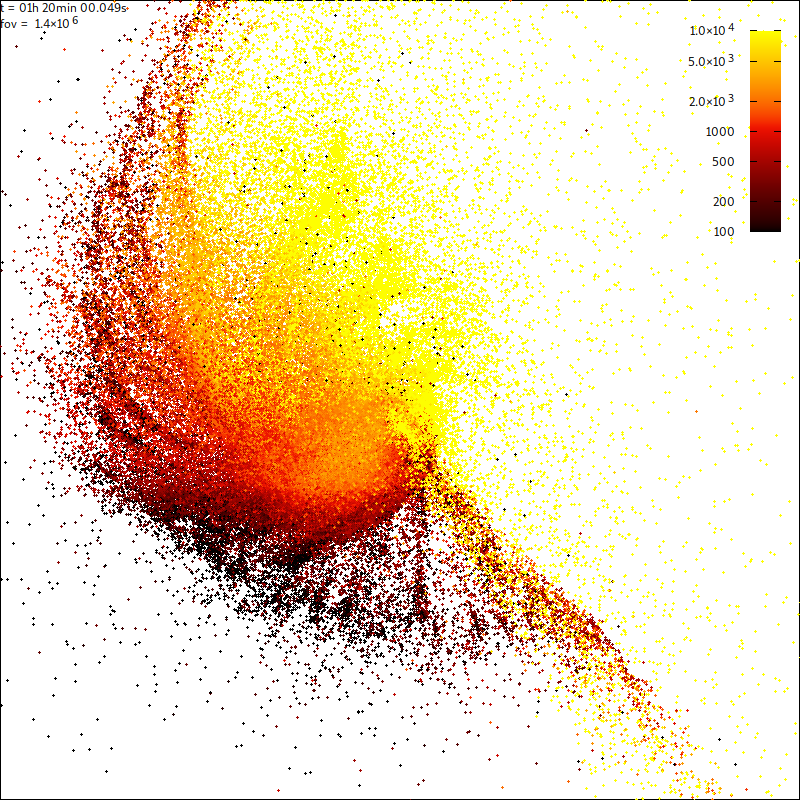}}
    \includegraphics[width=0.246\textwidth]{\detokenize{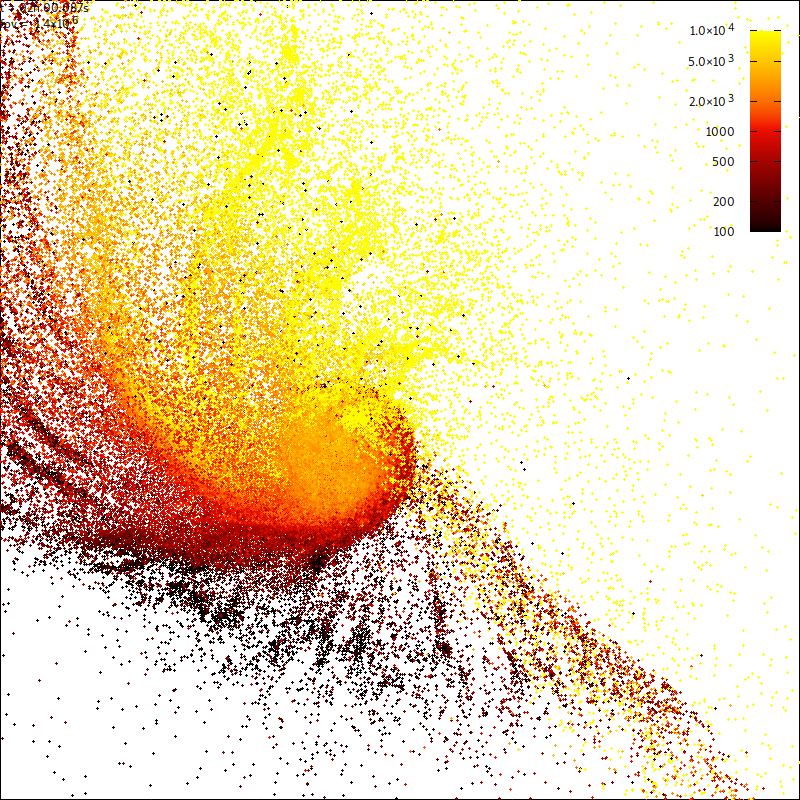}}
    \includegraphics[width=0.246\textwidth]{\detokenize{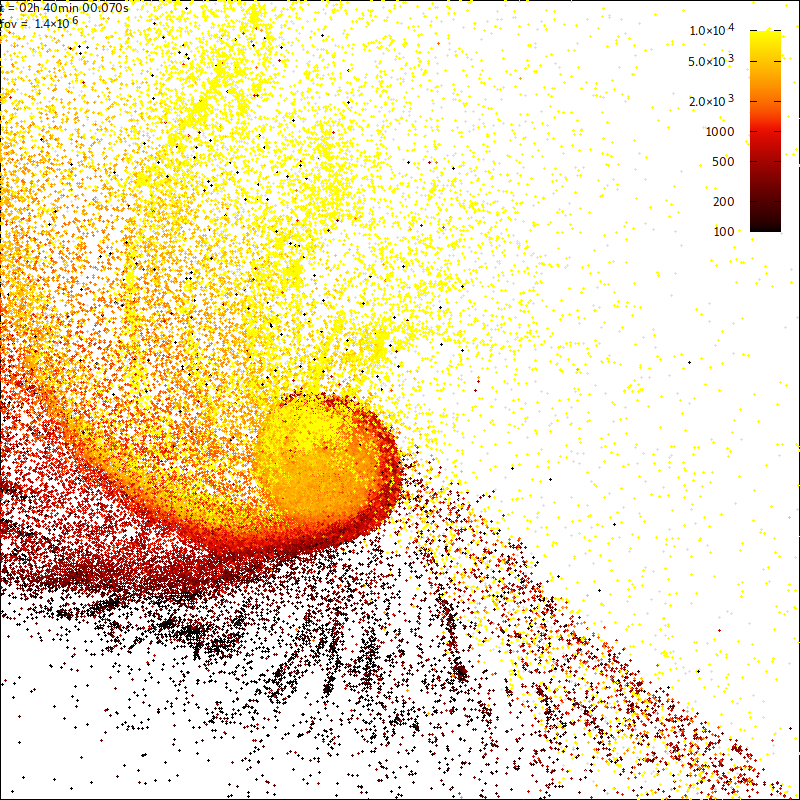}}

    \includegraphics[width=0.246\textwidth]{\detokenize{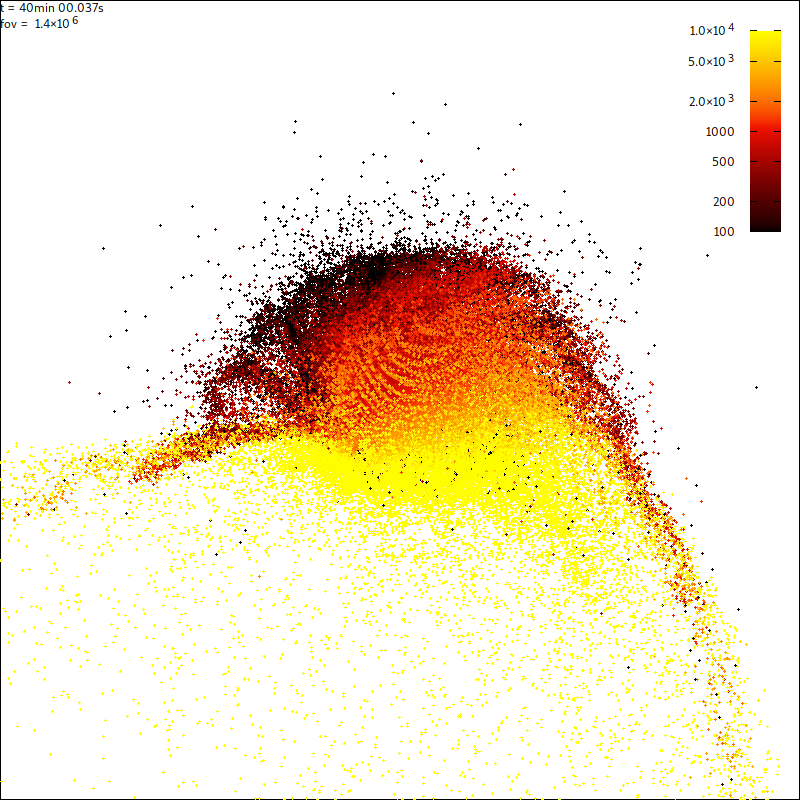}}
    \includegraphics[width=0.246\textwidth]{\detokenize{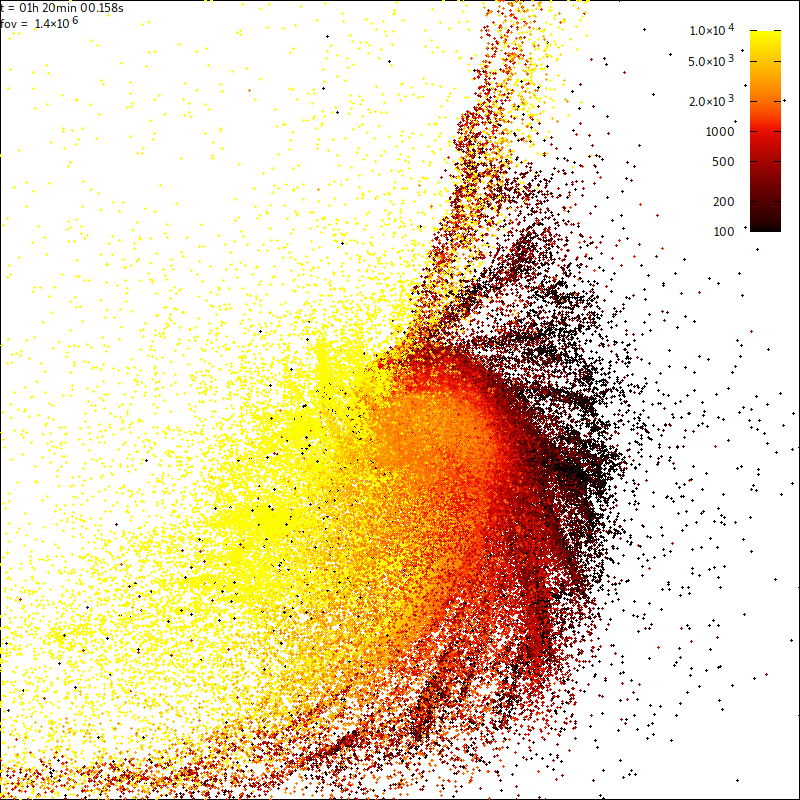}}
    \includegraphics[width=0.246\textwidth]{\detokenize{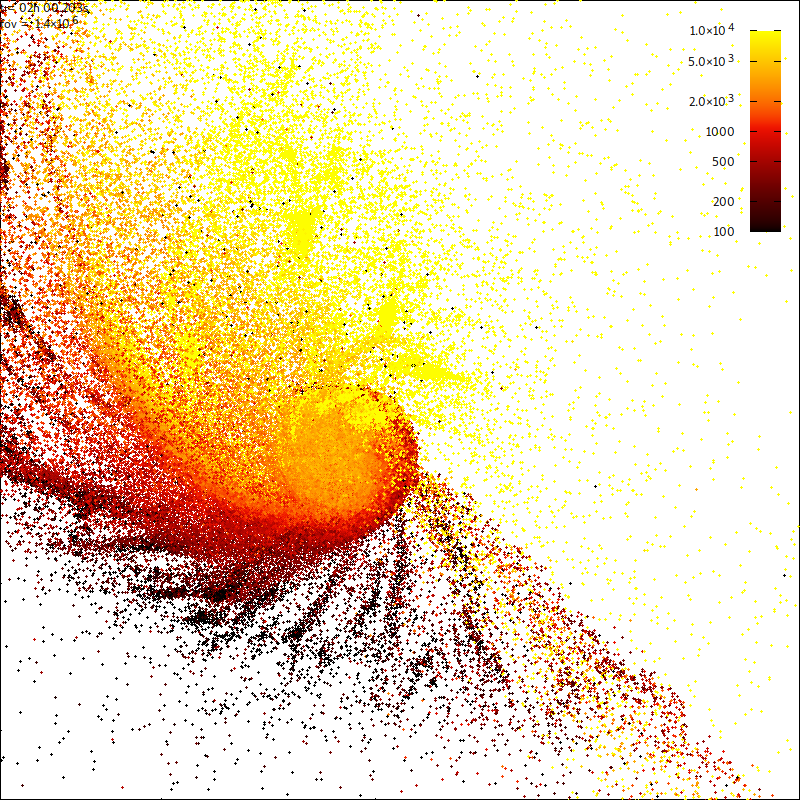}}
    \includegraphics[width=0.246\textwidth]{\detokenize{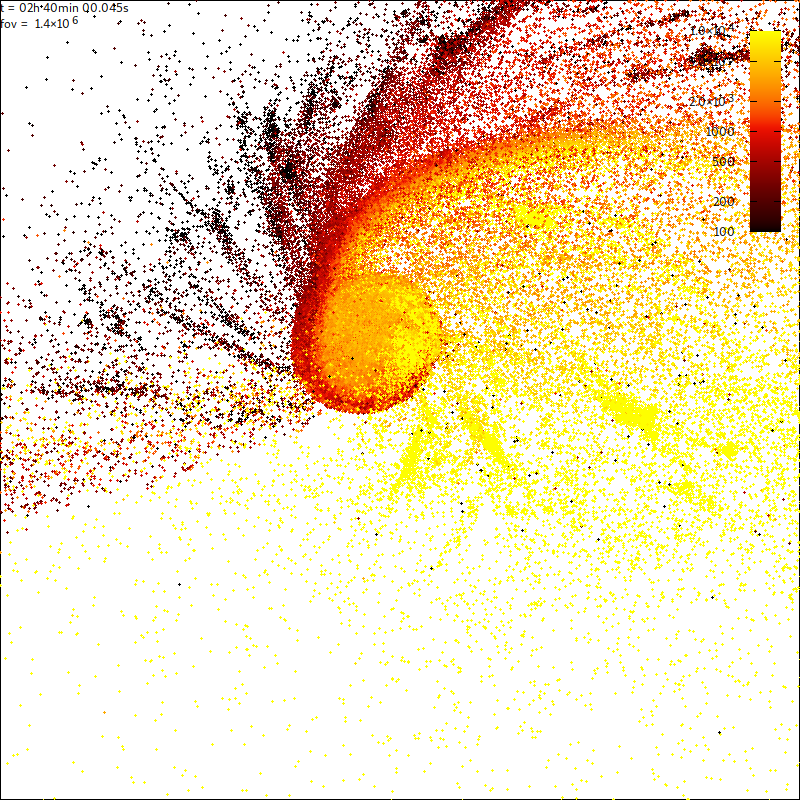}}
    \caption{Impact into $D_{\rm pb}=200\,\rm km$ target, computed with rotating target (upper row)
    and with rotating coordinate frame in which the target is stationary (lower row),
    plotted at times $t=40, 80, 120 \mbox{ and } 160\,\rm min$ after the impact.
    The color scale represents the specific energy of the particles (in SI units).
    }
    \label{fig:frame_comparison}
\end{figure*}

We can choose two different approaches to implementing the rotation of the target:
\begin{enumerate}
    \item rotate the particles around the center of the target,
    \item perform the simulation in the coordinate system co-rotating with the target.
\end{enumerate}

From a numerical point of view, the second approach is easier to handle,
as the particles of the target have initially zero velocities
and we thus avoid numerical problems with the bulk rotation 
outlined in Sec.~\ref{sec:equations}.
The rotation is taken into account by introducing inertial accelerations, 
i.e.~the last two terms of Eq.~(\ref{eq:motion}).

However, it only solves the issue partially; even though the target is stationary (in the co-rotating frame)
before the impact, the projectile can spin up the target and the impact can also eject rotating fragments.
To properly handle rotating bodies in SPH, it is necessary to introduce the correction tensor 
Eq.~(\ref{eq:correction_tensor}). This allows us to perform simulations in the inertial frame, 
which is a natural choice.

Ideally, these two approaches should produce identical results.
To test it, we ran two simulations with $D_{\rm pb}=200\,\rm km$ parent bodies rotating critically.
The results are plotted in Fig.~\ref{fig:frame_comparison}.
We observe some stochastic differences, but the spatial distribution of the ejected 
fragments is similar in both simulations.
This test confirms the consistency of both approaches.

\section{Implementation notes}
\label{sec:code}

\begin{figure*}
    \includegraphics[width=\textwidth]{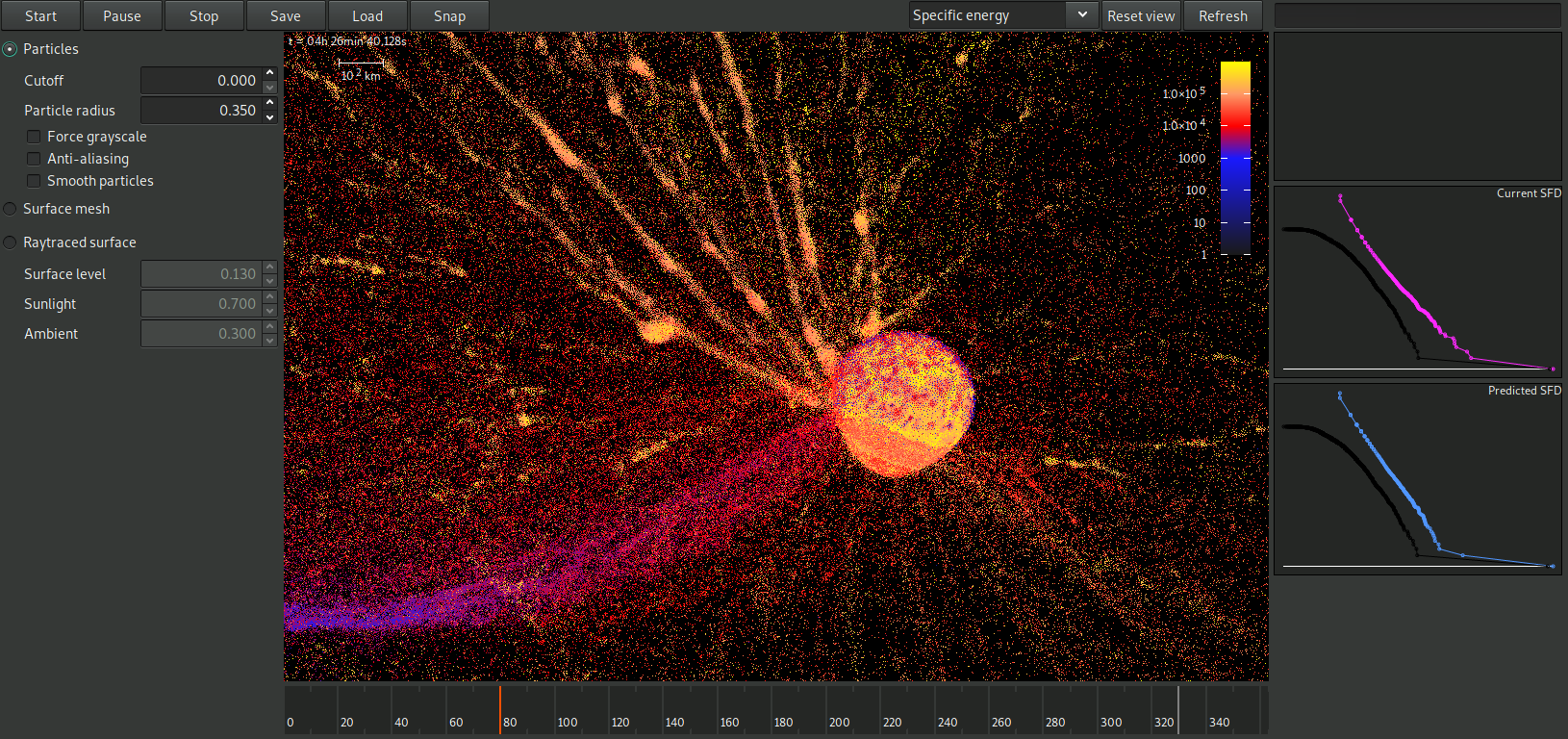}
    \caption{Screenshot of the graphical interface of our code.}
    \label{fig:screenshot}
\end{figure*}

The code \texttt{OpenSPH} used throughout this work is open-source,
available under MIT license.
As of \today, it can be be downloaded from \texttt{https://gitlab.com/sevecekp/sph}.
It includes a helpful graphical interface, allowing to interactively 
visualize the simulation, see Fig.~\ref{fig:screenshot}.
It is also a standalone viewer of \texttt{OpenSPH} output files 
and potentially files generated by other particle-based codes, 
provided their file formats are implemented.


Our code can be used as both an SPH solver and an N-body integrator,
as we separated computation of SPH derivatives and gravitational accelerations. 
In each time step, accelerations due to hydrodynamics and gravity are 
computed independently and summed up.
Even though we miss some optimization possibilities with this approach, 
it allows us to use the same code for both the fragmentation and 
the reaccumulation phase;
the hand-off is thus only an internal change of a solver,
replacing SPH hydrodynamics with a collision detection.

We use the Barnes-Hut algorithm to evaluate gravitational 
accelerations \citep{Barnes_Hut_1986}.
The code uses the same functionality in SPH and N-body solver, 
it only differs in the softening kernel $\phi$; in SPH, 
it is defined by Eq.~\ref{eq:softening_kernel},
while for N-body it corresponds to a homogeneous sphere.
Our implementation uses a k-d tree, which is also used to find particle 
neighbours.

In the fragmentation phase, we found that the time step criterion 
which uses stress tensor derivatives
(Eq.~\ref{eq:derivative_criterion}) is often 
unnecessarily restricting,
as the stress tensor changes rapidly inside the projectile.
However, we are not very interested in remnants of projectiles,
simulations can be thus sped up by applying the criterion 
only for particles of the target.
No such optimization is applied for CFL criterion, as it determines stability of the method
and it is thus essential for all particles.

In the reaccumulation phase, collided particles are merged, provided their 
relative velocity does not exceed the escape velocity $v_{\rm esc}$
(as in Eq.~\ref{eq:esc_velocity}) and the spin rate 
does not exceed the critical spin rate
$\omega_{\rm crit}$ (as in Eq.~\ref{eq:critical_freq}). 
In practice, both $v_{\rm esc}$  and  $\omega_{\rm crit}$ 
are multiplied by user-defined factors (i.e.~a merging limit).
It may be useful to tune it in such a way that a simplified N-body 
model of reaccumulation matches a full SPH simulation (in terms of resulting SFDs).

The total CPU time needed depends on a type of simulation. Generally, catastrophic impacts
take longer to compute than cratering impacts (quantities change more rapidly, 
hence smaller time steps are needed), 
smaller targets usually mean longer computation time (smaller SPH particles, which implies
smaller time steps due to the Courant criterion).
The code is parallelized using a custom thread pool, utilizing the native C++11 threads,
or optionally using Intel Thread Building Blocks library. For $D_{\rm km}=10\,\rm km$ target
and $N = 500{,}000$ particles,
a single simulation takes about 10 hours on a AMD Ryzen Threadripper 1950X 16-Core CPU. 

\bibliographystyle{elsarticle-harv}
\bibliography{references}

\end{document}

%% file: mass.tex
\begingroup
  \makeatletter
  \providecommand\color[2][]{%
    \GenericError{(gnuplot) \space\space\space\@spaces}{%
      Package color not loaded in conjunction with
      terminal option `colourtext'%
    }{See the gnuplot documentation for explanation.%
    }{Either use 'blacktext' in gnuplot or load the package
      color.sty in LaTeX.}%
    \renewcommand\color[2][]{}%
  }%
  \providecommand\includegraphics[2][]{%
    \GenericError{(gnuplot) \space\space\space\@spaces}{%
      Package graphicx or graphics not loaded%
    }{See the gnuplot documentation for explanation.%
    }{The gnuplot epslatex terminal needs graphicx.sty or graphics.sty.}%
    \renewcommand\includegraphics[2][]{}%
  }%
  \providecommand\rotatebox[2]{#2}%
  \@ifundefined{ifGPcolor}{%
    \newif\ifGPcolor
    \GPcolorfalse
  }{}%
  \@ifundefined{ifGPblacktext}{%
    \newif\ifGPblacktext
    \GPblacktexttrue
  }{}%
  \let\gplgaddtomacro\g@addto@macro
  \gdef\gplbacktext{}%
  \gdef\gplfronttext{}%
  \makeatother
  \ifGPblacktext
    \def\colorrgb#1{}%
    \def\colorgray#1{}%
  \else
    \ifGPcolor
      \def\colorrgb#1{\color[rgb]{#1}}%
      \def\colorgray#1{\color[gray]{#1}}%
      \expandafter\def\csname LTw\endcsname{\color{white}}%
      \expandafter\def\csname LTb\endcsname{\color{black}}%
      \expandafter\def\csname LTa\endcsname{\color{black}}%
      \expandafter\def\csname LT0\endcsname{\color[rgb]{1,0,0}}%
      \expandafter\def\csname LT1\endcsname{\color[rgb]{0,1,0}}%
      \expandafter\def\csname LT2\endcsname{\color[rgb]{0,0,1}}%
      \expandafter\def\csname LT3\endcsname{\color[rgb]{1,0,1}}%
      \expandafter\def\csname LT4\endcsname{\color[rgb]{0,1,1}}%
      \expandafter\def\csname LT5\endcsname{\color[rgb]{1,1,0}}%
      \expandafter\def\csname LT6\endcsname{\color[rgb]{0,0,0}}%
      \expandafter\def\csname LT7\endcsname{\color[rgb]{1,0.3,0}}%
      \expandafter\def\csname LT8\endcsname{\color[rgb]{0.5,0.5,0.5}}%
    \else
      \def\colorrgb#1{\color{black}}%
      \def\colorgray#1{\color[gray]{#1}}%
      \expandafter\def\csname LTw\endcsname{\color{white}}%
      \expandafter\def\csname LTb\endcsname{\color{black}}%
      \expandafter\def\csname LTa\endcsname{\color{black}}%
      \expandafter\def\csname LT0\endcsname{\color{black}}%
      \expandafter\def\csname LT1\endcsname{\color{black}}%
      \expandafter\def\csname LT2\endcsname{\color{black}}%
      \expandafter\def\csname LT3\endcsname{\color{black}}%
      \expandafter\def\csname LT4\endcsname{\color{black}}%
      \expandafter\def\csname LT5\endcsname{\color{black}}%
      \expandafter\def\csname LT6\endcsname{\color{black}}%
      \expandafter\def\csname LT7\endcsname{\color{black}}%
      \expandafter\def\csname LT8\endcsname{\color{black}}%
    \fi
  \fi
    \setlength{\unitlength}{0.0500bp}%
    \ifx\gptboxheight\undefined%
      \newlength{\gptboxheight}%
      \newlength{\gptboxwidth}%
      \newsavebox{\gptboxtext}%
    \fi%
    \setlength{\fboxrule}{0.5pt}%
    \setlength{\fboxsep}{1pt}%
\begin{picture}(10080.00,3456.00)%
    \gplgaddtomacro\gplbacktext{%
      \csname LTb\endcsname
      \put(1008,3110){\makebox(0,0)[l]{\strut{}$Q / \Qd  = 0.03$}}%
    }%
    \gplgaddtomacro\gplfronttext{%
      \csname LTb\endcsname
      \put(251,595){\makebox(0,0){\strut{}1}}%
      \put(1061,595){\makebox(0,0){\strut{}1.5}}%
      \put(1635,595){\makebox(0,0){\strut{}2}}%
      \put(2081,595){\makebox(0,0){\strut{}2.5}}%
      \put(1348,375){\makebox(0,0){\strut{}$P / P_{\rm crit}$}}%
      \put(190,1001){\makebox(0,0)[r]{\strut{}$-60$}}%
      \put(190,1280){\makebox(0,0)[r]{\strut{}$-40$}}%
      \put(190,1559){\makebox(0,0)[r]{\strut{}$-20$}}%
      \put(190,1838){\makebox(0,0)[r]{\strut{}$0$}}%
      \put(190,2117){\makebox(0,0)[r]{\strut{}$20$}}%
      \put(190,2396){\makebox(0,0)[r]{\strut{}$40$}}%
      \put(190,2675){\makebox(0,0)[r]{\strut{}$60$}}%
      \put(-272,1838){\rotatebox{-270}{\makebox(0,0){\strut{}$\phi [^\circ]$}}}%
    }%
    \gplgaddtomacro\gplbacktext{%
      \csname LTb\endcsname
      \put(3326,3110){\makebox(0,0)[l]{\strut{}$Q/\Qd = 0.1$}}%
    }%
    \gplgaddtomacro\gplfronttext{%
      \csname LTb\endcsname
      \put(2569,595){\makebox(0,0){\strut{}1}}%
      \put(3379,595){\makebox(0,0){\strut{}1.5}}%
      \put(3953,595){\makebox(0,0){\strut{}2}}%
      \put(4399,595){\makebox(0,0){\strut{}2.5}}%
      \put(3666,375){\makebox(0,0){\strut{}$P / P_{\rm crit}$}}%
    }%
    \gplgaddtomacro\gplbacktext{%
      \csname LTb\endcsname
      \put(5644,3110){\makebox(0,0)[l]{\strut{}$Q/\Qd = 0.3$}}%
    }%
    \gplgaddtomacro\gplfronttext{%
      \csname LTb\endcsname
      \put(4887,595){\makebox(0,0){\strut{}1}}%
      \put(5697,595){\makebox(0,0){\strut{}1.5}}%
      \put(6271,595){\makebox(0,0){\strut{}2}}%
      \put(6717,595){\makebox(0,0){\strut{}2.5}}%
      \put(5984,375){\makebox(0,0){\strut{}$P / P_{\rm crit}$}}%
    }%
    \gplgaddtomacro\gplbacktext{%
      \csname LTb\endcsname
      \put(7962,3110){\makebox(0,0)[l]{\strut{}$Q/\Qd = 1$}}%
    }%
    \gplgaddtomacro\gplfronttext{%
      \csname LTb\endcsname
      \put(7206,595){\makebox(0,0){\strut{}1}}%
      \put(8016,595){\makebox(0,0){\strut{}1.5}}%
      \put(8590,595){\makebox(0,0){\strut{}2}}%
      \put(9036,595){\makebox(0,0){\strut{}2.5}}%
      \put(8303,375){\makebox(0,0){\strut{}$P / P_{\rm crit}$}}%
      \put(9696,791){\makebox(0,0)[l]{\strut{}1}}%
      \put(9696,1692){\makebox(0,0)[l]{\strut{}2}}%
      \put(9696,2220){\makebox(0,0)[l]{\strut{}3}}%
      \put(9696,2594){\makebox(0,0)[l]{\strut{}4}}%
      \put(9696,2885){\makebox(0,0)[l]{\strut{}5}}%
    }%
    \gplbacktext
    \put(0,0){\includegraphics{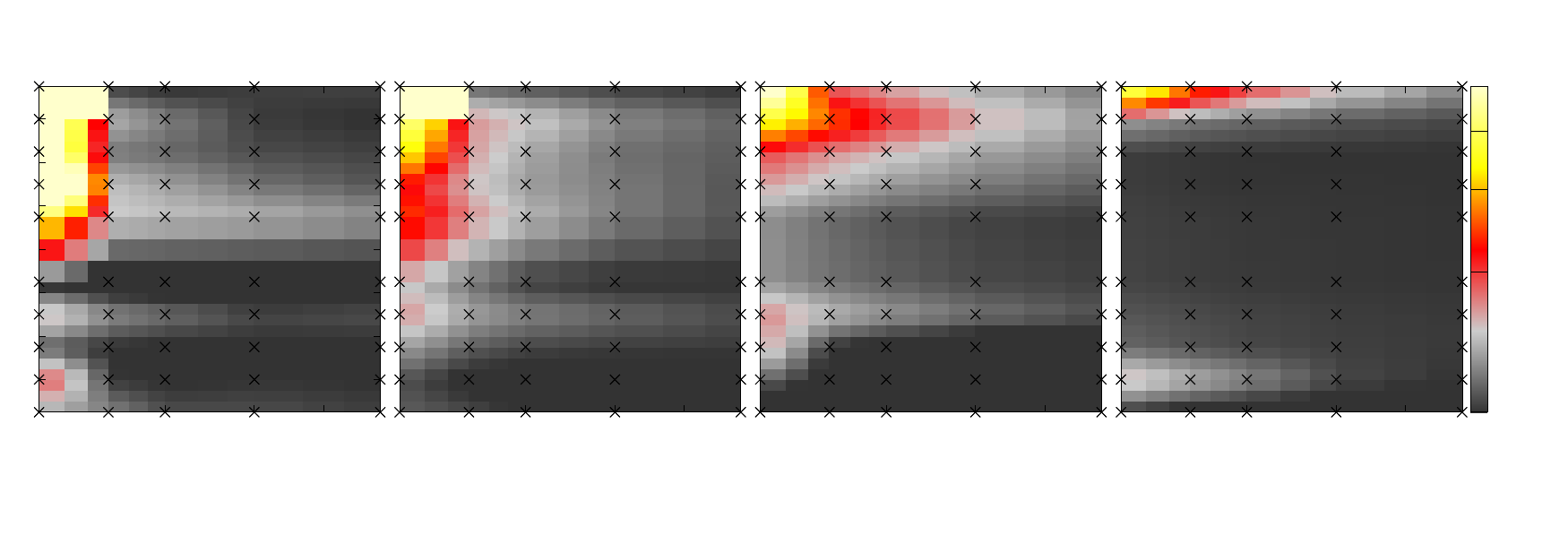}}%
    \gplfronttext
  \end{picture}%
\endgroup

%% file: period.tex
\begingroup
  \makeatletter
  \providecommand\color[2][]{%
    \GenericError{(gnuplot) \space\space\space\@spaces}{%
      Package color not loaded in conjunction with
      terminal option `colourtext'%
    }{See the gnuplot documentation for explanation.%
    }{Either use 'blacktext' in gnuplot or load the package
      color.sty in LaTeX.}%
    \renewcommand\color[2][]{}%
  }%
  \providecommand\includegraphics[2][]{%
    \GenericError{(gnuplot) \space\space\space\@spaces}{%
      Package graphicx or graphics not loaded%
    }{See the gnuplot documentation for explanation.%
    }{The gnuplot epslatex terminal needs graphicx.sty or graphics.sty.}%
    \renewcommand\includegraphics[2][]{}%
  }%
  \providecommand\rotatebox[2]{#2}%
  \@ifundefined{ifGPcolor}{%
    \newif\ifGPcolor
    \GPcolorfalse
  }{}%
  \@ifundefined{ifGPblacktext}{%
    \newif\ifGPblacktext
    \GPblacktexttrue
  }{}%
  \let\gplgaddtomacro\g@addto@macro
  \gdef\gplbacktext{}%
  \gdef\gplfronttext{}%
  \makeatother
  \ifGPblacktext
    \def\colorrgb#1{}%
    \def\colorgray#1{}%
  \else
    \ifGPcolor
      \def\colorrgb#1{\color[rgb]{#1}}%
      \def\colorgray#1{\color[gray]{#1}}%
      \expandafter\def\csname LTw\endcsname{\color{white}}%
      \expandafter\def\csname LTb\endcsname{\color{black}}%
      \expandafter\def\csname LTa\endcsname{\color{black}}%
      \expandafter\def\csname LT0\endcsname{\color[rgb]{1,0,0}}%
      \expandafter\def\csname LT1\endcsname{\color[rgb]{0,1,0}}%
      \expandafter\def\csname LT2\endcsname{\color[rgb]{0,0,1}}%
      \expandafter\def\csname LT3\endcsname{\color[rgb]{1,0,1}}%
      \expandafter\def\csname LT4\endcsname{\color[rgb]{0,1,1}}%
      \expandafter\def\csname LT5\endcsname{\color[rgb]{1,1,0}}%
      \expandafter\def\csname LT6\endcsname{\color[rgb]{0,0,0}}%
      \expandafter\def\csname LT7\endcsname{\color[rgb]{1,0.3,0}}%
      \expandafter\def\csname LT8\endcsname{\color[rgb]{0.5,0.5,0.5}}%
    \else
      \def\colorrgb#1{\color{black}}%
      \def\colorgray#1{\color[gray]{#1}}%
      \expandafter\def\csname LTw\endcsname{\color{white}}%
      \expandafter\def\csname LTb\endcsname{\color{black}}%
      \expandafter\def\csname LTa\endcsname{\color{black}}%
      \expandafter\def\csname LT0\endcsname{\color{black}}%
      \expandafter\def\csname LT1\endcsname{\color{black}}%
      \expandafter\def\csname LT2\endcsname{\color{black}}%
      \expandafter\def\csname LT3\endcsname{\color{black}}%
      \expandafter\def\csname LT4\endcsname{\color{black}}%
      \expandafter\def\csname LT5\endcsname{\color{black}}%
      \expandafter\def\csname LT6\endcsname{\color{black}}%
      \expandafter\def\csname LT7\endcsname{\color{black}}%
      \expandafter\def\csname LT8\endcsname{\color{black}}%
    \fi
  \fi
    \setlength{\unitlength}{0.0500bp}%
    \ifx\gptboxheight\undefined%
      \newlength{\gptboxheight}%
      \newlength{\gptboxwidth}%
      \newsavebox{\gptboxtext}%
    \fi%
    \setlength{\fboxrule}{0.5pt}%
    \setlength{\fboxsep}{1pt}%
\begin{picture}(10080.00,3456.00)%
    \gplgaddtomacro\gplbacktext{%
      \csname LTb\endcsname
      \put(1008,3110){\makebox(0,0)[l]{\strut{}$Q / \Qd  = 0.03$}}%
    }%
    \gplgaddtomacro\gplfronttext{%
      \csname LTb\endcsname
      \put(251,595){\makebox(0,0){\strut{}1}}%
      \put(640,595){\makebox(0,0){\strut{}2}}%
      \put(867,595){\makebox(0,0){\strut{}}}%
      \put(1029,595){\makebox(0,0){\strut{}}}%
      \put(1154,595){\makebox(0,0){\strut{}5}}%
      \put(1256,595){\makebox(0,0){\strut{}}}%
      \put(1343,595){\makebox(0,0){\strut{}}}%
      \put(1417,595){\makebox(0,0){\strut{}}}%
      \put(1483,595){\makebox(0,0){\strut{}}}%
      \put(1542,595){\makebox(0,0){\strut{}10}}%
      \put(1931,595){\makebox(0,0){\strut{}20}}%
      \put(2159,595){\makebox(0,0){\strut{}}}%
      \put(2320,595){\makebox(0,0){\strut{}}}%
      \put(2445,595){\makebox(0,0){\strut{}\hskip-5pt 50}}%
      \put(1348,375){\makebox(0,0){\strut{}$P / P_{\rm crit}$}}%
      \put(190,1001){\makebox(0,0)[r]{\strut{}$-60$}}%
      \put(190,1280){\makebox(0,0)[r]{\strut{}$-40$}}%
      \put(190,1559){\makebox(0,0)[r]{\strut{}$-20$}}%
      \put(190,1838){\makebox(0,0)[r]{\strut{}$0$}}%
      \put(190,2117){\makebox(0,0)[r]{\strut{}$20$}}%
      \put(190,2396){\makebox(0,0)[r]{\strut{}$40$}}%
      \put(190,2675){\makebox(0,0)[r]{\strut{}$60$}}%
      \put(-272,1838){\rotatebox{-270}{\makebox(0,0){\strut{}$\phi [^\circ]$}}}%
    }%
    \gplgaddtomacro\gplbacktext{%
      \csname LTb\endcsname
      \put(3326,3110){\makebox(0,0)[l]{\strut{}$Q/\Qd = 0.1$}}%
    }%
    \gplgaddtomacro\gplfronttext{%
      \csname LTb\endcsname
      \put(2569,595){\makebox(0,0){\strut{}1}}%
      \put(2958,595){\makebox(0,0){\strut{}2}}%
      \put(3185,595){\makebox(0,0){\strut{}}}%
      \put(3347,595){\makebox(0,0){\strut{}}}%
      \put(3472,595){\makebox(0,0){\strut{}5}}%
      \put(3574,595){\makebox(0,0){\strut{}}}%
      \put(3661,595){\makebox(0,0){\strut{}}}%
      \put(3735,595){\makebox(0,0){\strut{}}}%
      \put(3801,595){\makebox(0,0){\strut{}}}%
      \put(3860,595){\makebox(0,0){\strut{}10}}%
      \put(4249,595){\makebox(0,0){\strut{}20}}%
      \put(4477,595){\makebox(0,0){\strut{}}}%
      \put(4638,595){\makebox(0,0){\strut{}}}%
      \put(4763,595){\makebox(0,0){\strut{}\hskip-5pt 50}}%
      \put(3666,375){\makebox(0,0){\strut{}$P / P_{\rm crit}$}}%
    }%
    \gplgaddtomacro\gplbacktext{%
      \csname LTb\endcsname
      \put(5644,3110){\makebox(0,0)[l]{\strut{}$Q/\Qd = 0.3$}}%
    }%
    \gplgaddtomacro\gplfronttext{%
      \csname LTb\endcsname
      \put(4887,595){\makebox(0,0){\strut{}1}}%
      \put(5276,595){\makebox(0,0){\strut{}2}}%
      \put(5503,595){\makebox(0,0){\strut{}}}%
      \put(5665,595){\makebox(0,0){\strut{}}}%
      \put(5790,595){\makebox(0,0){\strut{}5}}%
      \put(5892,595){\makebox(0,0){\strut{}}}%
      \put(5979,595){\makebox(0,0){\strut{}}}%
      \put(6053,595){\makebox(0,0){\strut{}}}%
      \put(6119,595){\makebox(0,0){\strut{}}}%
      \put(6178,595){\makebox(0,0){\strut{}10}}%
      \put(6567,595){\makebox(0,0){\strut{}20}}%
      \put(6795,595){\makebox(0,0){\strut{}}}%
      \put(6956,595){\makebox(0,0){\strut{}}}%
      \put(7081,595){\makebox(0,0){\strut{}\hskip-5pt 50}}%
      \put(5984,375){\makebox(0,0){\strut{}$P / P_{\rm crit}$}}%
    }%
    \gplgaddtomacro\gplbacktext{%
      \csname LTb\endcsname
      \put(7962,3110){\makebox(0,0)[l]{\strut{}$Q/\Qd = 1$}}%
    }%
    \gplgaddtomacro\gplfronttext{%
      \csname LTb\endcsname
      \put(7206,595){\makebox(0,0){\strut{}1}}%
      \put(7595,595){\makebox(0,0){\strut{}2}}%
      \put(7822,595){\makebox(0,0){\strut{}}}%
      \put(7984,595){\makebox(0,0){\strut{}}}%
      \put(8109,595){\makebox(0,0){\strut{}5}}%
      \put(8211,595){\makebox(0,0){\strut{}}}%
      \put(8298,595){\makebox(0,0){\strut{}}}%
      \put(8372,595){\makebox(0,0){\strut{}}}%
      \put(8438,595){\makebox(0,0){\strut{}}}%
      \put(8497,595){\makebox(0,0){\strut{}10}}%
      \put(8886,595){\makebox(0,0){\strut{}20}}%
      \put(9114,595){\makebox(0,0){\strut{}}}%
      \put(9275,595){\makebox(0,0){\strut{}}}%
      \put(9400,595){\makebox(0,0){\strut{}\hskip-5pt 50}}%
      \put(8303,375){\makebox(0,0){\strut{}$P / P_{\rm crit}$}}%
      \put(9696,791){\makebox(0,0)[l]{\strut{}$-2$}}%
      \put(9696,1052){\makebox(0,0)[l]{\strut{}$-1.5$}}%
      \put(9696,1314){\makebox(0,0)[l]{\strut{}$-1$}}%
      \put(9696,1576){\makebox(0,0)[l]{\strut{}$-0.5$}}%
      \put(9696,1838){\makebox(0,0)[l]{\strut{}$0$}}%
      \put(9696,2099){\makebox(0,0)[l]{\strut{}$0.5$}}%
      \put(9696,2361){\makebox(0,0)[l]{\strut{}$1$}}%
      \put(9696,2623){\makebox(0,0)[l]{\strut{}$1.5$}}%
      \put(9696,2885){\makebox(0,0)[l]{\strut{}$2$}}%
      \put(10159,1838){\rotatebox{-270}{\makebox(0,0){\strut{}$\Delta\omega$ [rev/day]}}}%
    }%
    \gplbacktext
    \put(0,0){\includegraphics{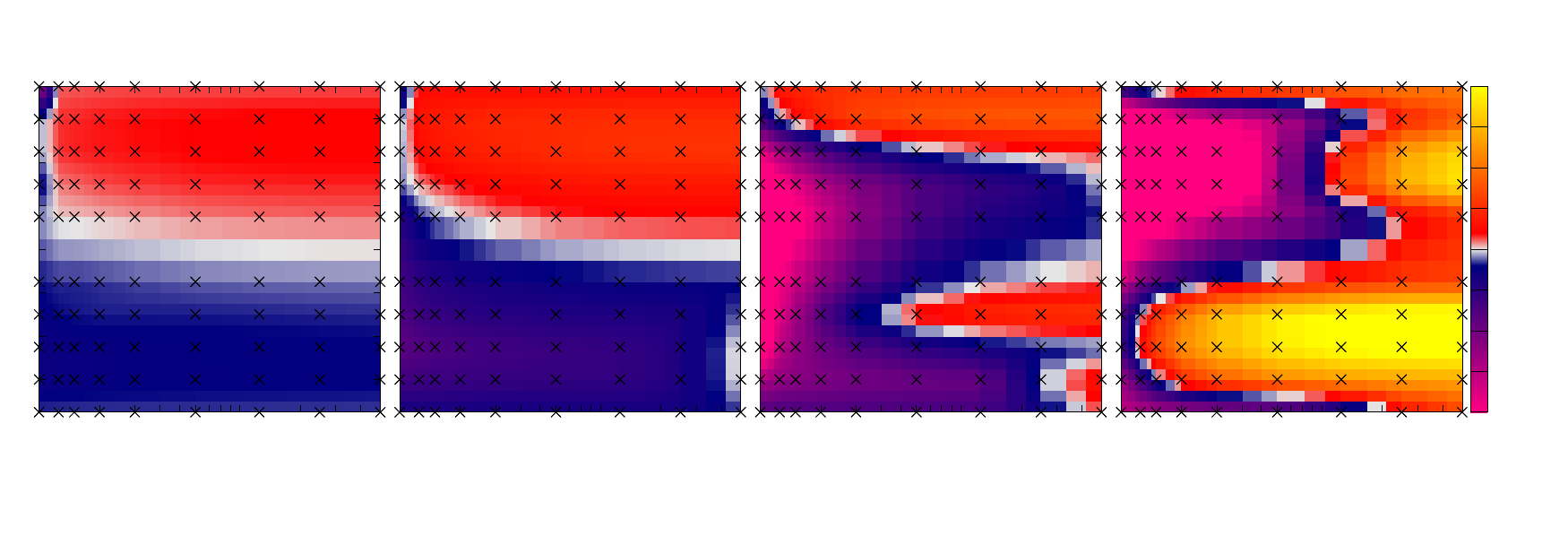}}%
    \gplfronttext
  \end{picture}%
\endgroup

%% file: gamma.tex
\begingroup
  \makeatletter
  \providecommand\color[2][]{%
    \GenericError{(gnuplot) \space\space\space\@spaces}{%
      Package color not loaded in conjunction with
      terminal option `colourtext'%
    }{See the gnuplot documentation for explanation.%
    }{Either use 'blacktext' in gnuplot or load the package
      color.sty in LaTeX.}%
    \renewcommand\color[2][]{}%
  }%
  \providecommand\includegraphics[2][]{%
    \GenericError{(gnuplot) \space\space\space\@spaces}{%
      Package graphicx or graphics not loaded%
    }{See the gnuplot documentation for explanation.%
    }{The gnuplot epslatex terminal needs graphicx.sty or graphics.sty.}%
    \renewcommand\includegraphics[2][]{}%
  }%
  \providecommand\rotatebox[2]{#2}%
  \@ifundefined{ifGPcolor}{%
    \newif\ifGPcolor
    \GPcolorfalse
  }{}%
  \@ifundefined{ifGPblacktext}{%
    \newif\ifGPblacktext
    \GPblacktexttrue
  }{}%
  \let\gplgaddtomacro\g@addto@macro
  \gdef\gplbacktext{}%
  \gdef\gplfronttext{}%
  \makeatother
  \ifGPblacktext
    \def\colorrgb#1{}%
    \def\colorgray#1{}%
  \else
    \ifGPcolor
      \def\colorrgb#1{\color[rgb]{#1}}%
      \def\colorgray#1{\color[gray]{#1}}%
      \expandafter\def\csname LTw\endcsname{\color{white}}%
      \expandafter\def\csname LTb\endcsname{\color{black}}%
      \expandafter\def\csname LTa\endcsname{\color{black}}%
      \expandafter\def\csname LT0\endcsname{\color[rgb]{1,0,0}}%
      \expandafter\def\csname LT1\endcsname{\color[rgb]{0,1,0}}%
      \expandafter\def\csname LT2\endcsname{\color[rgb]{0,0,1}}%
      \expandafter\def\csname LT3\endcsname{\color[rgb]{1,0,1}}%
      \expandafter\def\csname LT4\endcsname{\color[rgb]{0,1,1}}%
      \expandafter\def\csname LT5\endcsname{\color[rgb]{1,1,0}}%
      \expandafter\def\csname LT6\endcsname{\color[rgb]{0,0,0}}%
      \expandafter\def\csname LT7\endcsname{\color[rgb]{1,0.3,0}}%
      \expandafter\def\csname LT8\endcsname{\color[rgb]{0.5,0.5,0.5}}%
    \else
      \def\colorrgb#1{\color{black}}%
      \def\colorgray#1{\color[gray]{#1}}%
      \expandafter\def\csname LTw\endcsname{\color{white}}%
      \expandafter\def\csname LTb\endcsname{\color{black}}%
      \expandafter\def\csname LTa\endcsname{\color{black}}%
      \expandafter\def\csname LT0\endcsname{\color{black}}%
      \expandafter\def\csname LT1\endcsname{\color{black}}%
      \expandafter\def\csname LT2\endcsname{\color{black}}%
      \expandafter\def\csname LT3\endcsname{\color{black}}%
      \expandafter\def\csname LT4\endcsname{\color{black}}%
      \expandafter\def\csname LT5\endcsname{\color{black}}%
      \expandafter\def\csname LT6\endcsname{\color{black}}%
      \expandafter\def\csname LT7\endcsname{\color{black}}%
      \expandafter\def\csname LT8\endcsname{\color{black}}%
    \fi
  \fi
    \setlength{\unitlength}{0.0500bp}%
    \ifx\gptboxheight\undefined%
      \newlength{\gptboxheight}%
      \newlength{\gptboxwidth}%
      \newsavebox{\gptboxtext}%
    \fi%
    \setlength{\fboxrule}{0.5pt}%
    \setlength{\fboxsep}{1pt}%
\begin{picture}(10080.00,3456.00)%
    \gplgaddtomacro\gplbacktext{%
      \csname LTb\endcsname
      \put(1008,3110){\makebox(0,0)[l]{\strut{}$Q / \Qd  = 0.03$}}%
    }%
    \gplgaddtomacro\gplfronttext{%
      \csname LTb\endcsname
      \put(251,595){\makebox(0,0){\strut{}1}}%
      \put(640,595){\makebox(0,0){\strut{}2}}%
      \put(867,595){\makebox(0,0){\strut{}}}%
      \put(1029,595){\makebox(0,0){\strut{}}}%
      \put(1154,595){\makebox(0,0){\strut{}5}}%
      \put(1256,595){\makebox(0,0){\strut{}}}%
      \put(1343,595){\makebox(0,0){\strut{}}}%
      \put(1417,595){\makebox(0,0){\strut{}}}%
      \put(1483,595){\makebox(0,0){\strut{}}}%
      \put(1542,595){\makebox(0,0){\strut{}10}}%
      \put(1931,595){\makebox(0,0){\strut{}20}}%
      \put(2159,595){\makebox(0,0){\strut{}}}%
      \put(2320,595){\makebox(0,0){\strut{}}}%
      \put(2445,595){\makebox(0,0){\strut{}\hskip-5pt 50}}%
      \put(1348,375){\makebox(0,0){\strut{}$P / P_{\rm crit}$}}%
      \put(190,1001){\makebox(0,0)[r]{\strut{}$-60$}}%
      \put(190,1280){\makebox(0,0)[r]{\strut{}$-40$}}%
      \put(190,1559){\makebox(0,0)[r]{\strut{}$-20$}}%
      \put(190,1838){\makebox(0,0)[r]{\strut{}$0$}}%
      \put(190,2117){\makebox(0,0)[r]{\strut{}$20$}}%
      \put(190,2396){\makebox(0,0)[r]{\strut{}$40$}}%
      \put(190,2675){\makebox(0,0)[r]{\strut{}$60$}}%
      \put(-272,1838){\rotatebox{-270}{\makebox(0,0){\strut{}$\phi [^\circ]$}}}%
    }%
    \gplgaddtomacro\gplbacktext{%
      \csname LTb\endcsname
      \put(3326,3110){\makebox(0,0)[l]{\strut{}$Q/\Qd = 0.1$}}%
    }%
    \gplgaddtomacro\gplfronttext{%
      \csname LTb\endcsname
      \put(2569,595){\makebox(0,0){\strut{}1}}%
      \put(2958,595){\makebox(0,0){\strut{}2}}%
      \put(3185,595){\makebox(0,0){\strut{}}}%
      \put(3347,595){\makebox(0,0){\strut{}}}%
      \put(3472,595){\makebox(0,0){\strut{}5}}%
      \put(3574,595){\makebox(0,0){\strut{}}}%
      \put(3661,595){\makebox(0,0){\strut{}}}%
      \put(3735,595){\makebox(0,0){\strut{}}}%
      \put(3801,595){\makebox(0,0){\strut{}}}%
      \put(3860,595){\makebox(0,0){\strut{}10}}%
      \put(4249,595){\makebox(0,0){\strut{}20}}%
      \put(4477,595){\makebox(0,0){\strut{}}}%
      \put(4638,595){\makebox(0,0){\strut{}}}%
      \put(4763,595){\makebox(0,0){\strut{}\hskip-5pt 50}}%
      \put(3666,375){\makebox(0,0){\strut{}$P / P_{\rm crit}$}}%
    }%
    \gplgaddtomacro\gplbacktext{%
      \csname LTb\endcsname
      \put(5644,3110){\makebox(0,0)[l]{\strut{}$Q/\Qd = 0.3$}}%
    }%
    \gplgaddtomacro\gplfronttext{%
      \csname LTb\endcsname
      \put(4887,595){\makebox(0,0){\strut{}1}}%
      \put(5276,595){\makebox(0,0){\strut{}2}}%
      \put(5503,595){\makebox(0,0){\strut{}}}%
      \put(5665,595){\makebox(0,0){\strut{}}}%
      \put(5790,595){\makebox(0,0){\strut{}5}}%
      \put(5892,595){\makebox(0,0){\strut{}}}%
      \put(5979,595){\makebox(0,0){\strut{}}}%
      \put(6053,595){\makebox(0,0){\strut{}}}%
      \put(6119,595){\makebox(0,0){\strut{}}}%
      \put(6178,595){\makebox(0,0){\strut{}10}}%
      \put(6567,595){\makebox(0,0){\strut{}20}}%
      \put(6795,595){\makebox(0,0){\strut{}}}%
      \put(6956,595){\makebox(0,0){\strut{}}}%
      \put(7081,595){\makebox(0,0){\strut{}\hskip-5pt 50}}%
      \put(5984,375){\makebox(0,0){\strut{}$P / P_{\rm crit}$}}%
    }%
    \gplgaddtomacro\gplbacktext{%
      \csname LTb\endcsname
      \put(7962,3110){\makebox(0,0)[l]{\strut{}$Q/\Qd = 1$}}%
    }%
    \gplgaddtomacro\gplfronttext{%
      \csname LTb\endcsname
      \put(7206,595){\makebox(0,0){\strut{}1}}%
      \put(7595,595){\makebox(0,0){\strut{}2}}%
      \put(7822,595){\makebox(0,0){\strut{}}}%
      \put(7984,595){\makebox(0,0){\strut{}}}%
      \put(8109,595){\makebox(0,0){\strut{}5}}%
      \put(8211,595){\makebox(0,0){\strut{}}}%
      \put(8298,595){\makebox(0,0){\strut{}}}%
      \put(8372,595){\makebox(0,0){\strut{}}}%
      \put(8438,595){\makebox(0,0){\strut{}}}%
      \put(8497,595){\makebox(0,0){\strut{}10}}%
      \put(8886,595){\makebox(0,0){\strut{}20}}%
      \put(9114,595){\makebox(0,0){\strut{}}}%
      \put(9275,595){\makebox(0,0){\strut{}}}%
      \put(9400,595){\makebox(0,0){\strut{}\hskip-5pt 50}}%
      \put(8303,375){\makebox(0,0){\strut{}$P / P_{\rm crit}$}}%
      \put(9696,791){\makebox(0,0)[l]{\strut{}$-1$}}%
      \put(9696,1314){\makebox(0,0)[l]{\strut{}$-0.5$}}%
      \put(9696,1838){\makebox(0,0)[l]{\strut{}$0$}}%
      \put(9696,2361){\makebox(0,0)[l]{\strut{}$0.5$}}%
      \put(9696,2885){\makebox(0,0)[l]{\strut{}$1$}}%
      \put(10027,1838){\rotatebox{-270}{\makebox(0,0){\strut{}$\gamma$}}}%
    }%
    \gplbacktext
    \put(0,0){\includegraphics{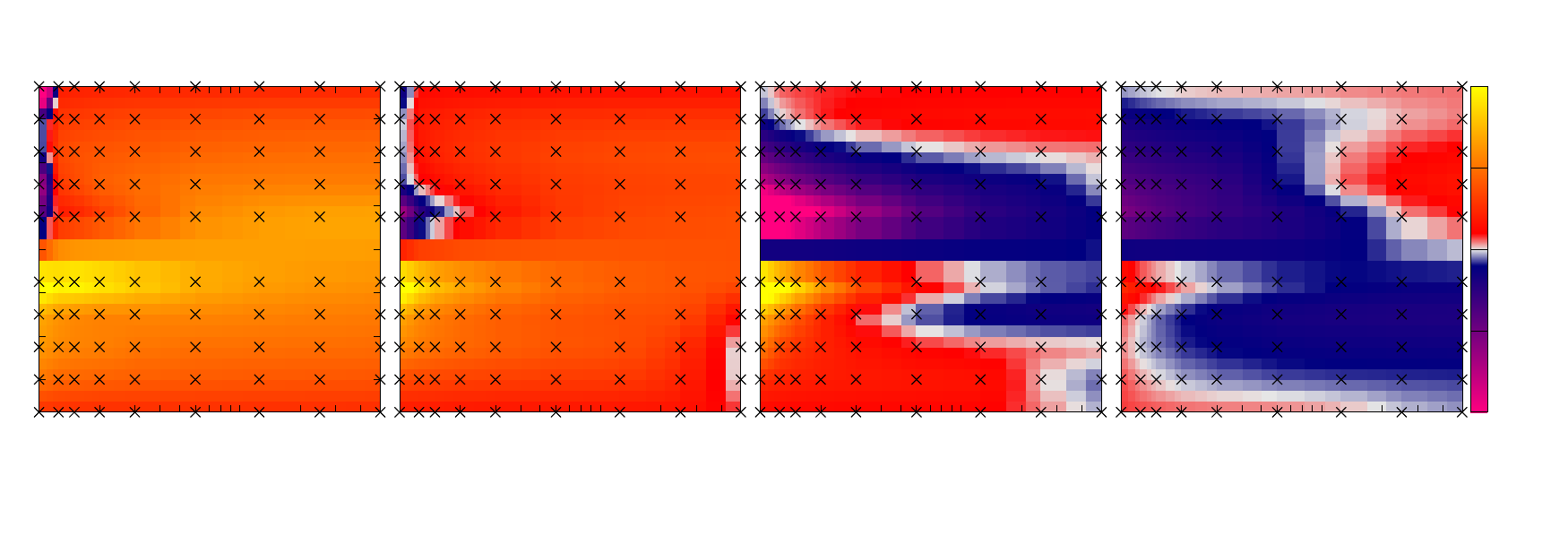}}%
    \gplfronttext
  \end{picture}%
\endgroup

%% file: integral_period.tex
\begingroup
  \makeatletter
  \providecommand\color[2][]{%
    \GenericError{(gnuplot) \space\space\space\@spaces}{%
      Package color not loaded in conjunction with
      terminal option `colourtext'%
    }{See the gnuplot documentation for explanation.%
    }{Either use 'blacktext' in gnuplot or load the package
      color.sty in LaTeX.}%
    \renewcommand\color[2][]{}%
  }%
  \providecommand\includegraphics[2][]{%
    \GenericError{(gnuplot) \space\space\space\@spaces}{%
      Package graphicx or graphics not loaded%
    }{See the gnuplot documentation for explanation.%
    }{The gnuplot epslatex terminal needs graphicx.sty or graphics.sty.}%
    \renewcommand\includegraphics[2][]{}%
  }%
  \providecommand\rotatebox[2]{#2}%
  \@ifundefined{ifGPcolor}{%
    \newif\ifGPcolor
    \GPcolorfalse
  }{}%
  \@ifundefined{ifGPblacktext}{%
    \newif\ifGPblacktext
    \GPblacktexttrue
  }{}%
  \let\gplgaddtomacro\g@addto@macro
  \gdef\gplbacktext{}%
  \gdef\gplfronttext{}%
  \makeatother
  \ifGPblacktext
    \def\colorrgb#1{}%
    \def\colorgray#1{}%
  \else
    \ifGPcolor
      \def\colorrgb#1{\color[rgb]{#1}}%
      \def\colorgray#1{\color[gray]{#1}}%
      \expandafter\def\csname LTw\endcsname{\color{white}}%
      \expandafter\def\csname LTb\endcsname{\color{black}}%
      \expandafter\def\csname LTa\endcsname{\color{black}}%
      \expandafter\def\csname LT0\endcsname{\color[rgb]{1,0,0}}%
      \expandafter\def\csname LT1\endcsname{\color[rgb]{0,1,0}}%
      \expandafter\def\csname LT2\endcsname{\color[rgb]{0,0,1}}%
      \expandafter\def\csname LT3\endcsname{\color[rgb]{1,0,1}}%
      \expandafter\def\csname LT4\endcsname{\color[rgb]{0,1,1}}%
      \expandafter\def\csname LT5\endcsname{\color[rgb]{1,1,0}}%
      \expandafter\def\csname LT6\endcsname{\color[rgb]{0,0,0}}%
      \expandafter\def\csname LT7\endcsname{\color[rgb]{1,0.3,0}}%
      \expandafter\def\csname LT8\endcsname{\color[rgb]{0.5,0.5,0.5}}%
    \else
      \def\colorrgb#1{\color{black}}%
      \def\colorgray#1{\color[gray]{#1}}%
      \expandafter\def\csname LTw\endcsname{\color{white}}%
      \expandafter\def\csname LTb\endcsname{\color{black}}%
      \expandafter\def\csname LTa\endcsname{\color{black}}%
      \expandafter\def\csname LT0\endcsname{\color{black}}%
      \expandafter\def\csname LT1\endcsname{\color{black}}%
      \expandafter\def\csname LT2\endcsname{\color{black}}%
      \expandafter\def\csname LT3\endcsname{\color{black}}%
      \expandafter\def\csname LT4\endcsname{\color{black}}%
      \expandafter\def\csname LT5\endcsname{\color{black}}%
      \expandafter\def\csname LT6\endcsname{\color{black}}%
      \expandafter\def\csname LT7\endcsname{\color{black}}%
      \expandafter\def\csname LT8\endcsname{\color{black}}%
    \fi
  \fi
    \setlength{\unitlength}{0.0500bp}%
    \ifx\gptboxheight\undefined%
      \newlength{\gptboxheight}%
      \newlength{\gptboxwidth}%
      \newsavebox{\gptboxtext}%
    \fi%
    \setlength{\fboxrule}{0.5pt}%
    \setlength{\fboxsep}{1pt}%
\begin{picture}(4608.00,3456.00)%
    \gplgaddtomacro\gplbacktext{%
    }%
    \gplgaddtomacro\gplfronttext{%
      \csname LTb\endcsname
      \put(695,648){\makebox(0,0){\strut{}1}}%
      \put(1265,648){\makebox(0,0){\strut{}2}}%
      \put(1599,648){\makebox(0,0){\strut{}}}%
      \put(1836,648){\makebox(0,0){\strut{}}}%
      \put(2019,648){\makebox(0,0){\strut{}5}}%
      \put(2169,648){\makebox(0,0){\strut{}}}%
      \put(2296,648){\makebox(0,0){\strut{}}}%
      \put(2405,648){\makebox(0,0){\strut{}}}%
      \put(2502,648){\makebox(0,0){\strut{}}}%
      \put(2589,648){\makebox(0,0){\strut{}10}}%
      \put(3159,648){\makebox(0,0){\strut{}20}}%
      \put(3493,648){\makebox(0,0){\strut{}}}%
      \put(3729,648){\makebox(0,0){\strut{}}}%
      \put(3913,648){\makebox(0,0){\strut{}50}}%
      \put(2304,428){\makebox(0,0){\strut{}$P / P_{\rm crit}$}}%
      \put(594,920){\makebox(0,0)[r]{\strut{}$300$}}%
      \put(594,1277){\makebox(0,0)[r]{\strut{}$400$}}%
      \put(594,1635){\makebox(0,0)[r]{\strut{}$500$}}%
      \put(594,1991){\makebox(0,0)[r]{\strut{}$600$}}%
      \put(594,2349){\makebox(0,0)[r]{\strut{}$700$}}%
      \put(594,2706){\makebox(0,0)[r]{\strut{}$800$}}%
      \put(132,1838){\rotatebox{-270}{\makebox(0,0){\strut{}$d_{\rm imp}$\,[m]}}}%
      \put(4285,791){\makebox(0,0)[l]{\strut{}$-1$}}%
      \put(4285,1314){\makebox(0,0)[l]{\strut{}$-0.5$}}%
      \put(4285,1838){\makebox(0,0)[l]{\strut{}$0$}}%
      \put(4285,2361){\makebox(0,0)[l]{\strut{}$0.5$}}%
      \put(4285,2885){\makebox(0,0)[l]{\strut{}$1$}}%
      \put(4748,1838){\rotatebox{-270}{\makebox(0,0){\strut{}$\Delta \omega$ [rev/day]}}}%
    }%
    \gplbacktext
    \put(0,0){\includegraphics{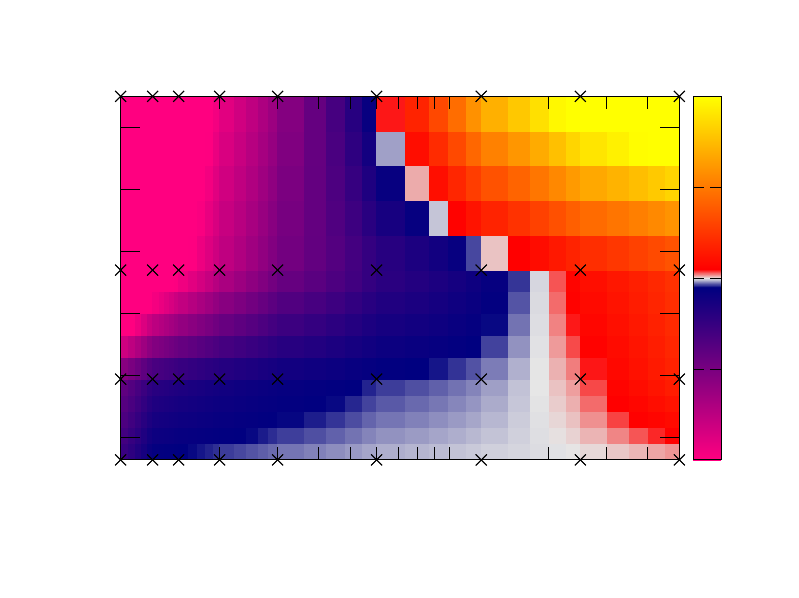}}%
    \gplfronttext
  \end{picture}%
\endgroup

%% file: integral_mass.tex
\begingroup
  \makeatletter
  \providecommand\color[2][]{%
    \GenericError{(gnuplot) \space\space\space\@spaces}{%
      Package color not loaded in conjunction with
      terminal option `colourtext'%
    }{See the gnuplot documentation for explanation.%
    }{Either use 'blacktext' in gnuplot or load the package
      color.sty in LaTeX.}%
    \renewcommand\color[2][]{}%
  }%
  \providecommand\includegraphics[2][]{%
    \GenericError{(gnuplot) \space\space\space\@spaces}{%
      Package graphicx or graphics not loaded%
    }{See the gnuplot documentation for explanation.%
    }{The gnuplot epslatex terminal needs graphicx.sty or graphics.sty.}%
    \renewcommand\includegraphics[2][]{}%
  }%
  \providecommand\rotatebox[2]{#2}%
  \@ifundefined{ifGPcolor}{%
    \newif\ifGPcolor
    \GPcolorfalse
  }{}%
  \@ifundefined{ifGPblacktext}{%
    \newif\ifGPblacktext
    \GPblacktexttrue
  }{}%
  \let\gplgaddtomacro\g@addto@macro
  \gdef\gplbacktext{}%
  \gdef\gplfronttext{}%
  \makeatother
  \ifGPblacktext
    \def\colorrgb#1{}%
    \def\colorgray#1{}%
  \else
    \ifGPcolor
      \def\colorrgb#1{\color[rgb]{#1}}%
      \def\colorgray#1{\color[gray]{#1}}%
      \expandafter\def\csname LTw\endcsname{\color{white}}%
      \expandafter\def\csname LTb\endcsname{\color{black}}%
      \expandafter\def\csname LTa\endcsname{\color{black}}%
      \expandafter\def\csname LT0\endcsname{\color[rgb]{1,0,0}}%
      \expandafter\def\csname LT1\endcsname{\color[rgb]{0,1,0}}%
      \expandafter\def\csname LT2\endcsname{\color[rgb]{0,0,1}}%
      \expandafter\def\csname LT3\endcsname{\color[rgb]{1,0,1}}%
      \expandafter\def\csname LT4\endcsname{\color[rgb]{0,1,1}}%
      \expandafter\def\csname LT5\endcsname{\color[rgb]{1,1,0}}%
      \expandafter\def\csname LT6\endcsname{\color[rgb]{0,0,0}}%
      \expandafter\def\csname LT7\endcsname{\color[rgb]{1,0.3,0}}%
      \expandafter\def\csname LT8\endcsname{\color[rgb]{0.5,0.5,0.5}}%
    \else
      \def\colorrgb#1{\color{black}}%
      \def\colorgray#1{\color[gray]{#1}}%
      \expandafter\def\csname LTw\endcsname{\color{white}}%
      \expandafter\def\csname LTb\endcsname{\color{black}}%
      \expandafter\def\csname LTa\endcsname{\color{black}}%
      \expandafter\def\csname LT0\endcsname{\color{black}}%
      \expandafter\def\csname LT1\endcsname{\color{black}}%
      \expandafter\def\csname LT2\endcsname{\color{black}}%
      \expandafter\def\csname LT3\endcsname{\color{black}}%
      \expandafter\def\csname LT4\endcsname{\color{black}}%
      \expandafter\def\csname LT5\endcsname{\color{black}}%
      \expandafter\def\csname LT6\endcsname{\color{black}}%
      \expandafter\def\csname LT7\endcsname{\color{black}}%
      \expandafter\def\csname LT8\endcsname{\color{black}}%
    \fi
  \fi
    \setlength{\unitlength}{0.0500bp}%
    \ifx\gptboxheight\undefined%
      \newlength{\gptboxheight}%
      \newlength{\gptboxwidth}%
      \newsavebox{\gptboxtext}%
    \fi%
    \setlength{\fboxrule}{0.5pt}%
    \setlength{\fboxsep}{1pt}%
\begin{picture}(4608.00,3456.00)%
    \gplgaddtomacro\gplbacktext{%
    }%
    \gplgaddtomacro\gplfronttext{%
      \csname LTb\endcsname
      \put(695,648){\makebox(0,0){\strut{}1}}%
      \put(1500,648){\makebox(0,0){\strut{}1.5}}%
      \put(2304,648){\makebox(0,0){\strut{}2}}%
      \put(3108,648){\makebox(0,0){\strut{}2.5}}%
      \put(2304,428){\makebox(0,0){\strut{}$P / P_{\rm crit}$}}%
      \put(594,920){\makebox(0,0)[r]{\strut{}$300$}}%
      \put(594,1277){\makebox(0,0)[r]{\strut{}$400$}}%
      \put(594,1635){\makebox(0,0)[r]{\strut{}$500$}}%
      \put(594,1991){\makebox(0,0)[r]{\strut{}$600$}}%
      \put(594,2349){\makebox(0,0)[r]{\strut{}$700$}}%
      \put(594,2706){\makebox(0,0)[r]{\strut{}$800$}}%
      \put(4285,791){\makebox(0,0)[l]{\strut{}$1$}}%
      \put(4285,1209){\makebox(0,0)[l]{\strut{}$1.2$}}%
      \put(4285,1628){\makebox(0,0)[l]{\strut{}$1.4$}}%
      \put(4285,2047){\makebox(0,0)[l]{\strut{}$1.6$}}%
      \put(4285,2466){\makebox(0,0)[l]{\strut{}$1.8$}}%
      \put(4285,2884){\makebox(0,0)[l]{\strut{}$2$}}%
      \put(4616,1838){\rotatebox{-270}{\makebox(0,0){\strut{}$\mu$}}}%
    }%
    \gplbacktext
    \put(0,0){\includegraphics{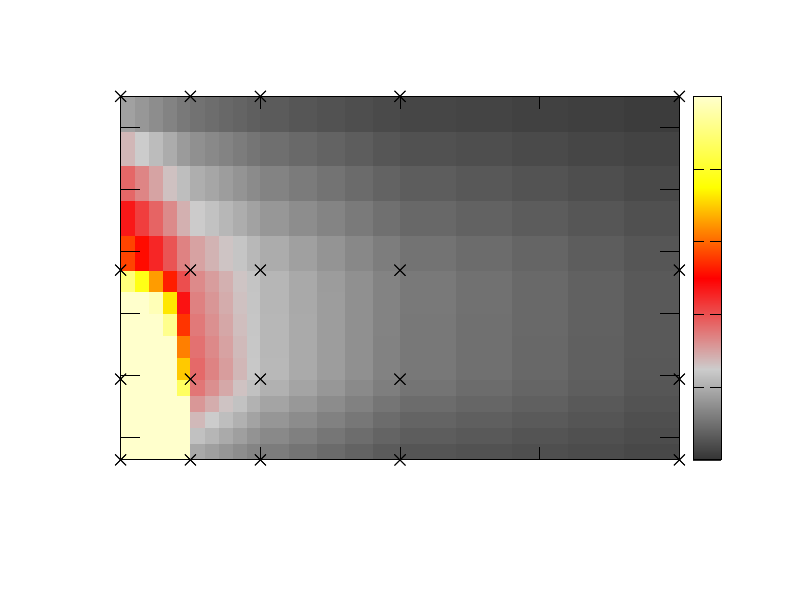}}%
    \gplfronttext
  \end{picture}%
\endgroup